\documentclass[fleqn,usenatbib]{mnras}

\usepackage{newtxtext,newtxmath}

\usepackage[T1]{fontenc}
\usepackage{ae,aecompl}

\usepackage{graphicx}	
\usepackage{amsmath}	
\usepackage{amssymb}	

\def\lsun{\hbox{L$_\odot$}}
\def\msun{\hbox{M$_\odot$}}
\def\mstar{\hbox{M$_\star$}}
\def\t4{\hbox{t$_{\rm 4}$}}

\def\msunyr{\hbox{M$_\odot$yr$^{-1}$}}
\def\dsfr{\hbox{M$_\odot$yr$^{-1}$kpc$^{-2}$}}

\def\cm3{\hbox{cm$^{-3}$}}

\title[Star clusters in the HiPEEC galaxies]{Star cluster formation in the most extreme environments:\\ Insights from the HiPEEC survey}

\author[Adamo et al.]{A.~Adamo$^{1}$\thanks{E-mail: angela.adamo@astro.su.se},
K.~Hollyhead$^{1}$,
M.~Messa$^{2}$,
J.~E.~Ryon$^{3}$,
V.~Bajaj$^{3}$,
A.~Runnholm$^{1}$,\newauthor
S.~Aalto$^{4}$,
D.~Calzetti$^{2}$,
J.~S.~Gallagher$^{5}$,
M. J. Hayes$^{1}$,
J.~M.~D. Kruijssen$^{6}$,
S.~K\"onig$^{4}$,\newauthor
S.~S.~Larsen$^{7}$,
J.~Melinder$^{1}$,
E.~Sabbi$^{3}$,
L.~J.~Smith$^{8}$, and G.~\"Ostlin$^{1}$\\
$^{1}$Department of Astronomy, Oscar Klein Centre, Stockholm University, AlbaNova, Stockholm SE-106 91, Sweden\\
$^{2}$Department of Astronomy, University of Massachusetts, Amherst, MA 01003, USA\\
$^{3}$Space Telescope Science Institute, 3700 San Martin Drive, Baltimore, MD 2121, USA\\
$^{4}$Department of Space, Earth and Environment, Onsala Space Observatory, Chalmers University of Technology, 43992 Onsala, Sweden\\
$^{5}$Department of Astronomy, University of Wisconsin, 475 N. Charter Street, Madison, WI 53706, USA\\
$^{6}$Astronomisches Rechen-Institut, Zentrum f\"ur Astronomie der Universit\"at Heidelberg, M\"onchhofstra\ss e 12-14, 69120 Heidelberg, Germany\\
$^{7}$Department of Astrophysics/IMAPP, Radboud University, PO Box 9010, 6500 GL Nijmegen, The Netherlands\\
$^{8}$Space Telescope Science Institute and European Space Agency, 3700 San Martin Drive, Baltimore, MD 2121, USA
}

\date{Accepted 2020 August 5. Received 2020 August 5; in original form 2020 July 21}

\pubyear{2019}

\begin{document}
\label{firstpage}
\pagerange{\pageref{firstpage}--\pageref{lastpage}}
\maketitle

\begin{abstract}
We present  the  \emph{Hubble imaging Probe of Extreme Environments and Clusters}, HiPEEC survey. We fit HST NUV to NIR broadband and H$\alpha$ fluxes, to derive star cluster ages, masses, extinctions and determine the star formation rate (SFR) of 6 merging galaxies. These systems are excellent laboratories to trace cluster formation under extreme gas physical conditions, rare in the local universe, but typical for star-forming galaxies at cosmic noon. We detect clusters with ages of 1-500 Myr and masses that exceed $10^7$ \msun. The recent cluster formation history and their distribution within the host galaxies suggest that systems like NGC34, NGC1614, NGC4194 are close to their final coalescing phase, while NGC3256, NGC3690, NGC6052 are at an earlier/intermediate stage. A Bayesian analysis of the cluster mass function in the age interval 1-100 Myr provides strong evidence in 4 of the 6 galaxies that an exponentially truncated power law better describes the observed mass distributions. For two galaxies, the fits are inconclusive due to low number statistics. We determine power-law slopes $\beta \sim-1.5$ to $-2.0$, and truncation masses, M$_c$, between $10^6$ and a few times $10^7$ \msun, among the highest values reported in the literature. Advanced mergers have higher M$_c$ than early/intermediate merger stage galaxies, suggesting rapid changes in the dense gas conditions during the merger. We compare the total stellar mass in clusters to the SFR of the galaxy, finding that these systems are among the most efficient environments to form star clusters in the local universe. 
\end{abstract}




\section{Introduction}
\label{sec:intro}

Rare major merger events and the more frequent minor mergers and interactions are key phases for galaxy evolution. In the $\Lambda$CDM cosmological framework, structures grow via hierarchical accretion of increasingly massive systems \citep[e.g.,][among many others]{white91, fa10, RG16, qu17}. In the local universe, observations of interacting/merging galaxies show that these galaxies experience enhancements in their star formation rates (SFRs) per unit mass with respect to control samples of no-interacting galaxies \citep[e.g.][among many others]{ellison13, patton13, knapen15}. It is during these interacting phases that  tidal torques may favour the transport of dense gas towards the centre of the systems fuelling both star formation but also accretion towards nuclear black holes \citep[e.g][and references therein]{koenig13, koenig14, H14}. Interacting/merging systems have higher molecular gas fraction and shorter depletion timescales than average main-sequence galaxies \citep[e.g][for a review]{saintonge12, KE12}. These findings suggest that merger events increase the efficiency at which molecular gas is formed thus resulting in starburst phases that displace these galaxies with respect to the local main sequence of star-forming galaxies \citep[e.g.][among many others]{brinch04,noeske07, elbaz18, popesso19}. A large fraction of these interacting systems have IR luminosities typical of (ultra)luminous IR galaxies, (U)LIRGs with $L_{\rm IR} > 10^{10}$($>10^{12}$) erg/s. Indeed, at least in the local universe, (U)LIRGs are associated with merging/interacting systems and in some cases powerful active galactic nuclei (AGN) \citep[e.g.][]{SM96, bellocchi13}.

In the local universe (U)LIRG systems are rare, but their number density increases with redshift ($z$). At the peak of the cosmic formation history ($z\sim2$), they represent the bulk of the galaxy population on the main sequence  \citep[e.g.][]{rodighiero11, magdis12, schreiber15}. High-redshift (U)LIRGs share many of the properties of their rare local analogues, like short depletion time scales and higher gas fractions \citep[][]{tacconi18}. However, high-$z$ (U)LIRGs are typically rotating disk galaxies, dominated by massive star-forming clumps, showing no sign of disturbed morphology or perturbed kinematics indicative of external interactions \citep{wis15}. The conditions for star formation have dramatically changed between cosmic noon and the local universe. Hence, studying rare local systems like (U)LIRGs may represent our only chance to probe star formation under extreme conditions typical of high-$z$ systems.

With the advent of the Hubble Space Telescope (HST), it has become possible to resolve large star-forming knots in merging/interacting systems into numerous compact young star clusters \citep[YSCs, e.g][]{whitmore95, holtz96, schw98, zepf99}. Their sizes and masses overlap with those of globular clusters (GCs), always believed to be relics of past assembly history confined to a much younger universe. Clearly, the discovery of such massive compact clusters, yet very young if compared to GCs, proved that conditions for GC like systems can also be found in the local universe. A major difference is that GC populations show peaked luminosity and mass distributions with an almost universal turn-over at $M_V\sim-7.2$ or $\log(M)=5.2$\msun \citep[see][and references therein]{BS06, jordan05} with some deviations \citep{huxor14}, whereas YSCs forming in merging/interacting galaxies have luminosity (mass) distributions that are well represented by power-law functions with slopes close to $-2$ \citep[e.g.][]{whitmore99} or slightly shallower \citep{randria13, linden17}. By using numerical approaches it has been investigated whether it is possible to reconcile the young and globular cluster mass functions via cluster disruption due to galactic tidal fields and/or tidal shocks by encounters with giant molecular clouds \citep[GMCs,][and references therein]{elmegreen10, kruijssen15, GR16}. Tidal fields alone cannot explain the almost universal turnover in the GC mass function \citep{vesperini03, kruijssen14, renaud17}. However, if merger/interaction dynamics favour the formation of the most massive YSCs we observe in the local universe \citep{bastian06, whitmore10}, the high pressures and gas densities nested within these galaxies \citep{johnson15, sun18} will also destroy clusters more efficiently as suggested by the disruption rates reported by e.g. \citet{whitmore10, linden17}, proposed by numerical works \citep{elmegreen10, kruijssen15}, and recovered in numerical simulations \citep{kruijssen12a, renaud15, pfeffer18}. Therefore, it remains still under investigation whether observed YSC mass functions can evolve into the GC ones. 

Another important similarity between YSC and GC populations is the presence of a possible truncation at the high mass end of their mass distribution in the form of a cut-off mass (M$_c$ or \mstar) if a Schechter function is used. \citet{jordan05} report that an evolved (accounting for stellar mass loss) Schechter function better describes the mass distribution at the high mass end of the GC mass function, with truncation masses, \mstar, that increase as a function of host luminosity (mass). Similarly YSC populations in local galaxies show evidence of a truncation in their mass function both in their directly measured luminosity functions and derived mass functions \citep[e.g.][]{gieles06b, larsen09, adamo15, johnson17, adamo17, messa18a}. Interestingly, already \citet{larsen09}, suggested that \mstar\, may change as a function of galactic environment. Indeed, changes in the recovered \mstar\, have been reported among galaxies \citep{johnson17, linden17}, and within galaxies \citep{adamo15, messa18b}. However, conclusive evidence for the presence of such truncation at the high mass end of the YSC mass function is still elusive, with works in the literature that report a pure power-law function as the preferred description for the YSC mass function in local galaxies, including starburst dwarfs and merger/interacting systems \citep[e.g.][]{chandar17, cook19, mok19}. From the theoretical perspective different models have been proposed that predict \mstar\, to depend on galactic physical properties and, thus, variations in \mstar\, from galaxy to galaxy \citep{RCK17, elmegreen18}.  

A parameter that describes cluster formation in galaxies is the fraction of stars forming in bound star clusters, referred to in the literature as Gamma, $\Gamma$, or cluster formation efficiency \citep[CFE,][]{bastian08}. A variety of observational evidence has pointed out a positive correlation between the fraction of stars forming in bound clusters (normalised to the total SFR in the entire galaxy or in regions of it) and $\Sigma_{SFR}$ \citep[e.g.][among many others]{goddard10, adamo11, adamo15, johnson16} which is still highly debated in the literature \citep[e.g.][]{KMK18, adamo20}. The $\Gamma$ vs. $\Sigma_{SFR}$ relation implies that galactic environments with higher SFR per unit area (or equivalently higher gas surface density or pressure) form a larger cluster population. The observed scatter in $\Gamma$ at given $\Sigma_{SFR}$ would suggest a dependence from other physical parameters. The fiducial model proposed by \citet{kruijssen12b} predicts the increasing trend observed in the data. The model finds that  $\Gamma$ depends not only on the gas surface density (traced by $\Sigma_{SFR}$) but also from the dynamical condition of the gas in the galaxy described by the Toomre parameter, $Q$, and the angular velocity of the gas. However, using a sample of 8 galaxies spanning a range of $\Sigma_{SFR}$ and galaxy morphologies, \citet{chandar17} suggest that the observed $\Gamma$ vs. $\Sigma_{SFR}$ relation is not real but  driven by mixing data sets that use different age intervals to measure $\Gamma$. When comparing $\Gamma$ at same age intervals, they observe that it is roughly constant in all galaxies and constantly decreases because clusters disrupt at very high rates. If this universal behaviour is systematically reproduced by other measurements, it would imply that cluster formation is decoupled from the galactic environment. 

In this work we will focus on a set of 6 galaxies that have been targeted with HST under the program \emph{Hubble imaging Probe of Extreme Environments and Clusters}, HiPEEC (GO 14066, PI Adamo). To understand and interpret the rapid evolution occurring in the cluster populations forming in galaxies experiencing enhanced episodes of star formation at any redshift, we need to probe the intensified duty cycle of gas consumption typically observed during merging phases in the local universe. HiPEEC is an UV-optical study of YSCs in a uniquely accessible sample of 6 starbursts located in the nearby ($D<80$ Mpc) universe. Our targets have SFRs higher than or comparable to the Antennae, and rich cluster populations. We will study the efficiency by which clusters form, the formation modes of the most massive clusters and their mass functions, their age and spatial distributions in the inner and outer regions of the galaxies, as well as obtain insights into the recent star formation histories of the host galaxies. 

\begin{figure*}
    \centering
    \includegraphics[angle=180,width=0.9\textwidth]{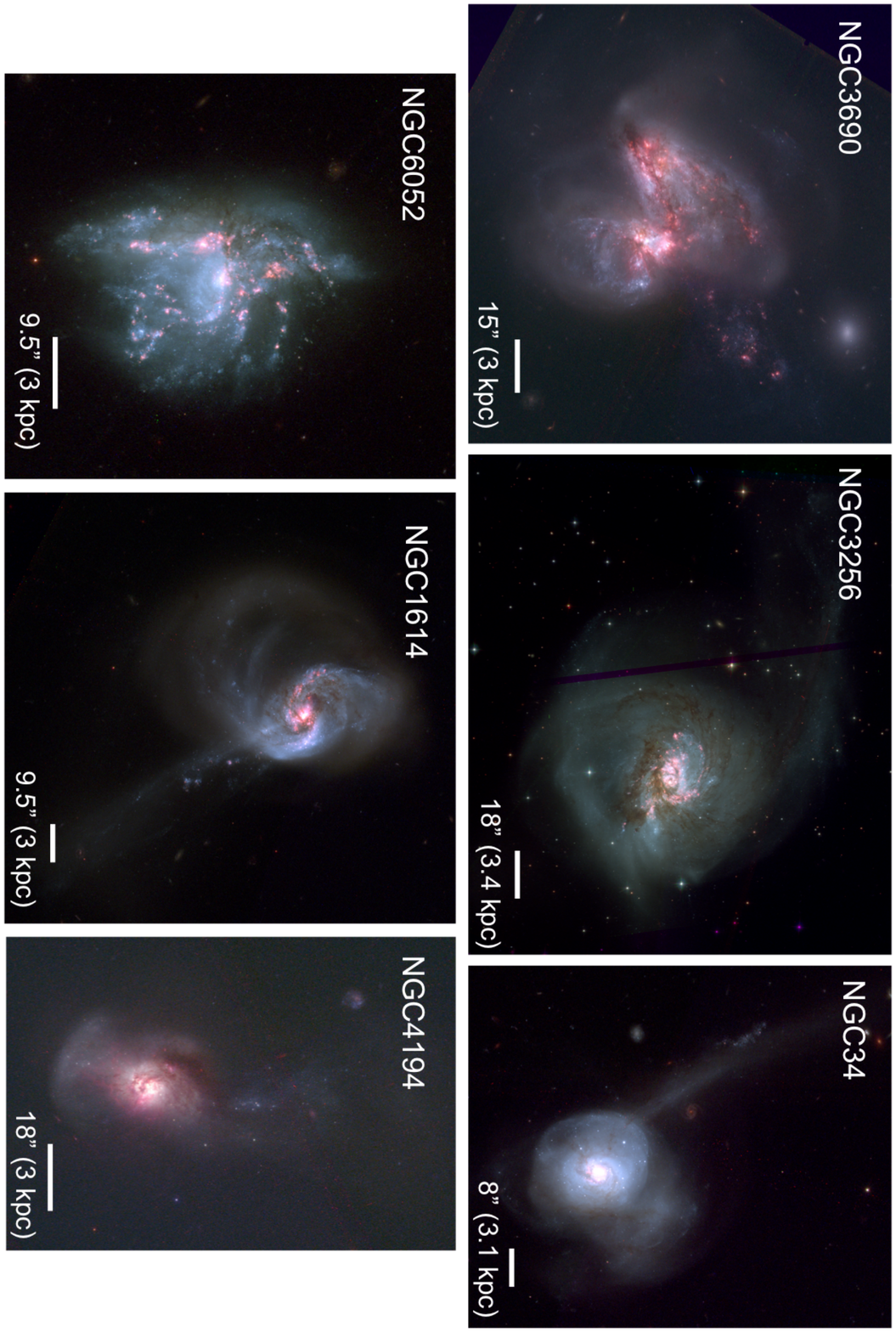}
    \caption{Color composite images of the HiPEEC sample. The reference system is rotated north-up and a scale bar of approximately 3 kpc is included in each panel. The narrow H$\alpha$ filter is always shown in the red channel.}
    \label{fig:fig1}
\end{figure*}

The paper is organised as follows. In Section~\ref{sec:obs} and \ref{sec:cats} we present the dataset and the cluster catalogue construction. In Section~\ref{sec:gen_sfr} we describe the continuum subtraction procedure to determine H$\alpha$ luminosity and derive  
SFR and $\Sigma_{\rm SFR}$. Cluster population physical properties, such as masses and ages are then used to map the merging phases of the galaxies in Section~\ref{sec:results}, as well as the cluster mass function and the cluster formation efficiency. We discuss and summarise our results in Section~\ref{sec:discussion}.

\section{Observations}
\label{sec:obs}

\subsection{The HiPEEC sample of galaxies}
\label{sec:hipeec}

The HiPEEC survey is designed to take full advantage of HST's capabilities to detect star clusters (within distances $\lesssim$ 80 Mpc), build their spectral energy distribution (SED), and obtain insights into the ISM conditions via observables like $\Sigma_{SFR}$ to investigate cluster formation during intense starburst phases. The aim is to extend star cluster population studies, like the Antennae ones \citep[e.g.,][]{whitmore99, whitmore10}, to other accessible merging/interacting systems. The HiPEEC sample has been drawn from the IRAS Revised Bright Galaxy Sample \citep[RBGS,][]{sanders03}, a catalogue containing bright IR systems up to a z of $\sim$0.09. Our sample is selected to include galaxies with total IR luminosity of L$_{\rm{IR}} > 10.8$ \lsun, which corresponds to a SFR$\geq 10$ \msunyr\,\citep[using the][conversion]{KE12}. A further selection has been made for face-on galaxies, with pre-existing ancillary UV-optical HST data to verify that the YSC populations are relatively unobscured. The selected targets are listed in Table~\ref{tab:galaxies} and 3-colour composites are shown in Figure~\ref{fig:fig1}. We leveraged the HST archive and completed the coverage of each target in at least 4 broad bands ($UBVI$) and a narrow band filter centred in H$\alpha$. For each target, we compile the complete list of the filters used and the respective exposure times in Table~\ref{tab:galaxies}. The inclusion of a data point ($U$ band)  below the Balmer break ($\sim4000$ \AA) and H$\alpha$ has proven to be effective in breaking the age vs extinction degeneracy when fitting the cluster SEDs \citep[e.g.][]{whitmore20, adamo10} and therefore improve cluster age-dating. In the text, we will refer to $UV$ if photometry is available at $\sim$2000 \AA; $U$ for photometry at $\sim$3000 \AA; $B \sim 4000$ \AA; $V \sim 5000$ \AA; H$\alpha$ for the narrow filter centred at the redshifted 6564.6 \AA\, emission line; $I$ for the reddest filter at about 8000 \AA.  Additionally, in Table~\ref{tab:galaxies}, we include an indication of the merging type and stage adopting the morphological classification scheme by \citet{haan11}. Our sample contains 3 major and 3 minor merger systems. In particular, NGC34 is the most advanced merger system, while NGC3690 is the target in the earliest phase, showing still the two distinguished interacting pair galaxies. According to this classification scheme, the remaining targets are already in advanced merger stages, with a single nucleus and perturbed morphologies.

In the analyses presented hereafter, we will analyse NGC3690 both as a single system and each component of the pair separately, which we name NGC3690A (North-East companion, conventionally referred to as IC694) and NGC3690B (South-West companion).
\begin{table*}
\centering
\begin{tabular}{l c c c c c l}
\centering
Galaxy & RA & Dec & Distance & merger & merger & Filters \\
     & [degrees] & [degrees] & (Mpc) & type &stage&  \\
\hline
NGC 34 & 2.777293 & -12.107314 & 80.6 & major & 6 & U/F275W$^{(a)}$ [2607s], U/F336W [8500s], A/F435W$^{(b)}$ [1260s]\\
 &  &  &  &  &  &   U/F555W [1940s], U/F665N [2800s], A/F814W$^{(b)}$ [720s]     \\
NGC 1614 & 68.499394 & -8.578883 & 65.5 & minor & 5 & U/F225W$^{(c)}$ [1400s], U/F336W [6510s], A/F435W$^{(b)}$ [1260s]\\
 &  &  &  &  &  &U/F555W [1450s], U/F665N$^{(h)}$ [3115s], A/F814W$^{(b)}$[720s]  \\
NGC 3256 & 156.963624 & -43.903748 & 38.5 & major & 5 & H/F330W$^{(d)}$ [13358s], A/F435W$^{(b)}$ [1320s]\\
 &  &  &  &  &  & A/F555W$^{(d)}$ [2552s], U/F665N$^{(e)}$ [1311s], A/F814W$^{(b)}$ [760s]   \\
NGC 3690 & 172.134583 & +58.561944 & 41.6 & major & 3 & U/F336W$^{(f)}$ [790s], U/F438W$^{(f)}$ [740s] \\
 &  &  &  &  &  & U/F555W [720s], U/F665N [740s], U/F814W$^{(f)}$ [1799s] \\
NGC 4194 & 183.539458 & +54.526833 & 39.0 & minor & 5 & U/F336W [2100s], U/F438W [1000s], U/F555W [600s]\\
 &  &  &  &  &  &  U/F665N [1467s], A/F814W$^{(g)}$ [2320s]  \\
NGC 6052 & 241.304125 & +20.542361 & 64.2 & minor & 5 &U/F336W [6480s], U/F438W [2370s], U/F555W [956s] \\
 &  &  &  &  &  &   U/F665N [1821s], U/F814W [1520s] \\


\end{tabular}
\caption{HiPEEC galaxy RA and Dec and distances according to NED; merger type where minor merger is defined as unequal mass progenitors with mass ratios $\geq 1:4$; merger stage according to \citet{haan11} where 1 are pre-merger separated galaxies with no tidal tails, 3 are ongoing merger with galaxies sharing common  envelopes, 4 late ongoing merger with one system and two visible nucleii (e.g. the Antennae system), 5 post-merger with single disturbed nucleus and strong tidal tail features, 6 post-merger with weak tidal tails; filters used in this work to perform cluster SED analysis and determine SFR (H$\alpha$). U stands for WFC3/UVIS detector, A for ACS/WFC detector, H for ACS/HRC detector. The total exposure time in each band are reported within brackets. Unless otherwise indicated data have been acquired under the program GO 14066, PI Adamo. $(a)$ GO 14593, PI Bastian; $(b)$ GO 10592, PI Evans; $(c)$ GO 13007, PI Armus; $(d)$ GO 9300 and 9735, PI Ford and Whitmore, respectively; $(e)$ GO 13333, PI Rich; $(f)$ GO 12295, PI Bond; $(g)$ GO 10769, PI Kaaret; $(h)$ GO 14095, PI Brammer.}
\label{tab:galaxies}
\end{table*}

\subsection{Data reductions}
The images of the targets (\verb+*flc.fits+ files) were downloaded from the MAST archive. All images taken in the same filter and visit were first drizzle combined together as images taken in the same visit are already well aligned to each other (but not aligned to images in other visits).  These visit-level drizzled images were then aligned using the Drizzlepac module \verb+TweakReg+, and were aligned to match the frame of the most recent image for that target.  The images were generally aligned to better than 10 mas.  The transformations derived via the alignment of the visit level drizzled images were then propagated back to the corresponding \verb+*flc+ images Drizzlepac module \verb+TweakBack+.  With all the \verb+flc+ images aligned, the final drizzled images could then be produced.  The images were drizzled North-up, and covering the same exact footprint on the sky as the visit-level drizzled images (i.e. the corners of all final images were at the same RAs/Decs).  All the data have been drizzled to the same resolution of 0.4''/pixel, i.e. the native pixel scale of the WFC3/UVIS detector. This pixel resolution correspond to a physical resolution range of $\sim$6 pc/px for the closest targets (NGC3256, NGC3690, NGC4194), 10 pc/px for NGC1614 and NGC6052, 12 pc/px for the most distant target, NGC34.

\section{Cluster catalogue products}
\label{sec:cats}

\subsection{HiPEEC pipeline: extraction and photometry}
\label{sec:pipeline}
For each HiPEEC galaxy, we have determined candidate cluster positions and extracted photometry via in-house software. The sequence of tasks follows closely the LEGUS pipeline {\sc legus\_clusters\_extraction\_v40} \citep{adamo17}, developed to extract and perform photometry of the cluster populations in the galaxies targeted by the Hubble treasury program Legacy ExtraGalactic UV Survey \citep[LEGUS][]{calzetti15}. However some modifications were required to adapt the cluster analysis for the change in distance of the HiPEEC galaxies. Within the distance range covered by the LEGUS galaxies, YSCs have light spread functions (or full width half maximum, FWHM) that are larger than stellar point spread functions (PSFs). This property was widely used to apply a concentration index (CI, difference of the source magnitude within two apertures at 1 and 3 px) selection of the cluster candidates \citep{adamo17}. In the HiPEEC pipeline we estimate the CI of all the extracted cluster candidates but we do not apply any initial CI selection since clusters at the HiPEEC distance are unresolved, i.e. have stellar PSF.  
We describe here below the general steps performed by the pipeline in sequential order:

\begin{itemize}
\item{Source Extractor \citep{sextractor} with optimised user-defined settings is used on the V band image to produce the positions of all the potential cluster candidates. The parameters are tuned for each galaxy, but we generally use small grid sizes (between 10 and 15 pixels) to create a smoothed background image that is subtracted before the extraction of point-sources is performed. Usually we include in the initial catalogue all the compact sources that have at least 3 contiguous pixels with a signal 5 $\sigma$ above the background of the frame. Standard deblending parameters are adopted.}
\item{Before photometry is performed, all the sources that are less than 6 px away from the edges of the frames or detector gaps are removed, because their photometry will be compromised. Sources that have distances smaller than 3 px from each other are considered duplicates, thus only one of them is retained in the catalogue.}
\item{Source extracted positions are fed into the {\sc phot} package in {\sc iraf}. Header information of each image is used to calculate ABmag zero points and the photometric error. Photometry is performed in all the available bands and a centring of the position is allowed within 1 px. Aperture photometry is performed using a radius of 3 px (0.12") and a local sky background subtraction is applied using an annulus of inner radius 5 px (0.2") and 1 px wide. Only candidates with magnitude errors equal or below 0.3 mag in $BVI$ bands are retained in the final catalogue.}
\item{CI are estimated by subtracting the $V$ band magnitude of the source at 1 and 3 px.}
\item{The final photometry accounts for loss due to a limited aperture and Galactic extinction. Aperture corrections are estimated using stellar encircled energies in each of the bands up to a radius of 20 px (0.8"). These corrections should be considered a lower limit to the real loss of light produced by using a fixed aperture, as in some cases cluster candidates are not stellar-like. We also correct the photometry in each band for Galactic extinction using NED values provided for each target and listed in Table~\ref{tab:ha}. The final photometry is estimated in ABmag. The final magnitude error is the square root of the sum in quadrature of the photometric error produced by {\sc phot} and the uncertainty on the Zero Point of 0.05 mag.}
\end{itemize}
The initial automatic catalogues contain 529 (NGC\,34), 1130 (NGC\,1614), 7239 (NGC\,3256), 1732 (NGC\,3690), 820 (NGC\,4194), 1786 (NGC\,6052) extracted sources with photometric error in $BVI$ bands less than 0.3 mag.

\begin{figure*}
    \centering
    \includegraphics[width=0.95\textwidth]{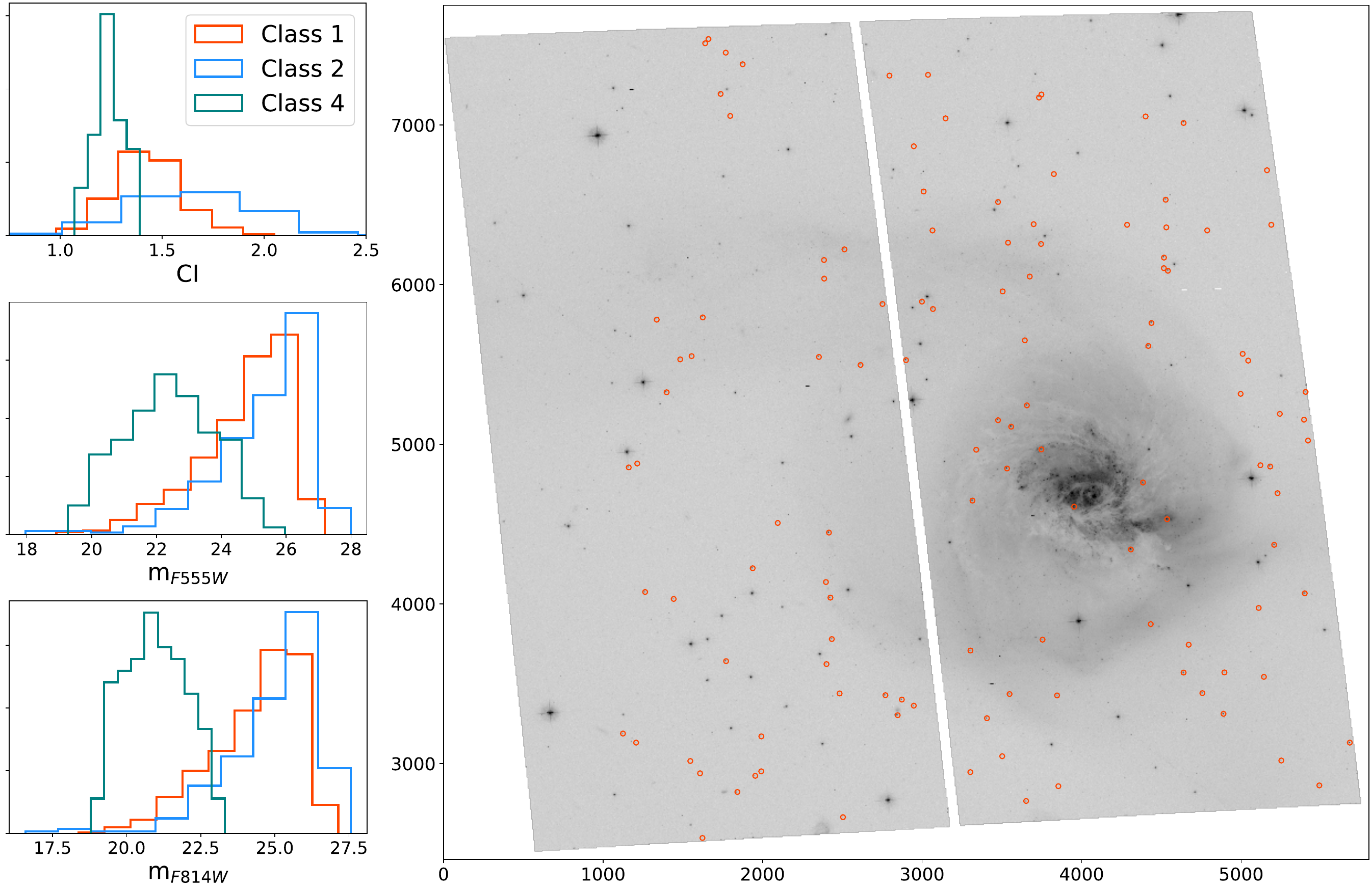}
    \caption{Normalised to unity area histograms of CI, and apparent magnitude in $V$ and $I$ bands of class 1, 2, and 4 systems in NGC~3256. Generally, class4 objects have stellar CI ($<$1.4 mag), and bright apparent magnitudes, especially in the $I$ band. Their distribution in the field of view of the galaxy (right panel) is quite homogeneous. Numerous bright foreground stars are also visible in the field of view, confirming the stellar nature of class 4.}
    \label{fig:fig2}
\end{figure*}

\subsection{Visual inspection of source catalogues}
\label{sec:selection}

Despite tailoring the source extraction procedure to each galaxy and implementing error cuts, the sources that have passed this initial selection can still include foreground stars or background galaxies. In previous studies, objects have been visually classified into several categories, including clusters and associations \citep[e.g.][]{whitmore10,bastian12, konstantopoulos13,
hollyhead16, adamo17, cook19}, which produces excellent results, but can be time consuming.

To improve the selection of potential cluster candidates we also implemented visual inspection of each catalogue. Each galaxy catalogue has been inspected by two people to reduce subjectivity in both the $V$ and $I$ band (the $V$ band is better for identifying younger clusters and the $I$ band for background galaxy contaminants, which are not clear in the $V$ band). A classification of 1 was given for a compact, point-like detection which we consider a cluster candidate, and 2 for foreground stars, extended, clumpy objects, which are very often background galaxies, and extended regions in between dusty lanes, that almost disappear in the I band (e.g., differential extinction in a bright background). The central knots of the galaxies have all been classified as 2. They have extended morphologies and some of them coincide with the position of the central AGN (only detected in optical in NGC34 and NGC3256). 

During visual inspection of NGC\,34, 40 new cluster candidates classified as 1 were manually added to the automatic catalogue \citep[e.g., also common  in the literature, see][]{whitmore20}, that therefore comprises 569 sources in total. NGC\,34 is our farthest galaxy, and therefore, the most challenging to analyse. No other target in the sample required manual addition of missed potential cluster candidates. 

The visual classification of one of the closest targets, NGC\,3256, represented a real challenge. The pipeline extracted > 7000 cluster candidates from the NGC~3256 images, which posed a visual inspection problem, due to the time required. For this galaxy we produced an initial classification using a machine learning algorithm. The program uses the {\sc sklearn} modules in {\sc python}, specifically a supervised neural network classifier ({\sc mplclassifier}) to identify objects as clusters (class 1), or non-clusters (class 2). The classifier was trained using cluster catalogues from the other 5 HiPEEC galaxies. The classification was based on images of individual sources cut down to 20x20 pixels from the whole galaxy fits file, as well as the CI index of the cluster and the $U-B$ and $V-I$ colours.  This classification was used as initial guess for visual inspection. During visual inspection, the line of sight towards NGC\,3256 appeared to be contaminated by foreground low-mass stars. Their stellar-like appearance prevented any automated approach, trained to classify systems on compactness, to distinguish them from cluster candidates. We visually classified in total 119 objects as low-mass foreground stars (i.e. M-dwarf candidates). These objects have been assigned class 4 in the final catalogue. In Figure~\ref{fig:fig2}, we show class 1, 2, and 4  histograms of the concentration index, magnitudes in $V$ and $I$ band, normalised to have an area of unity to favour the comparison. We also show the location of class 4 systems in the F555W frame, which appear homogeneously distributed in the field of view (shared area within the 4 broadbands). We estimate a number density of $\sim11/{\rm arcmin}^2$ class 4 objects. In general, their CI is very compact ($< 1.4$ mag) and their apparent magnitude brightens in the $I$ band, typical for M-dwarf stars. In the F555W field of view numerous bright foreground stars are also visible confirming that an overdensity of foreground stars is very likely toward this system. After visual classification, comparison with the machine learning classification resulted in an agreement of $\sim73\%$, very similar to the results reported in the literature \citep[e.g.][]{messa18a, grasha19, wei20}.

\subsection{Analysis of the completeness limits}
\label{sec:completeness}
To perform completeness tests on the cluster catalogues, we created mock cluster populations for each galaxy. A total of 50,000 mock clusters per galaxy have been used. The cluster masses were sampled from a power-law distribution $dN/dM \propto M^{-2.0}$ over a mass range $3\times10^3$ to $10^8$~M$_{\odot}$. The cluster extinctions E($B-V$) were sampled over the range 0.01 to 1.5 mag. The cluster ages were sampled from a continuous random distribution with an upper limit of 1 Gyr. The magnitude of each cluster was determined by matching its mass, age, and extinction to the single stellar population(SSP) Yggdrasil models (described in the next Section). The goal of this setup is to model consistently the change in luminosity across the $UBVI$ bands due to different combinations of cluster physical properties. For each target we perform completeness analyses in the 4 bands, $UBVI$ (see Section~\ref{sec:agemassfit}).    

To prevent overcrowding, the mock clusters were randomly placed in a polygonal region surrounding each galaxy in groups of 250. Each polygon was tailored to the shape of the galaxy, and the area of the polygon was matched to roughly twice the area containing 80\% of the H$\alpha$ flux (see Section~\ref{sec:hasfr}). The software package \texttt{BAOLAB} was used to properly scale empirically-derived PSFs to the magnitudes specified in the mock cluster catalogues and place them into the $UBVI$ images of each galaxy. The HiPEEC pipeline was run on each set of $UBVI$ images, and the resulting catalogues were combined. The fraction of recovered mock clusters as a function of input magnitude was calculated for both the inner and outer regions of each galaxy, as defined by the 80\% H$\alpha$ radius reported in Table~\ref{tab:ha}. The magnitude limits below which 90\% of the mock clusters were recovered are listed in Table~\ref{tab:completness}. 

In Appendix~\ref{app:completeness} we show the plots of the recovery fractions per magnitude bins in all the HiPEEC sample. We use the 90\% limits in the 4 reference bands in both inner and outer regions of each galaxy as reference for the cluster analysis. We notice here that a galaxy wide completeness test is not a valid assumption for realistic limits in such complex galactic environments, as already discussed by \citet{mulia2016, randria19}. The change is quite dramatic (up to $\sim$2 mag) in the shortest wavebands and, if not taken into account, it can severely affect cluster analyses. Moreover, the drop from 90 to 50\% completeness is not sharp, especially in the inner regions of the galaxies. Therefore, in Section~\ref{sec:results}, we will apply age and mass cuts that can allow to study the cluster populations in a homogeneous way, while compromising slightly on the mass-age-luminosity limits imposed by the completeness analysis.

\begin{table}
\centering
\begin{tabular}{c c c c}
\centering
Galaxy  &  Filter  &  Inner regions  &  Outer regions \\
\hline
NGC 0034	&	F336W	& 25.09	& 26.95	\\
			&	F435W	& 23.89	& 26.26	\\
			&	F555W	& 23.59	& 25.75	\\
			&	F814W	& 23.22	& 24.43	\\			
NGC 1614	&	F336W	& 24.31	& 26.24	\\
			&	F435W	& 23.26	& 25.50	\\
			&	F555W	& 23.11	& 25.16	\\
			&	F814W	& 22.78	& 24.10	\\
NGC 3256	&	F330W	& 23.45	& 25.08	\\
			&	F435W	& 22.58	& 24.86	\\
			&	F555W	& 22.30	& 24.41	\\
			&	F814W	& 21.90	& 23.34	\\
NGC 3690	&	F336W	& 24.12	& 26.01	\\
			&	F438W	& 23.52	& 25.61	\\
			&	F555W	& 23.27	& 24.76	\\
			&	F814W	& 22.57	& 23.47	\\
NGC 4194 	&	F336W	& 21.93	& 25.28	\\
			&	F438W	& 21.08	& 24.06	\\
			&	F555W	& 21.08	& 23.68	\\
			&	F814W	& 20.73	& 22.66	\\
NGC 6052 	&	F336W	& 24.30	& 26.12	\\
			&	F438W	& 23.82	& 25.57	\\
			&	F555W	& 23.54	& 25.16	\\
			&	F814W	& 23.35	& 24.05	\\
\end{tabular}
\caption{90\% Completeness limits reached in the 4 wide band filters used for cluster selection. Inner regions correspond to the area of the galaxies where 80\% of the H$\alpha$ flux has been estimated corresponding also to the regions with the highest crowding. We report also the completeness limits outside these regions. Both magnitude limits are considered in the cluster analysis}
\label{tab:completness}
\end{table}

\subsection{Fitting for the age and mass of the cluster candidates}
\label{sec:agemassfit}

After visual inspection of the sources contained in the catalogue of each galaxy, we used a minimum $\chi^2$ fitting routine as per \citet{adamo10,adamo17} to estimate the best age, extinction, and mass for each source that has entered the catalogue with a magnitude error of $\leq$0.3 mag in $BVI$ bands (corresponding to source detected with a signal to noise ratio better than 3). Fluxes in other bands ($UV$, $U$, H$\alpha$) are included in the fit if their photometric error is smaller than $0.3$ mag.   Yggdrasil SSP models\footnote{https://www.astro.uu.se/$\sim$ez/yggdrasil/yggdrasil.html} \citep{zack11} with a $10^6$ \msun\, single-burst star formation event and a Kroupa IMF \citep{kroupa2001} are used. The advantage of using Yggdrasil models is that they combine the SSP stellar continuum and absorption line spectra with the spectral emission (both continuum and emission lines) of the ionised gas, surrounding the stellar population. The spectral emission of the gas is obtained with Cloudy \citep{cloudy}, assuming a covering fraction of the ionising front of 50\%, a gas density of 100 cm$^{-3}$, and a filling factor of 0.01. Both stellar and nebular metallicity are fixed in this work to solar metallicity. We applied differential extinction (treating the gas and stellar components separately) using the Calzetti attenuation curve \citep{calzetti2000} to attenuate the flux across all wavelengths. The model spectra have been  redshifted to the appropriate distance for each galaxy (see Table~\ref{tab:ha} for the adopted redshifts). We then convolve the models with the bandpasses relevant for each galaxy using {\sc pysynphot} in {\sc python}. The model grid covers a range of ages (0-14 Gyr) and a range of extinctions ($E(B-V) =$ 0--1.5 mag).

A minimum $\chi^2$ fitting was then performed between the modelled and the observed SEDs. The solution that provided the smallest $\chi^2$, resulted in the best age, extinction, and mass. Errors were also estimated within the solutions that had a resulting $\chi^2 \leq \chi^2_{\rm best}+2.3$\footnote{This condition depends on the number of free parameters. We strictly constrain two free parameters, age and extinction. The cluster mass is not a free parameter as it results from the normalisation factor between the observed and modelled SED.}, which correspond to the 68\% confidence levels around the best solution \citep{adamo17}.

There is a clear degeneracy in the SSP models between age and extinction, clearly visible in the colour--colour diagrams (see Figure~\ref{fig:ccplot}). In order to disentangle this degeneracy and give accurate estimates of the age and extinction we require that our cluster candidates have a photometric error better than 0.3 mag (3$\sigma$ detection or better) in a fourth band, $U$, to provide information on the source SED below the Balmer break at $\sim4000$. When available, detection in the narrow H$\alpha$ filter, as well as $NUV$ band, have been included if their photometric errors are smaller than 0.3 mag. In the last decade, numerous surveys have shown the effectiveness of including information in the NUV and in the H$\alpha$, especially for very young ages ($< 10$ Myr) below the Balmer break at $\sim4000$ \AA\,\citep[e.g.][among many others]{hayes05, calzetti15}. 

\begin{table}
\centering
\begin{tabular}{c c c c}
\centering
Galaxy  &  Extracted  &  Class 1  &  Final selection \\
\hline
NGC 0034	&	569	& 301	& 243	\\
NGC 1614	&	1130	& 556	& 460	\\
NGC 3256	&	7239	& 2811	& 917	\\
NGC 3690	&	1732	& 868	& 765	\\
NGC 4194 	&	820	& 454	& 380	\\
NGC 6052 	&	1786	& 679	& 580	\\
\end{tabular}
\caption{Summary of the cluster catalogues for the HiPEEC sample. We list the number of automatically extracted sources from HST images with detection in $BVI$ bands, the number of compact objects classified as potential cluster candidates by visual inspection. Final number of class 1 cluster candidates, with detection in $UBVI$ bands and SED reduced $\chi^2$ better than 10.}
\label{tab:catalogue}
\end{table}

In recent years, stochastically populated SSP models \citep[e.g. SLUG][]{krumholz15} have been implemented to analyse the observed SEDs of young star clusters in local galaxies. These models take into account that in low mass clusters (typically below 10$^4$ \msun), the stellar IMF is not fully populated at the high mass end, causing variations in the observed SED that are not accounted for in deterministic SSP models \citep[e.g.][]{fouesneau12}. However, differences in the recovered cluster physical properties using stochastic or deterministic SSP models disappear at cluster masses above 10$^4$ \msun\, \citep{krumholz15}. Due to the completeness limits of our datasets, we will limit our analysis to cluster masses above  $5\times10^4$ \msun, hence, beyond the mass regime affected by stochastic effects.

\subsection{Cluster candidates selection and physical properties}
\label{sec:fincat}
The final selection of cluster candidates in each galaxy is done by cross-correlating the photometric, morphological, and SED fitting properties of the sources contained in each catalogue. To carry out our analysis we select only objects that are detected with a photometric error better than 0.3 mag (3$\sigma$ detection or better) in those that we consider the 4 reference bands, $UBVI$. Moreover, we include only sources that have a morphological class 1, and a reduced $\chi^2$ smaller than 10. In Table~\ref{tab:catalogue}, we summarise the number of cluster candidates that satisfied each step in the selection process in each galaxy. Similarly to what is observed in the LEGUS sample (Kim et al in prep), only about 30-50\% of the initially extracted sources are potential clusters.

\begin{figure*}
    \centering
    \includegraphics[width=0.95\textwidth]{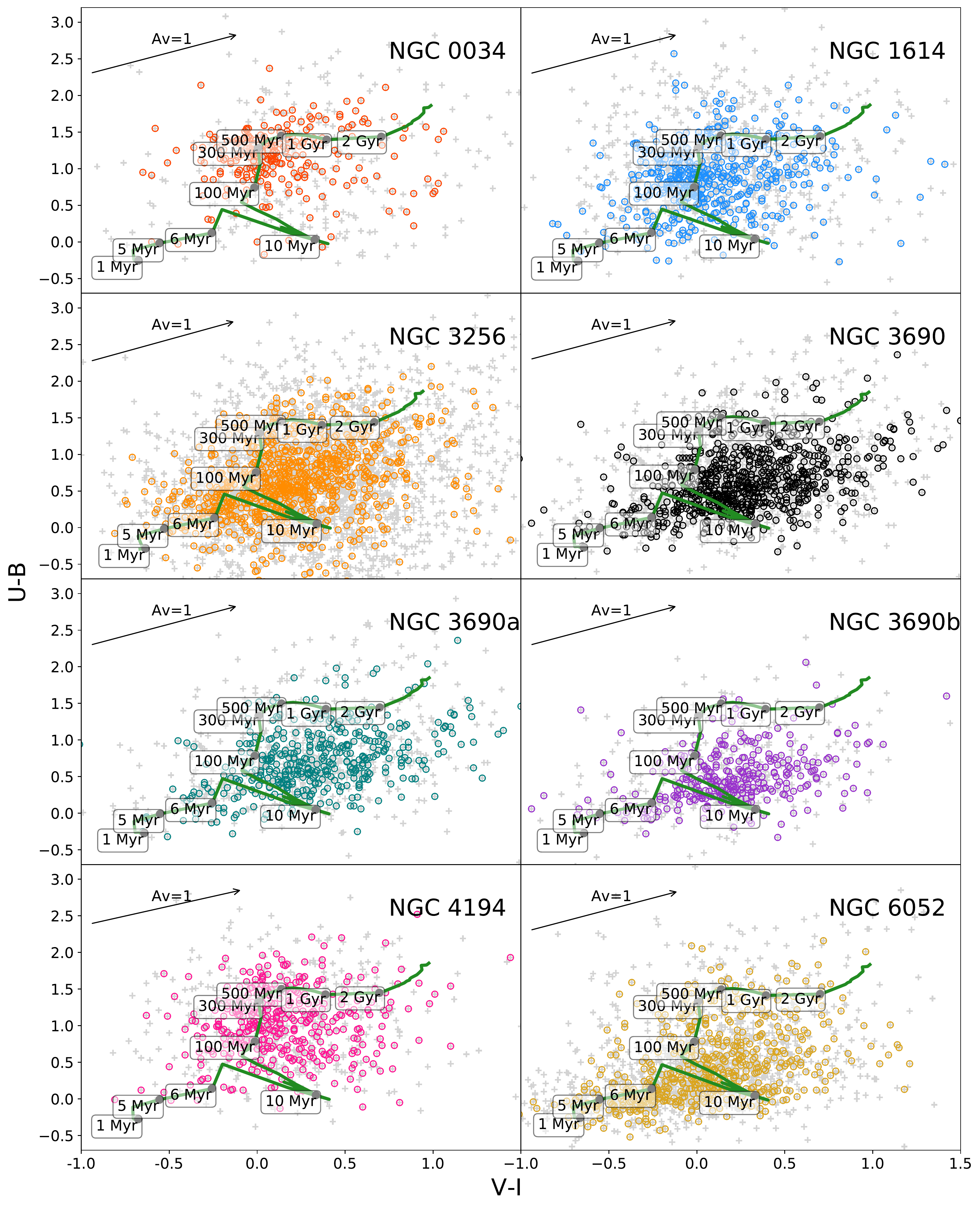}
    \caption{Colour-colour plots for each of the galaxies in the sample. The green line is the model track using the Yggdrasil cluster evolution models \protect\citep{zack11} for each with E(B-V)=0, Z=0.02 (solar) and assuming instantaneous starburst. Several ages along this track are labelled. The arrow in the top left corner is the extinction vector for each galaxy, the length being equal to A$_{\rm v}$ of 1. Grey crosses are used for sources that enter the catalogue (thus detected in $UBVI$) and are classified as class 1 after visual inspection. Coloured circles show the location in the colour-colour diagram of the cluster candidates of each galaxy, i.e. class 1 sources, detected in the $UBVI$ bands with a $\sigma < 0.3$ mag, and with a reduced $\chi_{\rm best}^2 \leq 10$. The galaxy NGC3690 is also separated into the two interacting systems, NGC3690A (or IC694 according to NED) and NGC3690B (or NGC3690 according to NED). See main text.}
    \label{fig:ccplot}
\end{figure*}

In Figure~\ref{fig:ccplot}, we show the colour properties of the objects contained in the source catalogue of each target, before (grey crosses) and after (coloured circles) the selection criteria above are applied. The application of these selection criteria clearly removes the large scatter of the automatic extracted sources. The populations of cluster candidates (coloured circles) in each galaxy have $U-B$ vs $V-I$ colours that follow the SSP tracks, show a large reddening, and also a variety of age distributions as outlined by the location of peak densities along the model tracks. 

\begin{figure*}
    \centering
    \includegraphics[width=0.95\textwidth]{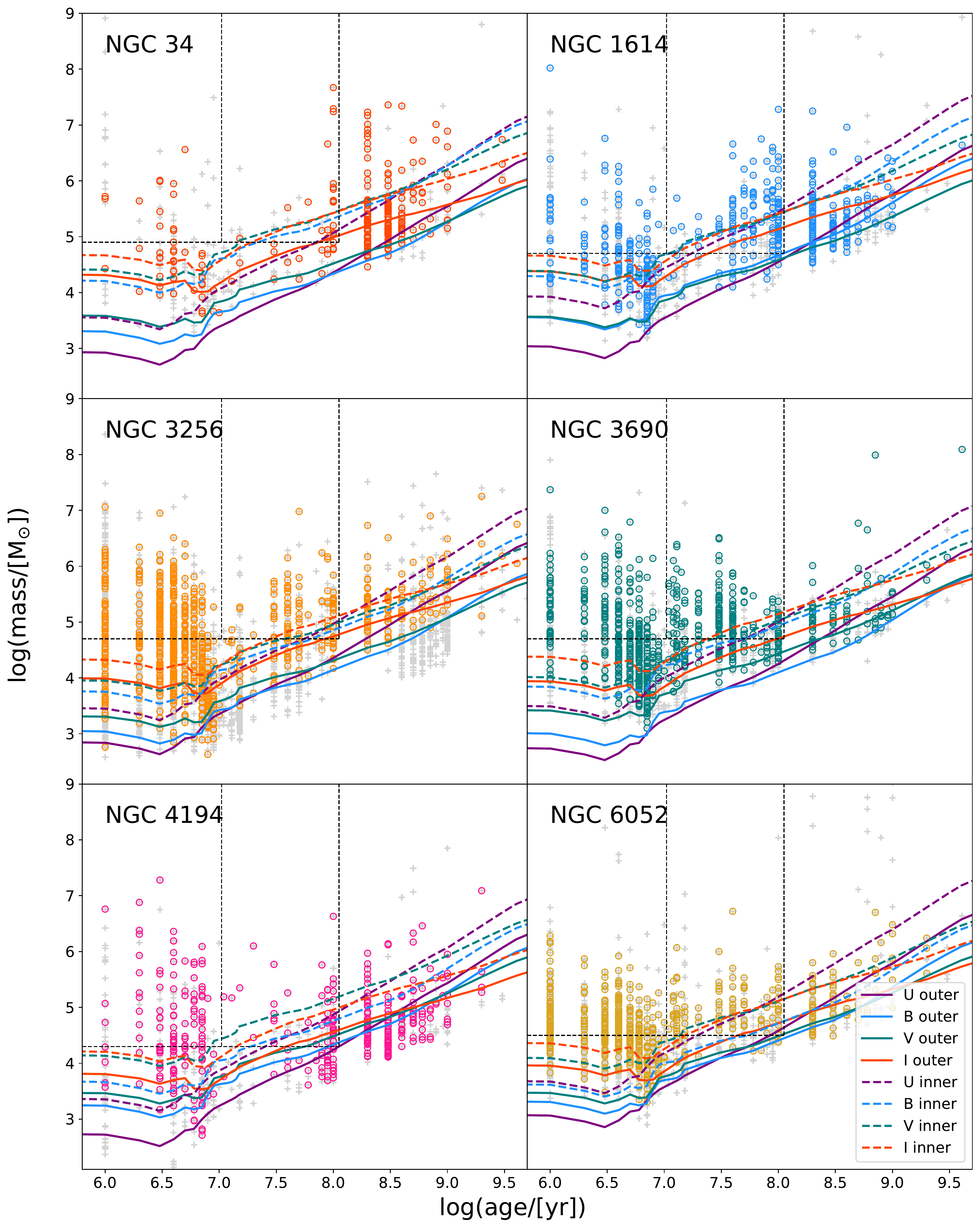}
    \caption{Age-mass diagrams of the HiPEEC galaxies. Grey crosses are used for sources that enter the catalogue (thus detected in $UBVI$) and are classified as class 1 after visual inspection. Coloured circles show the location in the colour-colour diagram of the cluster candidates of each galaxy, i.e. class 1 sources, detected in the $UBVI$ bands with a $\sigma > 0.3$ mag, and with a reduced $\chi^2 \leq 10$. Yggdrasil cluster evolution models \protect\citep{zack11} are used to convert 90\% detection limits into a mass limit as a function of age for the inner (dashed lines) and outer (solid lines) regions of each galaxy as defined in the main text.}
    \label{fig:agemassall}
\end{figure*}

Similarly, we show in Figure~\ref{fig:agemassall} the age-mass diagrams of the automatic extracted sources and of the cluster candidates of each galaxy. The magnitude values corresponding to 90 \% completeness limits (listed in Table~\ref{tab:completness}) in the 4 broadbands are converted to mass limit as a function of age and included for each galaxy. Detection limits are quite different among filters and galaxies, and within the central areas of the galaxies. In general from the age distributions we see that all the galaxies have extended episodes of cluster formation, probably related to the length of the interacting/merging phase. Except for NGC34 (already in an advanced merger phase), all galaxies have a large number of clusters with ages lower than 10 Myr. As expected a large number of clusters more massive than $10^5$ \msun\, are observed in all the systems. In all the targets except NGC6052, we observe that clusters can reach masses of $10^7$ \msun\, as also observed in the Antennae and other merging systems in the local universe \citep{whitmore10, linden17, randria19}. To overcome incompleteness in the detection at the lower mass end and enable homogeneous analyses of the CFE and mass function distribution in all galaxies we apply mass and age limits for the selection of clusters. They are listed in Table~\ref{tab:MF} and shown as dashed black lines in each age-mass panel. The detection limits are not as strict at the oldest age interval ($\sim$ 50--100 Myr) where the completeness, especially in the inner regions, is lower than 90 \%. However, we discussed in Section~\ref{sec:completeness} that the completeness rate does show a slow decline in the recovery rates, suggesting that we are not including a large incompleteness factor in our analysis. In  Section~\ref{sec:massfunc}, we further discuss whether stricter mass limits would change the outcome of the mass function analysis.

\section{H$\alpha$ as tracer of recent star formation rate}
\label{sec:gen_sfr}
To trace the global star formation properties of the HiPEEC galaxies we use H$\alpha$ maps extracted from  the HST narrow band imaging. The goal is to compare the properties of the cluster populations detected at optical wavelengths and the strength and rate of star formation as traced by an optical comparable tracer, i.e. H$\alpha$ emission produced in these dusty galactic environments.

\subsection{H$\alpha$ continuum subtraction and selected area}
\label{sec:halpha}

The HST narrow band filter centred on the H$\alpha$ emission is approximately 100 \AA\, wide, implying that two important corrections need to be taken into account when producing continuum subtracted line maps. The first item is the subtraction of the [NII]6550,6585 doublet
 emission contamination that enters the filter transmission; secondly, the over-subtraction of the H$\alpha$ flux due to underlying absorption from evolved stellar populations. To take these corrections into account we use the Lyman Alpha eXtraction  Software \citep[LaXs][Melinder et al. in prep.]{hayes09, ostlin14}. LaXs has been optimised to perform continuum subtraction using a pixel--to--pixel SED fit algorithm. The SED analysis relies on the combination of two stellar populations, an old extinction-free stellar population fixed at 5 Gyr (with variable normalisation) and a younger one with age, extinction, and normalisation left as free parameters. The input parameters are the redshift, Galactic extinction, [NII]6585/H$\alpha$ ratio of each galaxy (listed in Table~\ref{tab:ha}). Broadband $UBVI$ and the narrow H$\alpha$ filter images are used as input to LaXs. The best fit to the observed SED provides the continuum at the position of the H$\alpha$ emission and the correction for stellar H$\alpha$ absorption wings. The final H$\alpha$ emission at each pixel is then corrected by contamination of the [NII] doublet emission, corrected for the filter transmission at the location of the three lines of interest.  Before the analysis, the images have been cropped to the regions containing most of the $V$ band light of the galaxy. Moreover, even if the data are homogeneously drizzled to the same pixel resolution, some small variations in the PSF remain among frames. To overcome these small variations, the data have been smoothed with a gaussian function of FWHM of 2 px. 
 
\begin{figure*}
    \centering
    \includegraphics[width=0.95\textwidth]{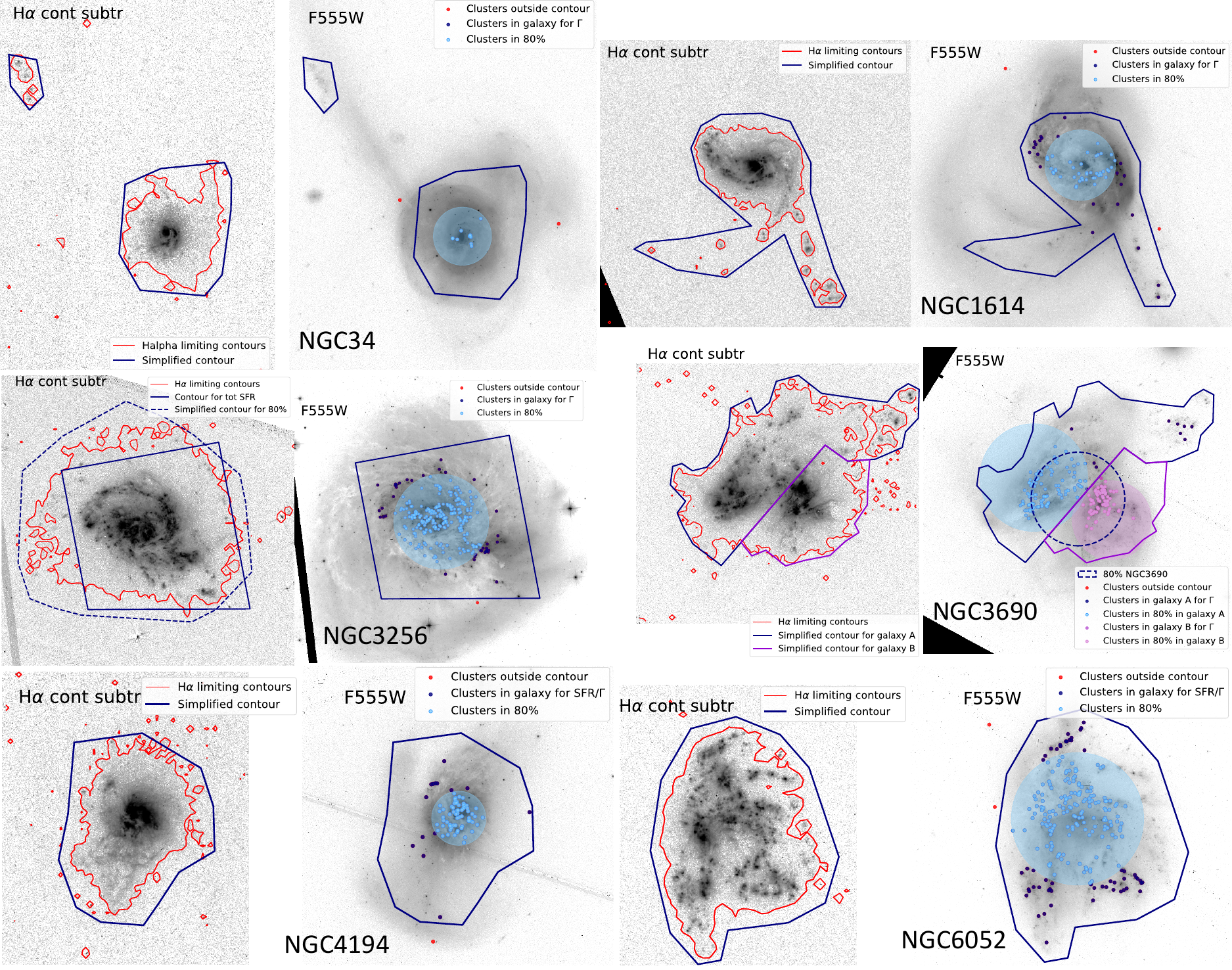}
    \caption{H$\alpha$ continuum subtracted maps of the HiPEEC galaxies compared to their $V$ band emission. The left panel shows the H$\alpha$ emission with 3$\sigma$ contours in red (the detection limits are listed in Table~\ref{tab:ha}). The simplified polygonal contours shown in solid blue lines are drawn to contain the 3$\sigma$ contours and used to estimate the total H$\alpha$ luminosity, the SFR, and the $\Sigma_{\rm SFR}$. The right panel shows the F555W image with the simplified contours and light blue shaded circle corresponding to the region containing 80\% of the total H$\alpha$ flux. We include the positions of the clusters used to estimate $\Gamma$ (see Section~\ref{sec:gamma}). Clusters that meet the selection criteria (age$\leq10$ Myr and M$\geq$M$_{\rm lim}$) but lie  outside the regions where SFR has been estimated are plotted as red dots. In the case of NGC3690 system, we show the division into NGC3690A (IC694, blue and cyan contours) and NGC3690B (NGC3690, purple and pink contours). We use the combined simplified contours of NGC3690A and NGC3690B to estimate the 80\% area of the whole system delimited by the blue dashed line circle.  }
    \label{fig:ds9cont}
\end{figure*}

The resulting H$\alpha$ continuum subtracted images are shown in Figure~\ref{fig:ds9cont} along with the corresponding $V$ bands. For each galaxy, we determine the value of the background and build the contours that enclose flux detection above a 3$\sigma$ detection limit (see Table~\ref{tab:ha}). The contours are overlaid in red on the H$\alpha$ continuum subtracted panels shown in Figure~\ref{fig:ds9cont}. In general, we observe that the H$\alpha$ emission is mostly concentrated in the inner regions of our galaxies, with clumpy emission along the main morphological features of the galaxy. Clumpy H$\alpha$ emission is also found in the tidal features of most of the galaxies. Noteworthy is the filamentary H$\alpha$ emission in the south-west of NGC4194, the only galaxy where we observed filamentary H$\alpha$ emission consistent with galactic scale winds and outflows of ionised gas. Similar morphologies in H$\alpha$ emission are typically observed in  nearby starburst galaxies, for example as in M82 and NGC253 \citep[][]{west09, west11}. Indeed, these features are well linked with the extreme nature of the starburst occurring in NGC4194, as it will be discussed below, using the cluster properties as a tracer.  

In order to investigate cluster formation as a function of their position in the galaxy (inner vs. outer regions and tidal features) we define the area in the galaxy where 80\% of H$\alpha$ luminosity is produced. To determine the 80\% area we use surface brightness profiles. We fix the centre of each galaxy, except in NGC3690, to coincide with the brightest H$\alpha$ emission. In the NGC3690 system the centres have been selected such that we could separate the two pair galaxies, and that it would correspond to the centre of the system (see position of the cyan, pink, blue dashed circles, respectively in the right panel of Figure~\ref{fig:ds9cont}). We simplify the H$\alpha$ 3 $\sigma$ (red) contours as the polygonal areas shown in Figure~\ref{fig:ds9cont} by the blue contours. We use the Python package {\sc shapely} to select pixels that are within these simplified contours, and sum the counts. These measurements correspond to the total H$\alpha$ flux of the galaxy. In NGC34 and NGC3256, we mask the central pixels applying an upper flux cut, in order to avoid contamination from the AGN. In the case of NGC3256 we also masked bright and poorly subtracted foreground stars. If present, negative flux pixels or without detected flux are also excluded in all the targets (especially at the edges, in proximity of the most external blue contours). We then build H$\alpha$ surface brightness radial profiles and determine the area that contains 80\% of the flux. We convert that area to a circular radius. In Table~\ref{tab:ha} we list all the physical properties we use to run the LaXs pipeline, as well as the 80\% radii. In Figure~\ref{fig:ds9cont}, we show the position  of clusters younger than 10 Myr and more massive than the mass limit (reported in Table~\ref{tab:MF}) overlaid on the $V$band images of the galaxies. These clusters will be included in our calculations of $\Gamma$ and used to derive the YSC mass function in the age range 1-10 Myr, as explained in Section~\ref{sec:massfunc} and Section~\ref{sec:gamma}. We notice that in the case of the NGC3690 pair, we determine the 80\% area of the total system using the combined outer contours of each interacting galaxy (see Figure~\ref{fig:ds9cont}). We also attempted to separate the two systems into NGC3690A and B. Determined values for the whole system and of each interacting companion are reported in Table~\ref{tab:ha} and \ref{tab:sfr}.

\subsection{H$\alpha$ star formation rates}
\label{sec:hasfr}

In star-forming galaxies, H$\alpha$ emission is directly linked to the ionising radiation produced by massive stars (M$>$10 \msun) and therefore sensitive to short time scales (1-10 Myr), proportional to the life-time of these stars \citep[e.g][]{KE12}. However during these time scales, extinction will affect H$\alpha$ emission in a similar fashion as it affects the YSCs that are responsible for ionising a large fraction of the HII regions. Different methods in the literature have been proposed to account for the missing contribution to the total SFR estimates by ionising radiation absorbed by the dust. \citet{calzetti07} provide a relation that uses MIR emission either at 8 or 24 $\micron$ to correct for the hidden SFR missed by H$\alpha$. However, in our case, the Spitzer images of our targets have too low resolution and quality to allow us to apply a similar surface selection as done with the H$\alpha$ maps. Moreover this correction also includes the very embedded phases of star formation, a phase we are not able to probe in our cluster population since our detection is based at optical wavelengths. We therefore opt to correct the H$\alpha$ fluxes by the median extinction determined in clusters with ages below 10 Myr. The correction at the H$\alpha$ wavelength has been estimated using the starburst extinction formulation \citep{calzetti2000}, similarly to that used for the cluster SED analysis\footnote{Under the assumption that stellar continuum and ionised gas emission are affected by the same reddening}.  This method has the advantage of using similar extinction corrections obtained at comparable wavelengths and time scales for both the cluster population and the tracer of star formation, therefore facilitating the comparison between the two. Total extinction corrected H$\alpha$ luminosities are converted into SFR using the \citet{KE12} relation assuming a Kroupa IMF, as done in the SED analysis of the clusters. Because of the small photometric errors resulting in a few percent error on the H$\alpha$ flux, and considering the uncertainties on the extinction correction applied, we decide to apply a 10\% error to the total estimated SFR, as we believe this to be a more realistic assumption of the true error.

\begin{figure*}
    \centering
    \includegraphics[width=0.95\textwidth]{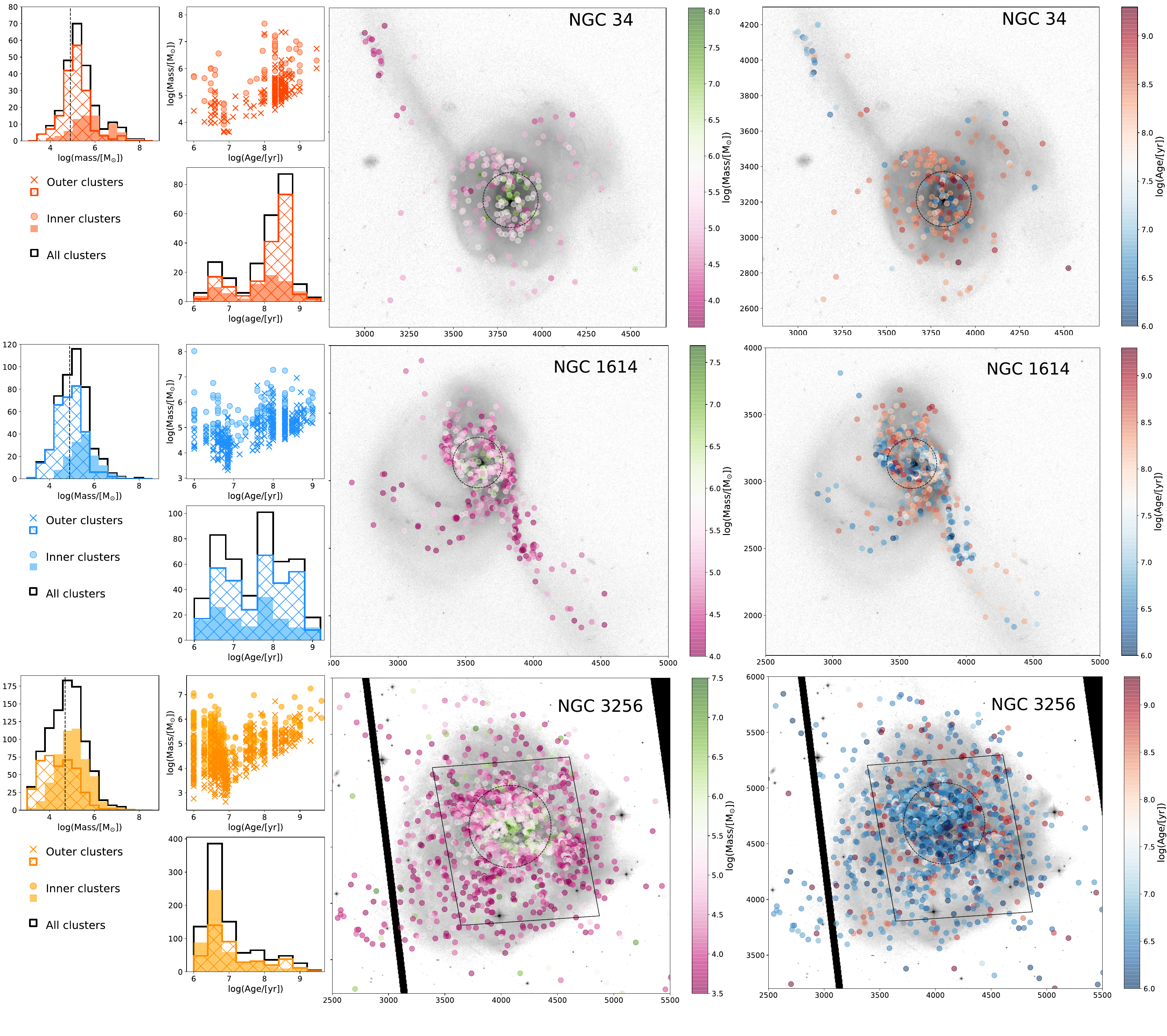}
    \caption{Cluster physical properties as a function of galactic position. For each target we show in the left panel the statistical properties (age and mass distributions, age--mass diagram) of the clusters that satisfy the selection criteria (black solid histograms) and of those located inside (solid filled histograms, filled dots) or outside (grid histograms, crosses) the R$_{80}$ (indicated by the dotted circle in the central and right panel). The vertical black dashed line in the mass histogram shows the mass limit used for the cluster mass function analysis and reported in Table~\ref{tab:MF}. The central panel shows the position of the clusters within their host galaxies, colour--coded accordingly to their determined mass (age). We use as background image the $V$ band, plotted in log scales and with the north rotated up. The sizes of the circle in kiloparsec are listed in Table~\ref{tab:ha}.}
    \label{fig:agemass1}
\end{figure*}

The $\Sigma_{\rm SFR}$ within the 80\%  radius and total $\Sigma_{\rm SFR}$ are obtained by dividing the SFR by the area enclosed by the circular aperture and the simplified contours, respectively. In the case of NGC3256, the $U$ band coverage is not as extended as in the other bands (we show the footprint of the U band coverage in the bottom panels of Figure~\ref{fig:agemass1}). In Figure~\ref{fig:ds9cont} we show the 3$\sigma$ and the simplified contours of NGC3256 obtained from the H$\alpha$ emission map which are used to estimate the radius of the circular region containing 80\% of the H$\alpha$ flux. However, since we require clusters to be detected in $UBVI$ bands for the $\Gamma$ and mass function analyses, we determine total SFR and $\Sigma_{\rm SFR}$ using the intersection of the $U$ band footprint and the 3 $\sigma$ H$\alpha$ detection boundary as showed by the polygonal solid line in the stamps of NGC3256.

\begin{table*}
\centering
\begin{tabular}{l c c c c c c}
\centering

Galaxy & redshift & MW $E(B-V)$ & NII$_{6584}$/H$\alpha$ & $E(B-V)_{cl}$  & 3$\sigma_{H\alpha}$& R$_{80\%}$ \\
   &  & mag & & mag & [$10^{-18}$ erg/s/cm$^2$]& [kpc] \\
\hline
NGC 34 & 0.01962 & 0.027 & 1.08$(a)$ & 0.68 & 0.1 & 2.44  \\
NGC 1614 & 0.01594 & 0.154 & 0.53$(a)$ & 0.62 & 0.3 & 2.38 \\
NGC 3256 & 0.00935 & 0.11 & 0.33$(b)$ & 0.61 & 0.5 & 2.74 \\
NGC 3690 & 0.01041 &0.17 & 0.39$(a)$ & 0.56 & 0.1 & 3.56 \\
NGC 3690A & 0.01041 &0.17 & 0.39$(a)$ & 0.65 & 0.1 & 4.10 \\
NGC 3690B & 0.01041 &0.17 & 0.39$(a)$ & 0.45 & 0.1 & 3.00 \\
NGC 4194 & 0.00834 & 0.016 & 0.47$(a)$ & 0.70 & 0.5 & 1.02\\
NGC6052 & 0.01581 & 0.076 & 0.16$(a)$ & 0.38 & 0.1 & 3.98 \\
\hline
\end{tabular}
\caption{HiPEEC galaxy properties. Redshift and Milky Way extinction are obtained from NED. NII$_{6584}$/H$\alpha$ are from $(a)$ \citet{MK06}, $(b)$ \citet{moran99}. $E(B-V)_{cl}$ is the median extinction in clusters younger than 10 Myr and masses larger than the minimum mass cut listed in Table~\ref{tab:sfr}. These extinction values have been used to correct the observed H$\alpha$ fluxes of each galaxy. R$_{80\%}$ is the radius that simplifies the area of the galaxy containing 80\% of the H$\alpha$ emission. A reminder that NGC3690A is also referred to as IC694, while NGC 3690B is the south-west companion.}
\label{tab:ha}
\end{table*}

\begin{table*}
\centering
\begin{tabular}{l c c c c c }
\centering
Galaxy & SFR & SFR(80\%)& Area & $\Sigma_{\rm SFR}$ & $\Sigma_{\rm SFR}$(80\%)\\
 &[\msunyr] & [\msunyr] & kpc$^2$ & [\dsfr] & [\dsfr]  \\
\hline
NGC 34	& 5.8(0.6) &	4.7(0.5) &	87.03	& 0.07 (0.01) &	0.26(0.02)	\\
NGC 1614	& 27.4(2.7) & 22.6(2.3) & 80.94 &	0.34(0.03)&	1.54(0.15)	\\
NGC 3256	& 44.6(4.5)& 37.7(3.77) & 78.06	& 0.57(0.06) & 1.52(0.15)	\\
NGC 3690A & 28.9(2.9)& 25.4(2.5) &	104.52 & 0.28 (0.03) &0.49(0.05)	\\
NGC 3690B & 18.5(1.8) & 17.4(1.7) &	45.74	& 	0.40(0.04) & 0.62(0.06) \\
NGC 3690& 47.4(4.9) & 37.6(3.7) &	150.26  & 	0.31(0.03) & 0.95(0.09) \\
NGC 4194	& 13.6(1.4)	& 11.2(1.1) & 29.26 &	0.46(0.05) & 3.59(0.36) \\
NGC 6052 & 15.3(1.5) &	11.6(1.1) & 121.72 & 0.13(0.01) & 0.23(0.02) \\

\hline
\end{tabular}
\caption{SFR and SFR per unit area estimated for the total galaxy (delimited by the solid blue lines in Figure~\ref{fig:ds9cont}) and within the circular area defined by R$_{80\%}$. The area reported in the fourth column is the total area within the simplified 3$\sigma_{H\alpha}$ contours shown in Figure~\ref{fig:ds9cont}. A reminder that NGC3690A is also referred to as IC694, while NGC 3690B is the south-west companion.}
\label{tab:sfr}
\end{table*}

In Table~\ref{tab:sfr}, we list the SFR and SFR per unit area, $\Sigma_{\rm SFR}$, for the whole galaxy, and estimated within the circular area defined by R$_{80\%}$. 

We have searched the literature for SFR values of the HiPEEC galaxies to compare with our total SFR estimates. \citet{U2012} report UV and FIR SFR for 3 of our targets, NGC34, NGC1614, NGC3690. The values quoted for UV and FIR, respectively, are 2.5 vs. 31.6 \msunyr\, for NGC34, 3.2 vs. 47.9  \msunyr\, for NGC1614, 8.9 vs. 89.1 \msunyr\, for the NGC3690 system. The values have been estimated using a Chabrier IMF assumption which is very close to the Kroupa IMF calibration used in this work. Considering that UV is significantly more affected by extinction and that FIR offers an extinction free view of the total SFR of the galaxy, our derived values obtained from extinction corrected H$\alpha$ emission are well within these ranges. The values reported in the literature for NGC4194 change quite widely. From FUV and optical long-slit HST spectroscopy,\citet{w04,h06} report a SFR of about $7$ \msunyr\, for the central star-forming knots and a total SFR for the galaxy between 30 and 40 \msunyr. When compared to our values, we need to account for a correction from Salpeter to Kroupa IMF (a factor of 0.68), and that their H$\alpha$ luminosity has been corrected for $E(B-V)$ values $\geq 0.8$ mag, therefore higher than the value adopted here of 0.7 mag derived from the clusters SEDs. Therefore, our H$\alpha$ luminosity is at least 0.74 smaller than their extinction corrected values. Thus, our SFR values are within a factor of 2 of those reported by \citet{h06}. Interestingly, our SFR values are very close to the one reported by \citet{storchi94} obtained with integrated spectroscopy. They report 38 \msunyr\, corresponding to 13.2 \msunyr, after conversion from Salpeter to Kroupa IMF and $H_0$ from 50 to 70 km s$^{-1}$ Mpc$^{-1}$, adopted in this work. \citet{storchi94} also reports SFR measurements for NGC6052 of 11.3 \msunyr which would correspond to $\sim 4$ \msunyr\, after the $H_0$ correction. This value is about 4 times lower than our reported value for this galaxy. The difference in this case arises partially from missing flux outside the slit (notice the difference between the area of the H$\alpha$ emission between NGC4194, compact, and NGC6052, extended) and from the extinction correction applied to H$\alpha$, $E(B-V)=0.2$ mag versus our 0.38 mag, that results in a factor of 1.7 in the correction. Finally, our extinction corrected H$\alpha$ SFR for NGC3256 are in very good agreement with SFR values reported in \citet{goddard10}, and more recently by \citet{michi20}, obtained from FIR luminosity and H$\beta$ fluxes corrected for extinction, respectively. We notice that our value is about 30\% lower than the one reported in \citet{chandar17}, obtained by correcting H$\alpha$ with 24 $\mu$m  emission. 

\begin{figure*}
    \centering
    \includegraphics[width=0.95\textwidth]{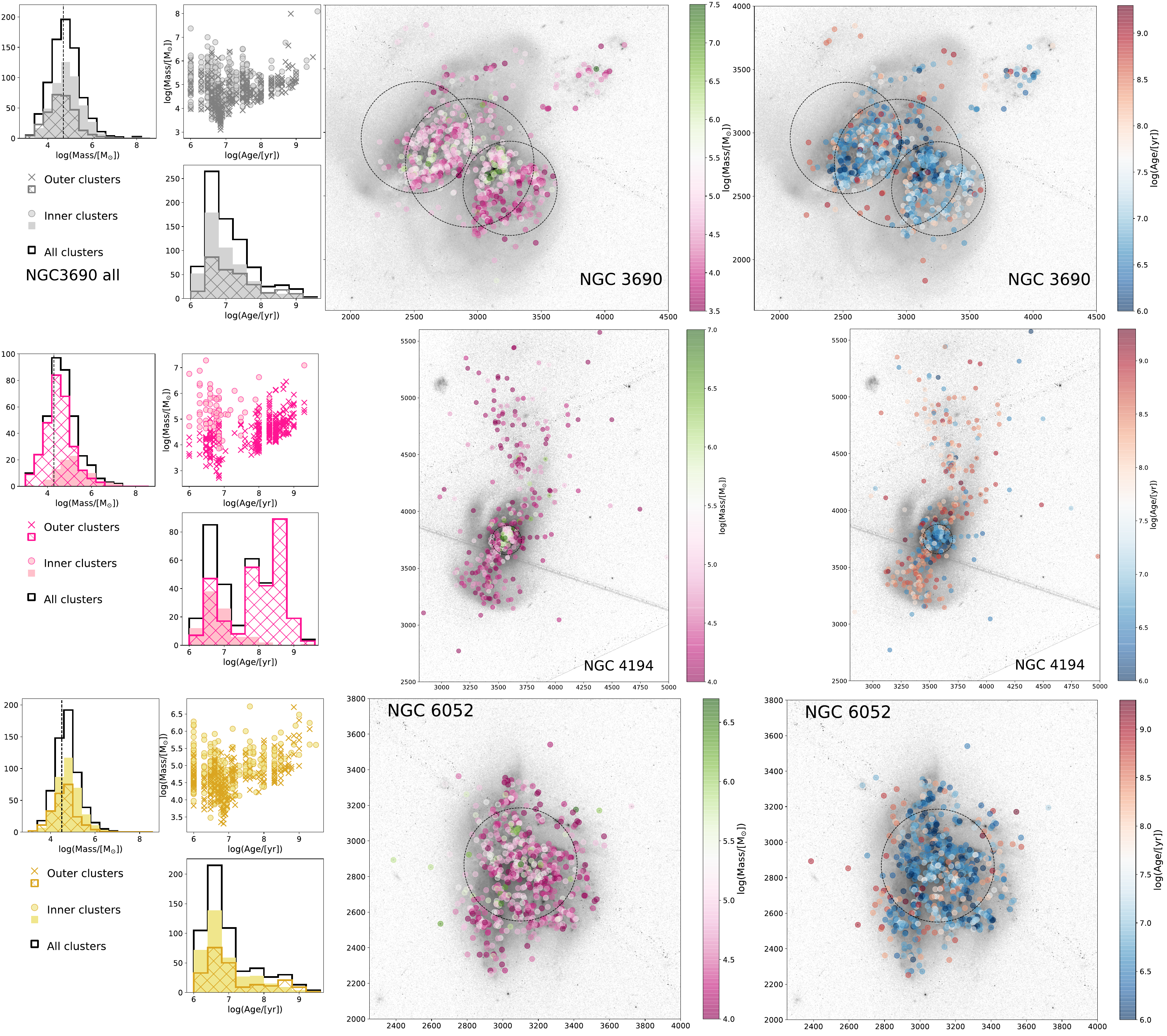}
    \caption{Cluster physical properties as a function of galactic position. See caption of Figure~\ref{fig:agemass1}.}
    \label{fig:agemass2}
\end{figure*}

\section{Results}
\label{sec:results}

\subsection{Cluster populations as a function of position in the galaxy}
\label{sec:cluster_pos}
In this section, we focus on the general properties of the cluster populations as a function of position within their host systems. In Figures~\ref{fig:agemass1} and \ref{fig:agemass2}, we show on the left column the age-mass diagram, accompanied by age and mass histograms of the candidate clusters located within (labelled as inner clusters) or outside (outer clusters) the 80\% radius of each target. In the central and right panels, we show the positions of the cluster candidates within each galaxy ($V$ band frame) colour-coded accordingly to their masses and ages. The circular dashed line, consistent with the area where 80\% of H$\alpha$ is observed, shows at which location we separate inner and outer clusters. Within the mass completeness limits (discussed in Section~\ref{sec:fincat} and shown in the mass histograms as vertical dashed line), we find that generally the most massive clusters are located in the inner region of the host galaxies. This is especially true for clusters younger than 10 Myr in all targets. In some galaxies like NGC34, NGC1614, NGC4194, the most massive inner clusters can be up to 2 orders of magnitude more massive than the outer cluster population, while in the other targets the difference can reach up to a factor of 10. In the outer regions, clusters younger than 10 Myr are mostly located in tidal features, streams and distorted spiral arms. Interestingly at ages larger than 10 Myr, we do not observe any pronounced difference in the mass ranges covered by the inner and outer clusters.  Similar trends between cluster ages and position within their host galaxies have been found in the numerical simulations by \citet{kruijssen12a}.  There the authors observe that between the last passage and the final coalescing phase very young clusters are found in spiral arms, tidal streams and inner regions, while older clusters are more randomly distributed because they are ejected away from their original orbits by the interaction.

The cluster age histograms contain interesting clues about the recent star formation history of the host galaxy. While detection limits severely affect the total number of clusters recovered at age ranges above 10 Myr, we clearly see significant differences in the age distributions of clusters across the 6 targets. 

We will discuss here the main features of each target.

\subsubsection{NGC34}
\label{n34}
 NGC34 has already reached the final coalescing phase. Morphologically, the galaxy shows stellar shells far from the central regions, and streams of stars and gas that has led to formation of lower mass clusters. The pronounced age peak at $\sim$ 600 Myr suggests that the final merger event has taken place between 300 and 600 Myr ago, in agreement with a spectroscopic analysis performed by \citet{ss2007}. It is during that rapid merging phase that the galaxy has formed some of the most massive YSCs observed in the local universe \citep{bastian13}. From spectral fitting \citet{cz2014} reports that cluster 1 in NGC34 is about $10^7$ \msun\, and has an age of 100$\pm$30 Myr. Our SED fit produces a best solution in very good agreement with previous published results for this cluster (\# 183 in our catalogue). We find an age of 100$^{+90}_{-10}$ Myr, a mass of 1.9$^{+0.4}_{-0.1}\times10^7$ \msun, and an internal extinction of $E(B-V)=0.14^{+0.04}_{-0.11}$ mag. The molecular gas conditions in the central region of NGC34 are consistent with being heated by the AGN rather than a nuclear starburst \citep[e.g.][]{xu14}, whom report a rotating molecular gas disk of 200 pc in radius (corresponding to 16 px in our dataset). We do not detect any cluster within this central region. Young clusters with ages below 10 Myr are mainly associated with visible spiral structures in the inner region of the galaxy.

\begin{figure*}
    \centering
    \includegraphics[width=0.48\textwidth]{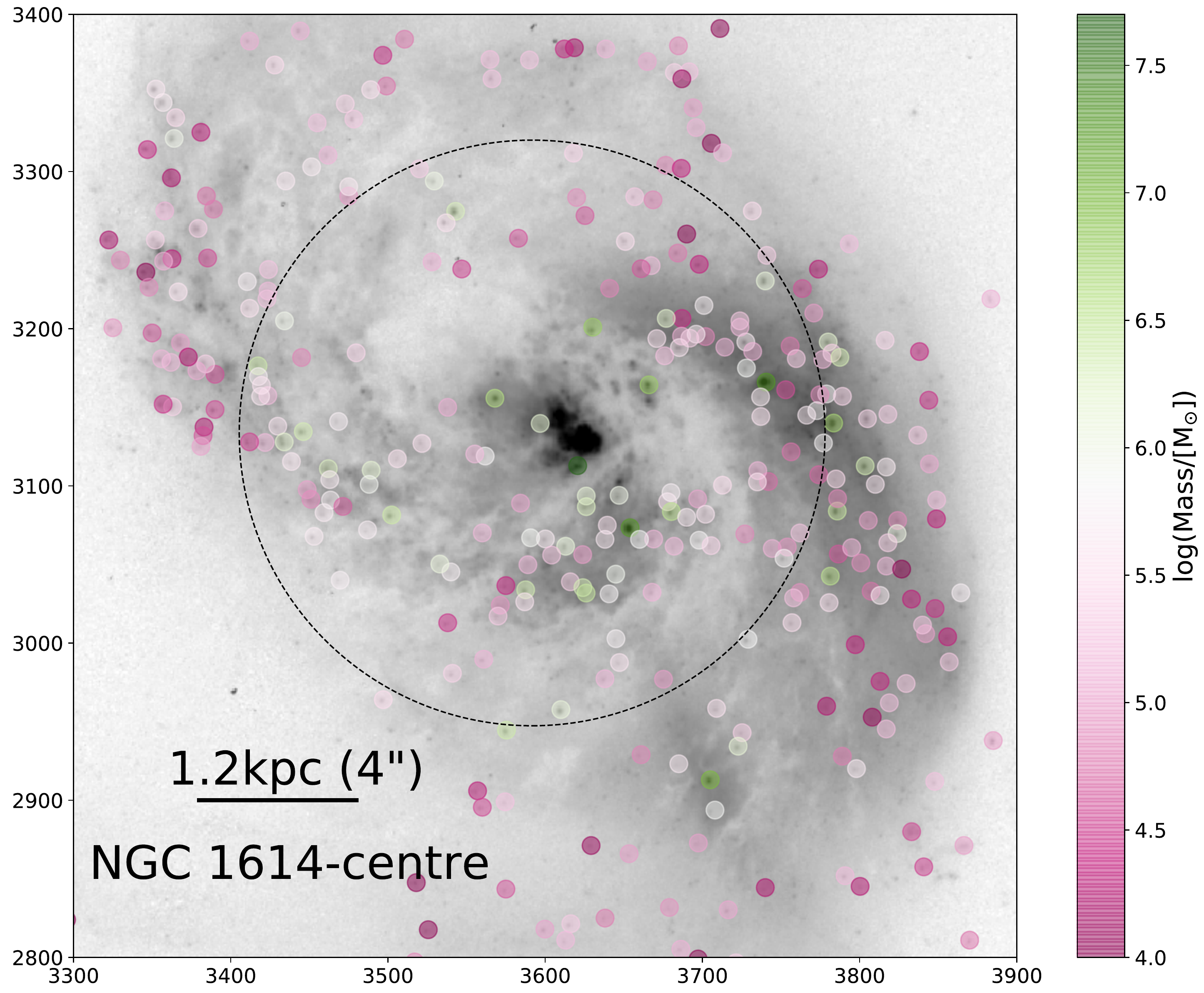}
    \includegraphics[width=0.48\textwidth]{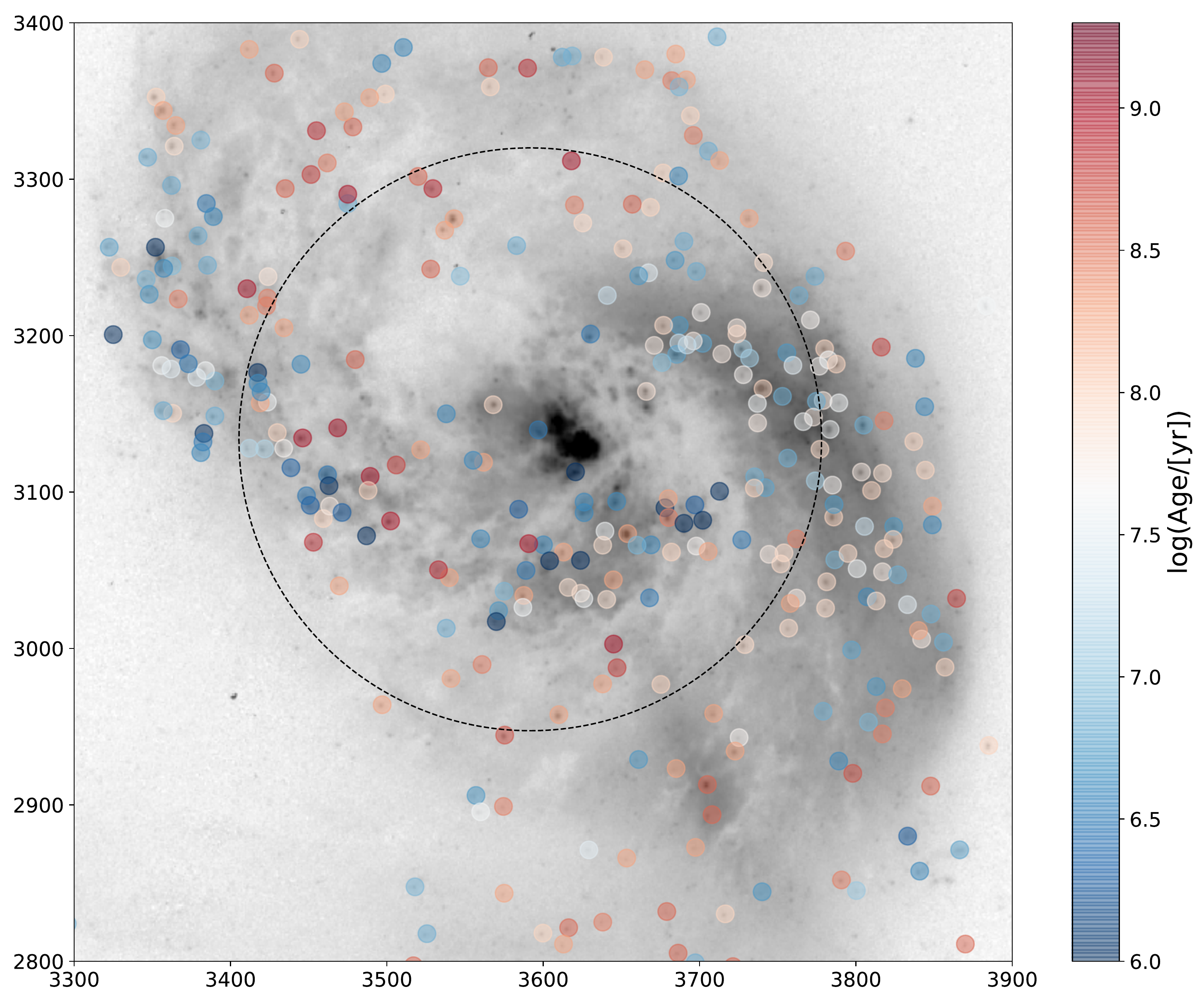}\\
    \includegraphics[width=0.48\textwidth]{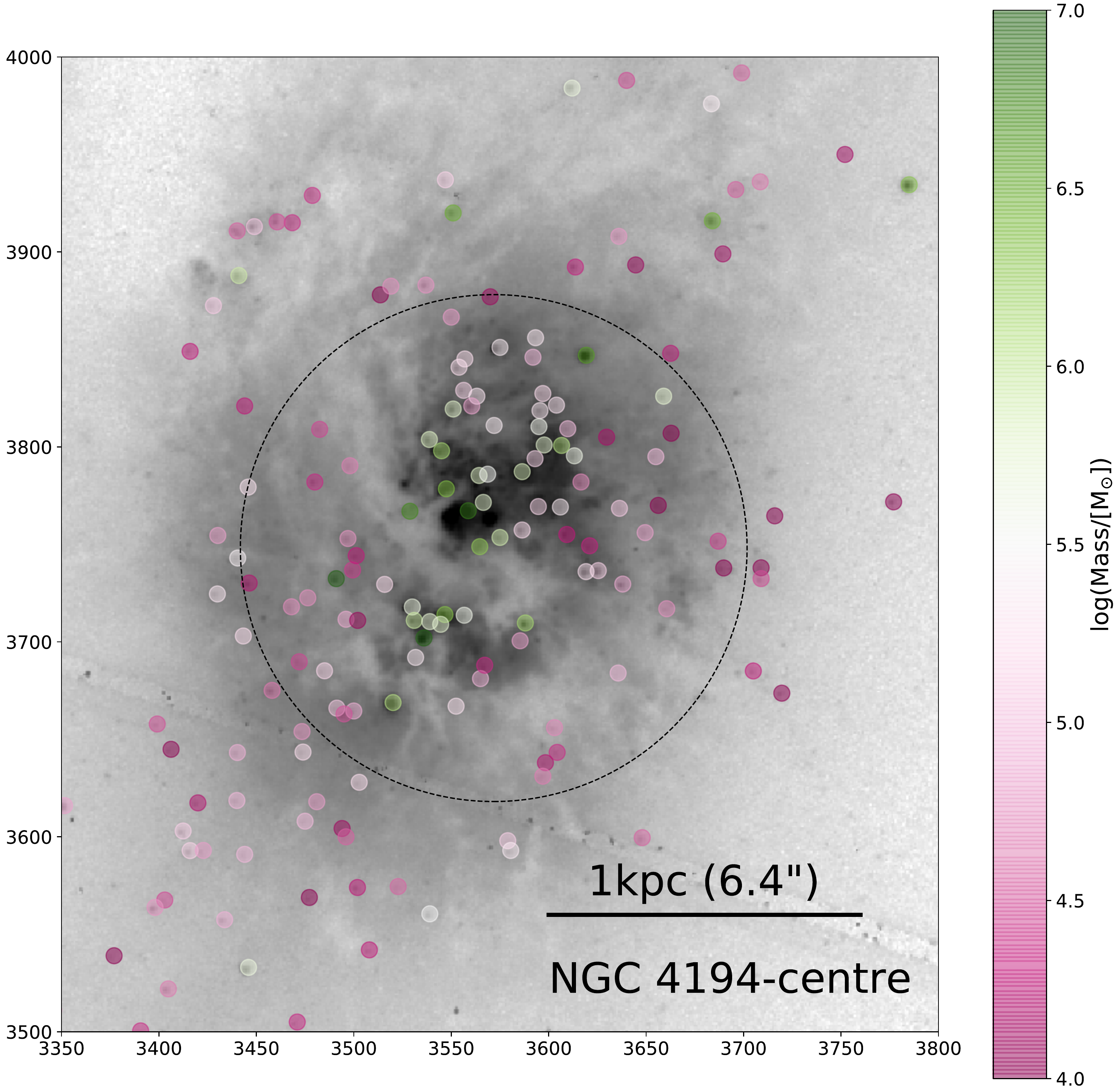}
    \includegraphics[width=0.48\textwidth]{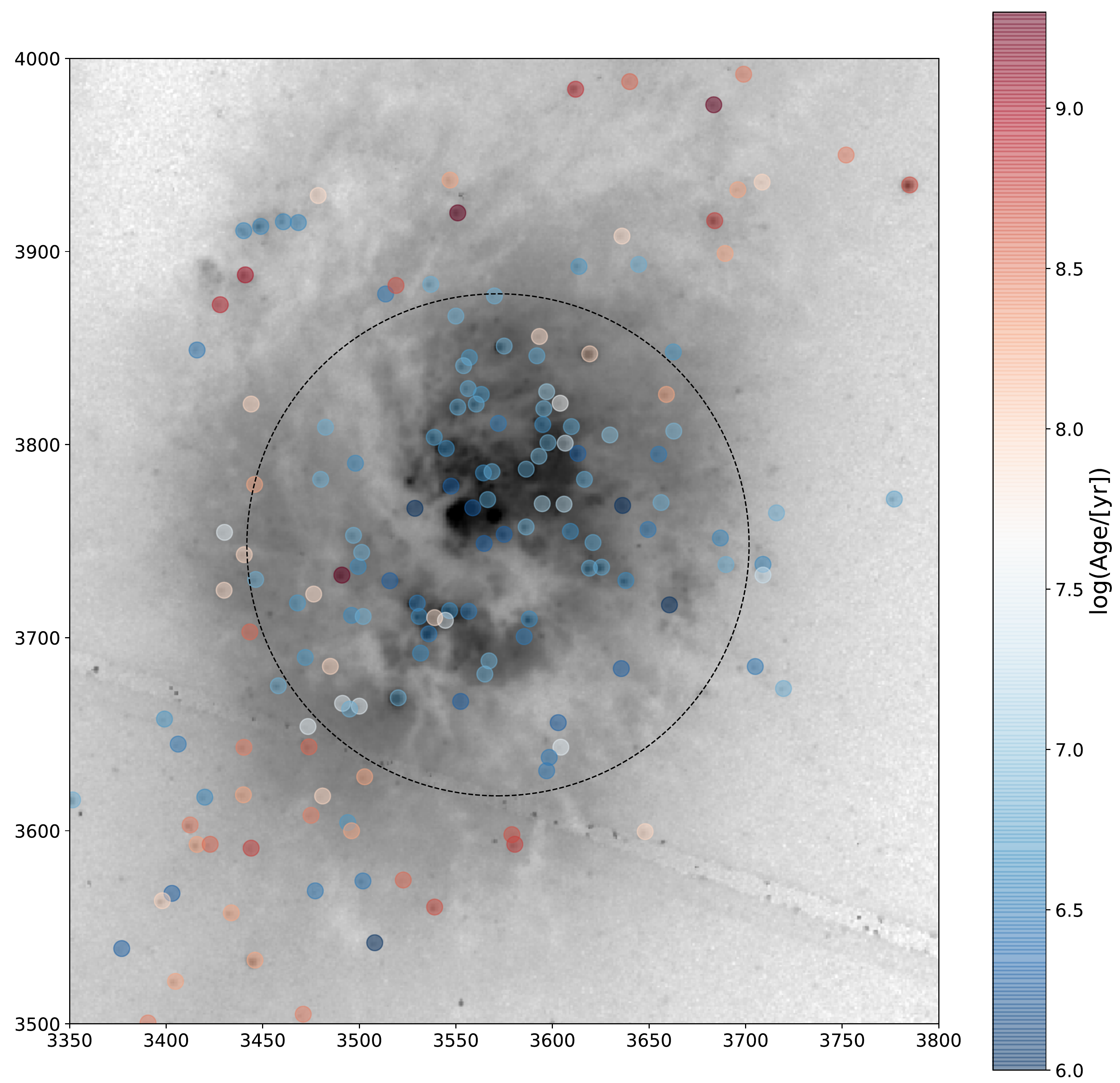}\\
    \caption{A zoom-in cutout of the inner kiloparsec of NGC1614 and NGC4194. The cluster positions are overlaid on the $V$band image colour-coded accordingly to their mass (left) and age (right). Extended clumps and nuclear regions have been flagged during visual inspection. The circular dashed lines delimit the region contained with R(80\%)}
    \label{fig:agemass_centre}
\end{figure*}

\subsubsection{NGC1614 and NGC4194}
\label{n1614}
NGC1614 and NGC4194 show a significant enhancement in cluster formation between ages of 100 and 600 Myr and are still forming clusters at elevated rates. Both systems are well known to have merger induced nuclear starburst rings \citep{AH2001, koenig13, koenig14}. By studying high-spatial resolution ($\sim$ 100 pc) mapping of the $^{12}$CO(2-1) transition, \citet{koenig13, koenig14} report the detection of tens of giant molecular associations, barely resolved in size and with masses above $10^7$ \msun, corresponding to a total mass in molecular gas clumps of 4 and 8 $\times 10^8$ \msun\ located within the central region of $\sim$ 300 and 230 pc in radius in NGC4194 and NGC1614, respectively. In   Figure~\ref{fig:agemass_centre}, we zoom within the inner regions of these two galaxies. We find that clusters selected at optical wavelengths are offset from the central dusty regions where these molecular clumps have been detected. While it is plausible that these gas clumps are the progenitors of the very massive clusters (masses above 10$^6$ \msun) detected in these inner regions, it is also difficult to provide a direct link as optical and millimetre wavelengths provide different temporal shots of the star formation happening in these systems. 

The combination of $^{12}$CO maps and denser gas tracers, also outside the nuclear regions, help to build a more complete picture of the feeding on the current starburst in both systems. Molecular gas with lower densities is transported from larger galactic scales to the nuclear ring and farther within the central starburst, where it reaches densities of $10^3-10^4$ \msun/pc$^3$ and high-pressure, typical of merging systems \citep{koenig16, koenig18}. As a result of these dense gas streams, several young and massive star clusters are observed within the central kpc of these targets. 

The right panels of Figure~\ref{fig:agemass_centre} show the age distributions of these massive clusters.  In NGC1614, we observe a quite large age gradient, with the youngest star clusters located in the south-east arm of galaxy. The north-west arm hosts mainly clusters older than 10 Myr. This region is also the brightest in the optical images published by \citet{linden17}, while in H$\alpha$ is not particularly bright (Figure~\ref{fig:ds9cont}), confirming that it is a dust-free more evolved region of the galaxy. This age gradient in the starburst of NGC1614 was also reported in other studies \citep{AH2001, SM2004}, suggesting that the starburst has been ongoing for at least ~100 Myr, and if we use the cluster population as a tracer it may have started already 500-600 Myr ago. On the other hand, the massive clusters, within 1 kpc of the centre of NGC4194, are all very massive and with young ages ($<10$ Myr), as already reported by previous studies \citep{w04, h06}. Outside this region, clusters are forming with lower masses and in a much smaller number. Morphological signatures, especially stellar shells in the southern part of the galaxy suggest that the massive progenitor of NGC4194 was an early type galaxy that has merged with a lower mass, gas rich component. Because of the confinement of the starburst to the centre of the galaxy, accompanied by a very young massive cluster population which shows no age gradient, the current starburst is very likely responsible for driving ionised gas winds observed in the H$\alpha$ maps.

\subsubsection{NGC3256, NGC3690, and NGC6052}
\label{n3690}
The remaining systems, NGC3256, NGC3690, and NGC6052 have increasing number of clusters with ages younger than 100 Myr and peaked below 10 Myr, suggesting a recent enhancement in their SFR. Very similar conclusions have been reached by different studies of the cluster population of NGC3256 \citep{goddard10, mulia2016} and of the NGC3690 system by \citet{randria19}. NGC3256 is well know to host a very powerful nuclear outflow powered by a nuclear starburst \citep{sakamoto14}. We see a clear enhancement of massive and very young cluster formation towards the inner regions of this galaxy, although the nuclear starburst and associated molecular outflow remain hidden at the wavelengths probed in our study. 

NGC3690 is the only system in our sample, where the two galaxies are still traceable, similarly to that observed in the more famous Antennae galaxies. In Figure~\ref{fig:agemass2} we see that the most massive clusters are currently forming in the peak of the H$\alpha$ luminosities of each component, which is close to the nuclear regions of NGC3690A and B. However, it is important to notice that also in this system, our H$\alpha$ tracer and optical study of the cluster population is not able to provide information about the extreme compact nuclear starburst of NGC3690A which coincides with the peak emission in dense gas and radio continuum \citep{aalto1997} detected in the galaxy. This starburst is almost completely extincted in the optical as is revealed by an H$\alpha$ depression where the CO emission peaks on the nucleus. As already observed by \citet{randria19}, whom include high resolution near-IR data in their analysis, we do not see a large enhancement of cluster formation in the overlapping region between the two systems of NGC3690. As a comparison, a large fraction of very massive clusters is detected in the overlap region of the Antennae system \citep{whitmore10}. That region is also where most of the dense gas and cold dust emission is detected in the Antennae \citep{whitmore14}. With a gas fraction of $\sim$30\% \citep{larson16}, it is very likely that cluster formation will drift towards the overlapping region of the NGC3690 system in its future evolution. This phase is also predicted by numerical simulations \citep{kruijssen12a, renaud15}, which find enhancement in star and cluster formation in the overlapping region because of the high gas pressure exerted by the merger during the final coalescing phase. 

NGC6052 shows a strongly perturbed morphology and distorted arms, morphologically dominated by star-forming clumps and stellar clusters. Clusters younger than 10 Myr are detected along the main body of the system. This uniform age distribution is supported by the MIR imaging and spectroscopic study of the galaxy performed by \citet{whelan07}. They find that the 16$\mu$m emission and MIR emission lines in the detected knots are consistent with being produced by  
ionisation from a very young stellar population ($<$6 Myr).

\subsection{Cluster Mass functions}
\label{sec:massfunc}

\subsubsection{Formalism}
The analysis of the cluster mass function has been performed using the same Bayesian inference method developed by \citet{johnson17} and implemented in \citet{messa18b}. We refer to \citet{messa18b} for the formalism used to define the likelihood function in the form of a Schechter and power-law function. We use the Bayes' theorem to derive the posterior probability distribution function of the truncation mass, M$_{c}$, and the slope, $\beta$, in the case of the Schechter function, and only $\beta$ in the case of a pure power law function. Flat priors are used for the slope (fixed to $-1<\beta<-3$) and the truncation mass. In the latter case, M$_{c}$ is estimated between the mass interval defined by M$_{min}$, listed in Table~\ref{tab:MF} for each target, and M$_{max}=10^9$ \msun. For both functions, the integral used to normalize the likelihood function is calculated up to an upper mass of M$_{up}=10^9$ \msun. The latter assumption implies that the maximum observed mass is M$_{max, obs}<<M_{up}$ ensuring that the power-law function does not have an upper limit. The posterior probability distributions have been sampled with the Python package {\tt emcee} \citet{emcee13}, which implements a Markov Chain Monte Carlo (MCMC) sampler from \citet{2010CAMCS...5...65G}. We use 100 walkers, each producing 600 step chains, and we discard the first 100 burn-in steps of each walker. This results in 50000 sampling values for each fit. As discussed in \citet{adamo20}, the advantage of this method is that it does not depend on the way the mass distributions are binned, therefore removing any bias caused by the least populated upper mass bins. The best fitted values are respectively the median values of the marginalised posterior probability distribution function for each of the Schechter function parameters, accompanied by a 1$\sigma$ confidence interval (the 16th to 84th percentile range of the marginalised posterior). For the power-law function, we report the median and 1$\sigma$ interval of the recovered slope values. 

\begin{figure}
    \centering
    \includegraphics[width=8.5cm]{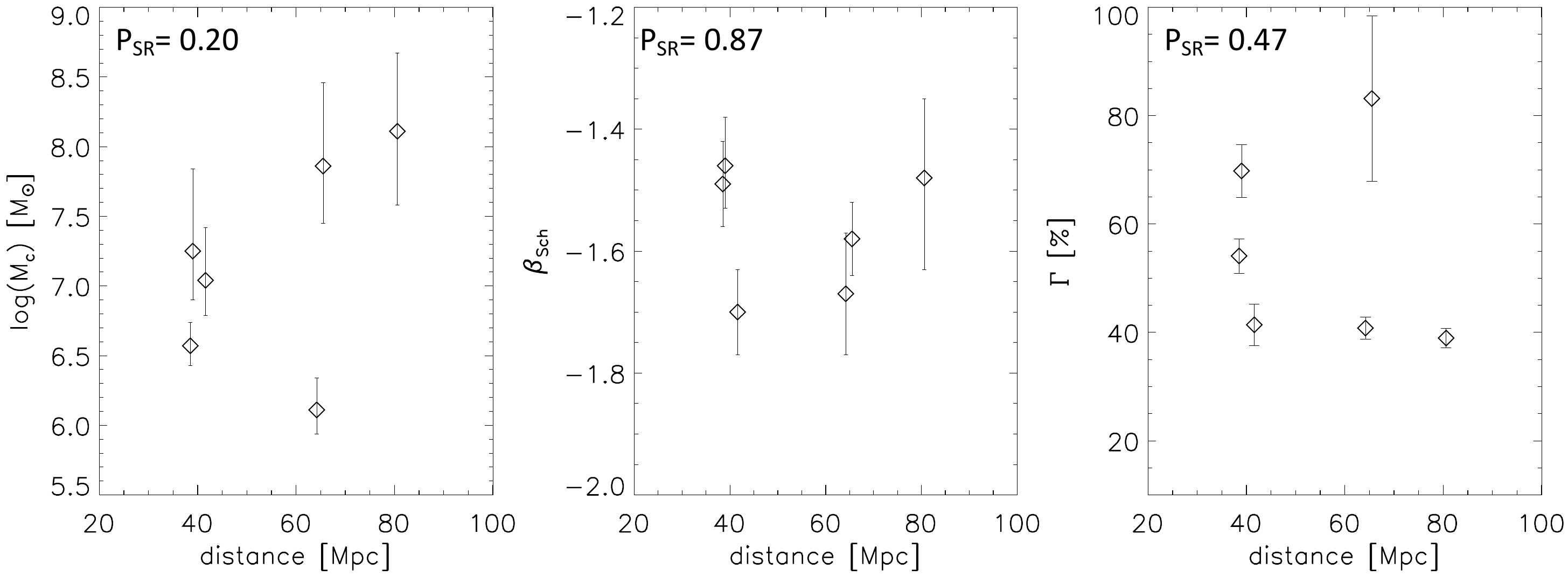}
    \caption{The recovered M$_c$, slope $\beta_{Sch}$, and $\Gamma$ values for the HiPEEC sample plotted as a function of the distance of the galaxies. The high probabilities, p$_{SR}$, associated to the Spearman's rank correlation analysis and reported in each panel, suggest that there is no significant correlation between the recovered cluster physical properties and the distance of the galaxies.}
    \label{fig:SR}
\end{figure}

\begin{figure*}
    \centering
    \includegraphics[width=0.32\textwidth]{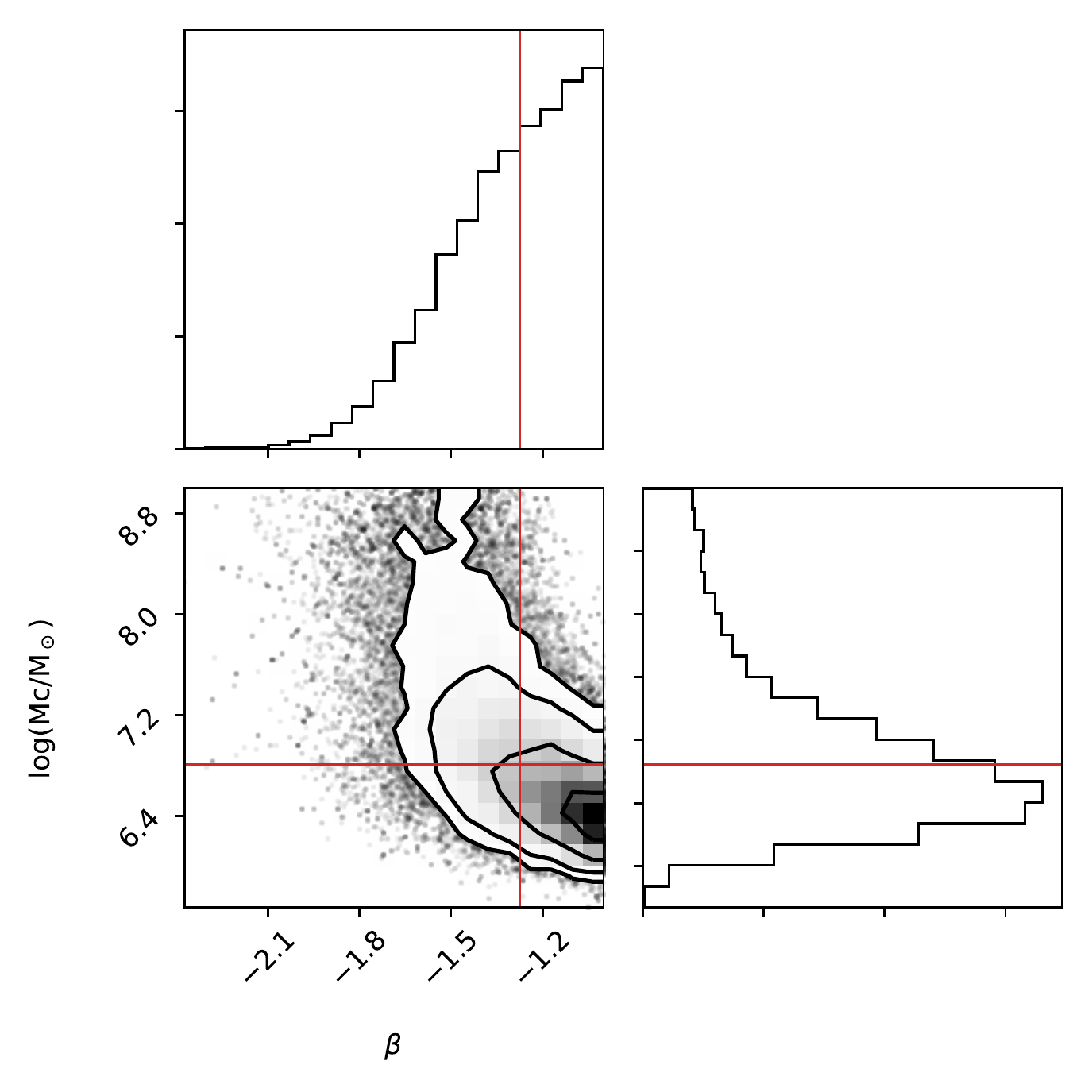}
    \includegraphics[width=0.32\textwidth]{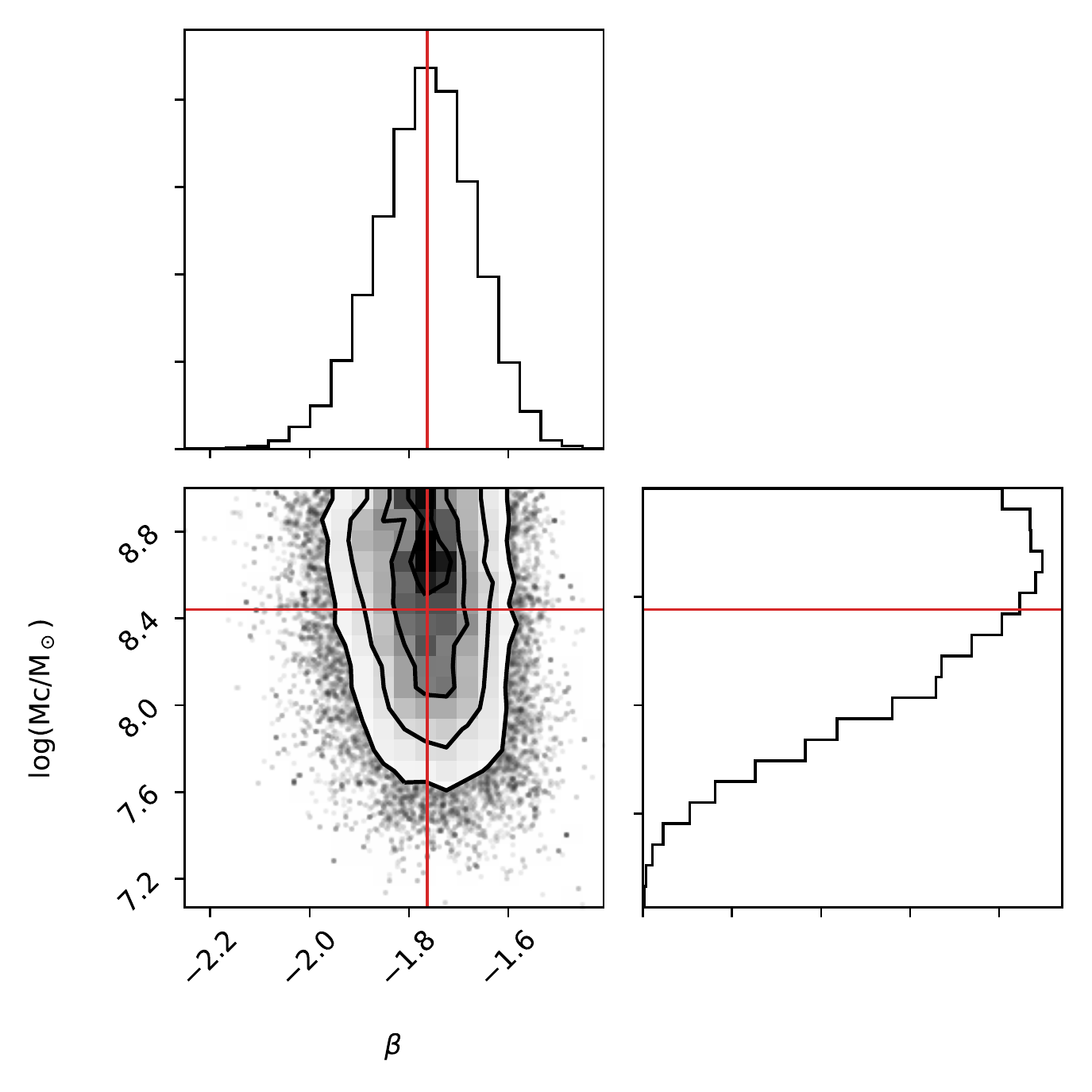}
    \includegraphics[width=0.32\textwidth]{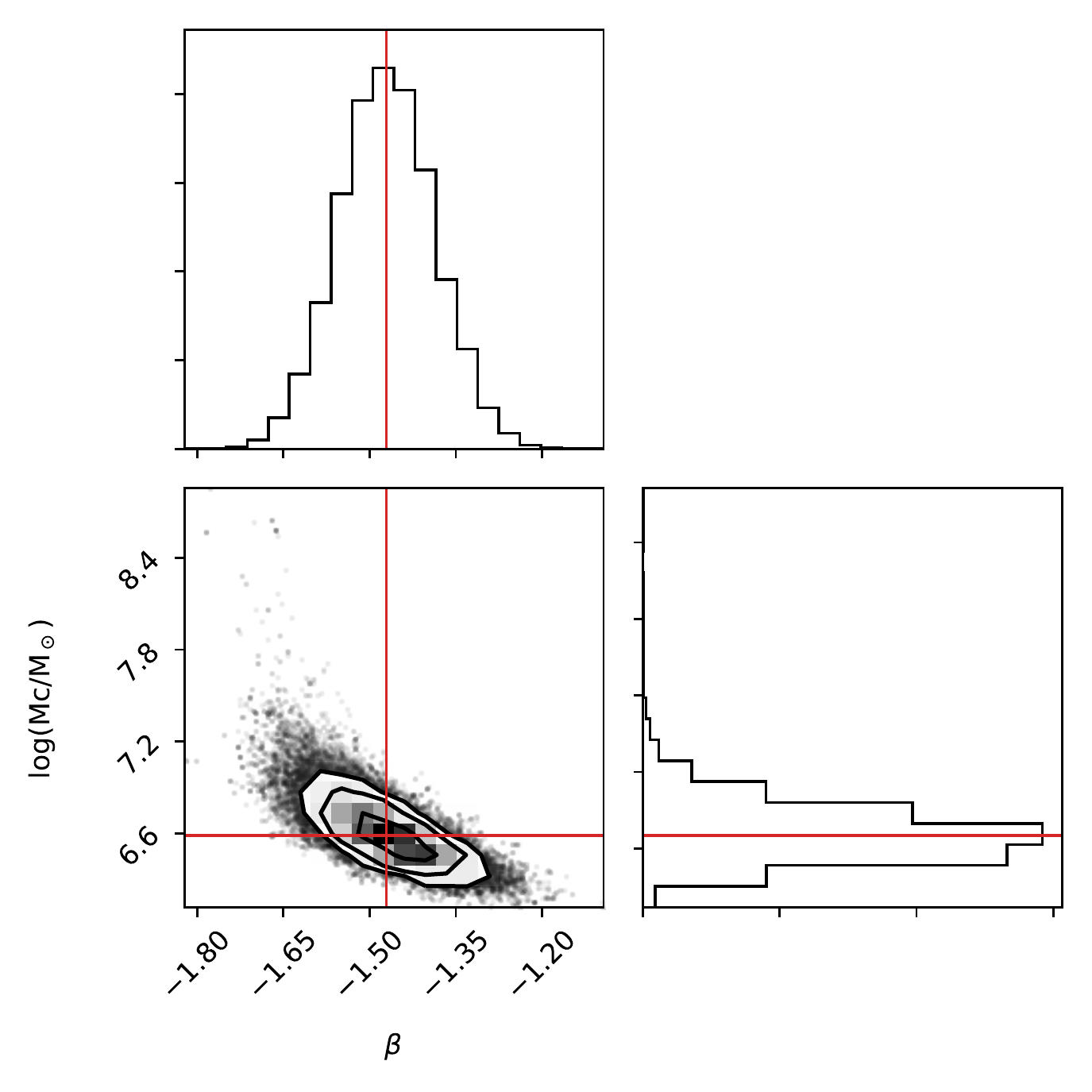}\\
    \includegraphics[width=0.32\textwidth]{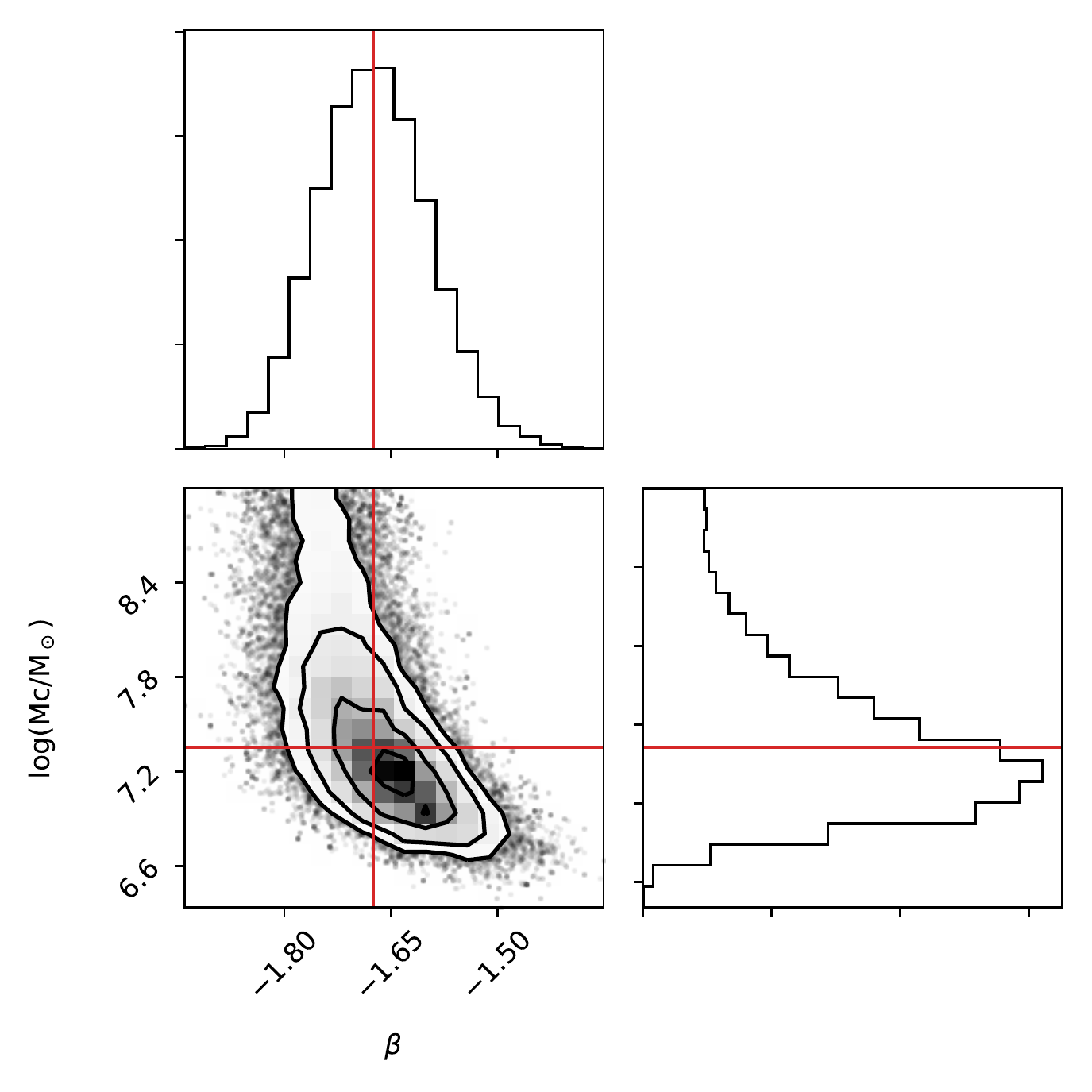}
    \includegraphics[width=0.32\textwidth]{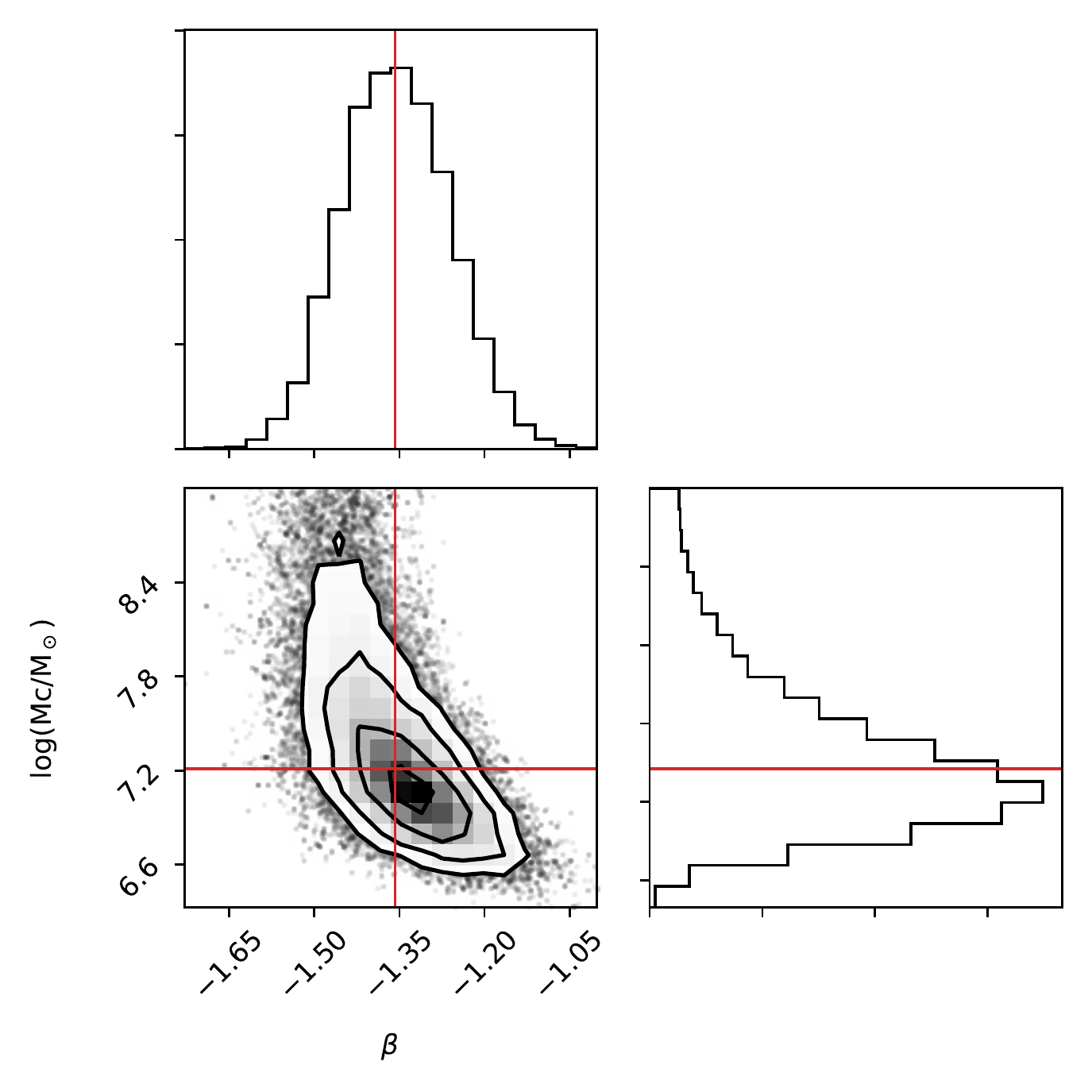}
    \includegraphics[width=0.32\textwidth]{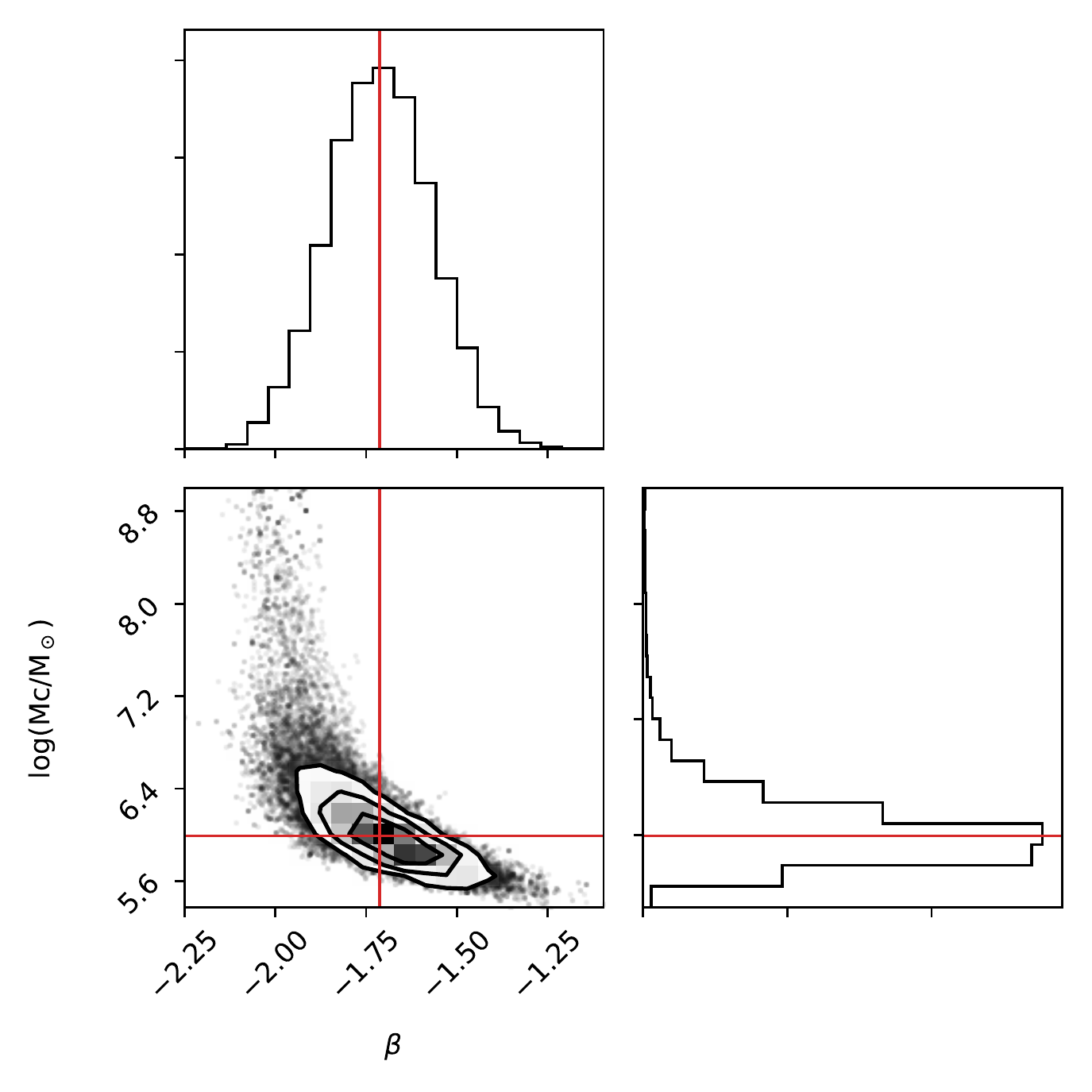}\\
        \caption{Bayesian Schechter function fit to the cluster mass function of NGC34 (top left), NGC1614 (top centre), NGC3256 (top right), NGC3690 (bottom left), NGC4194 (bottom centre), NGC6052 (bottom right). We used clusters masses above M$_{min}$ and age range 1--10 Myr. The corner plots show the parameter space visited by the walkers in our cluster mass function fitting analysis. The contours represent the 1, 2 and 3 sigma values of the density distribution around the two fitted parameters, M$_c$ and $\beta_{\rm Sch}$. The solid red lines show the median values of the mass and slope distributions reported in Table ~\ref{tab:MF}. We observe that when the fit does not converge the median is a poor representation of the peak value.}
    \label{fig:cornerplots_app}
\end{figure*}

\begin{figure*}
    \centering
    \includegraphics[width=0.32\textwidth]{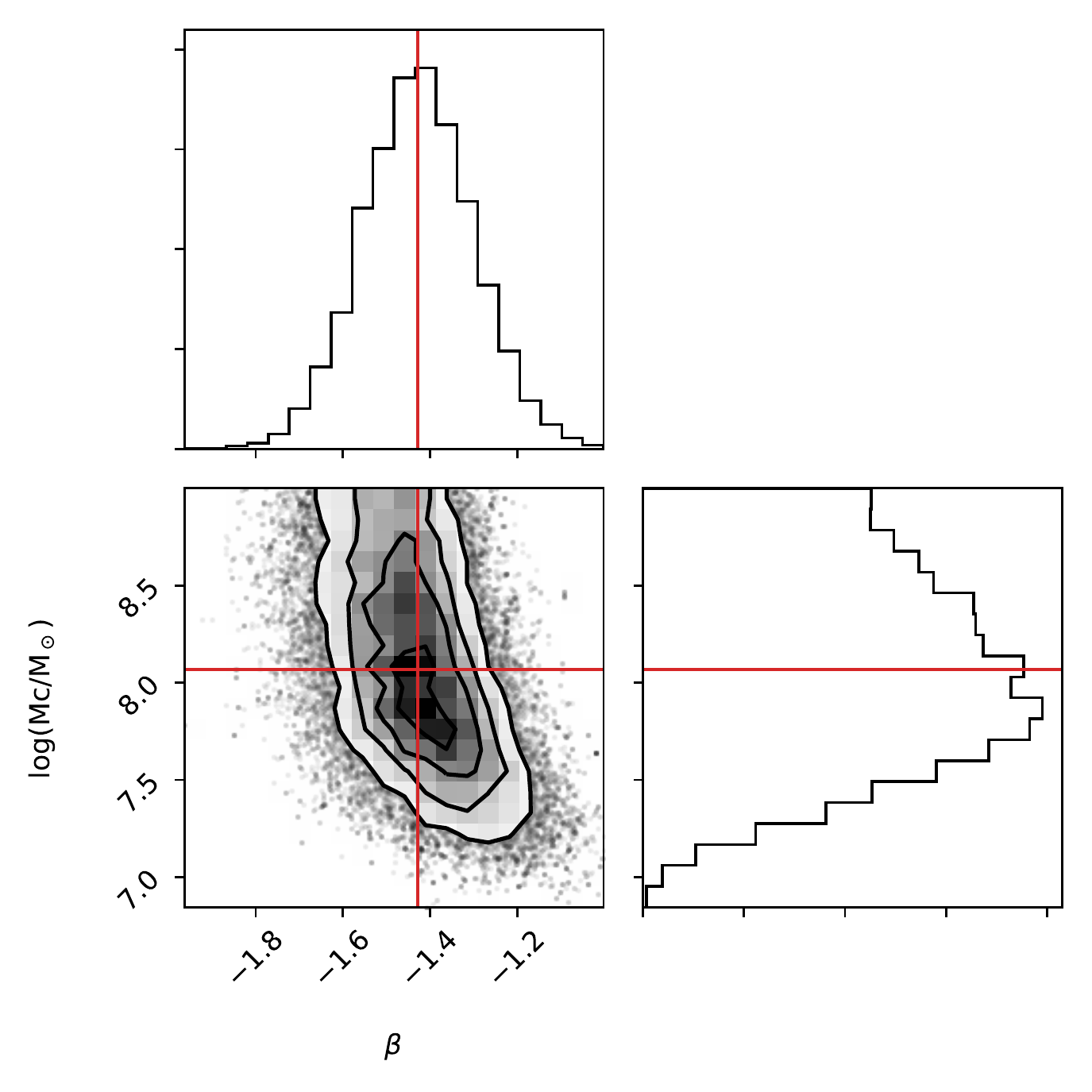}
    \includegraphics[width=0.32\textwidth]{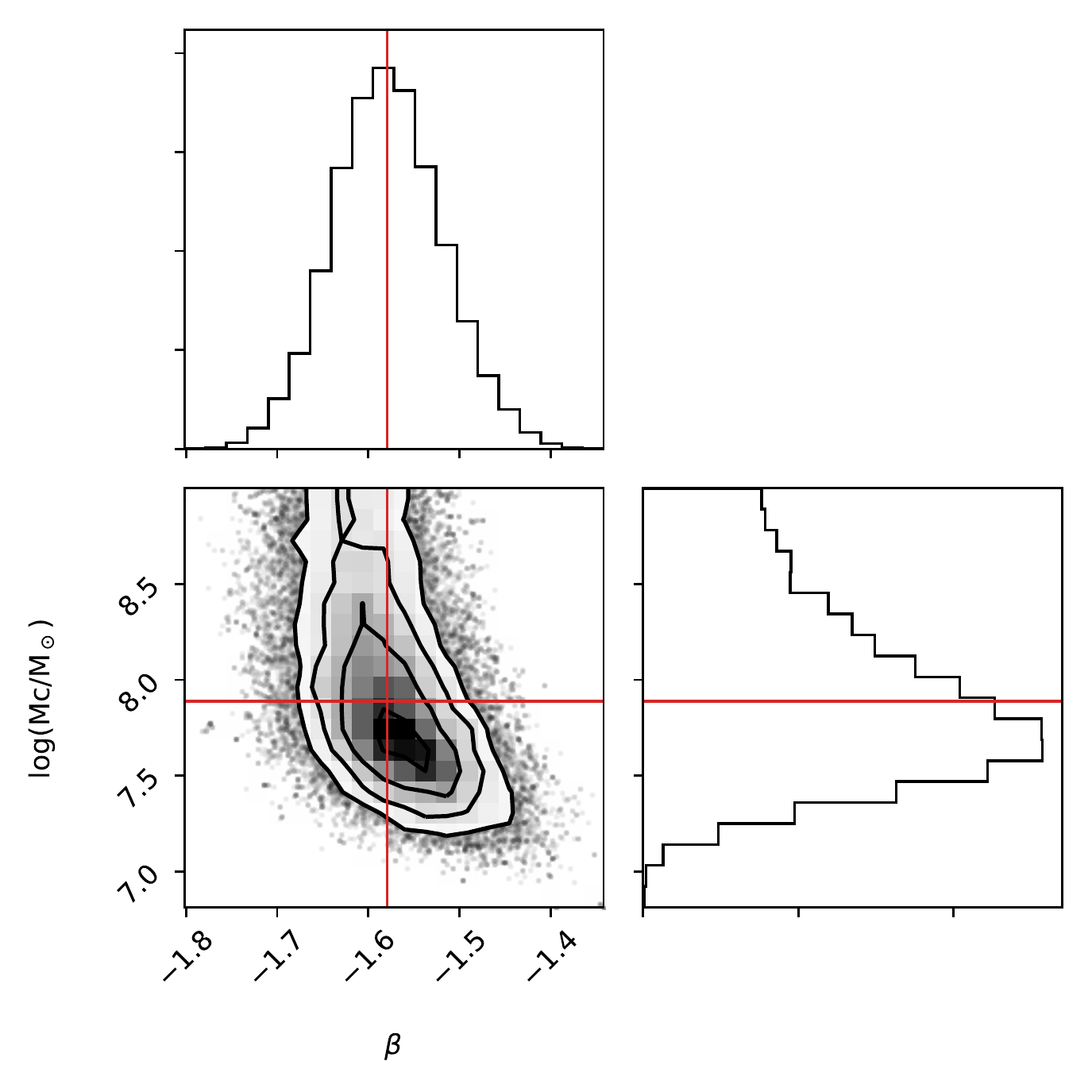}
        \includegraphics[width=0.32\textwidth]{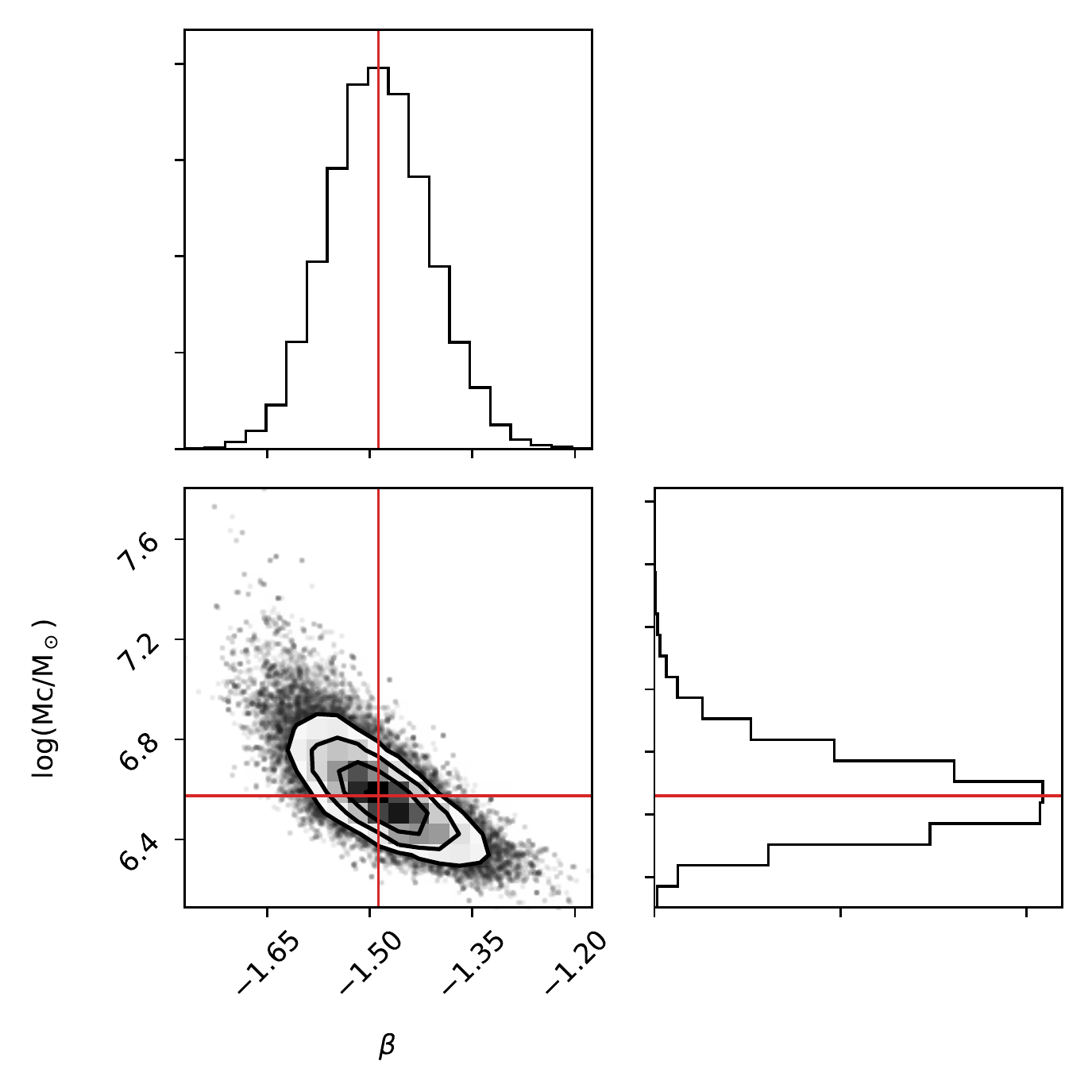}\\
    \includegraphics[width=0.32\textwidth]{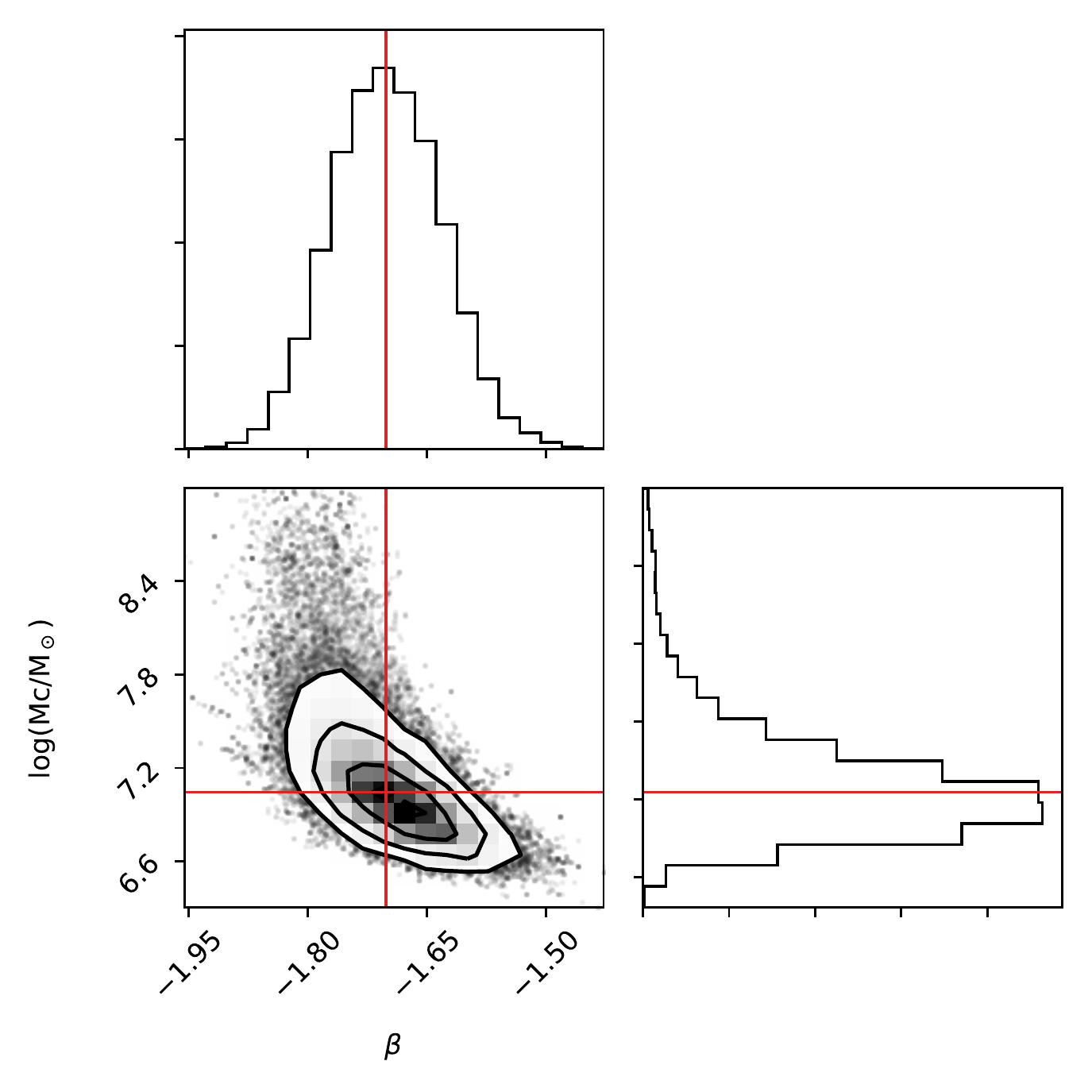}
     \includegraphics[width=0.32\textwidth]{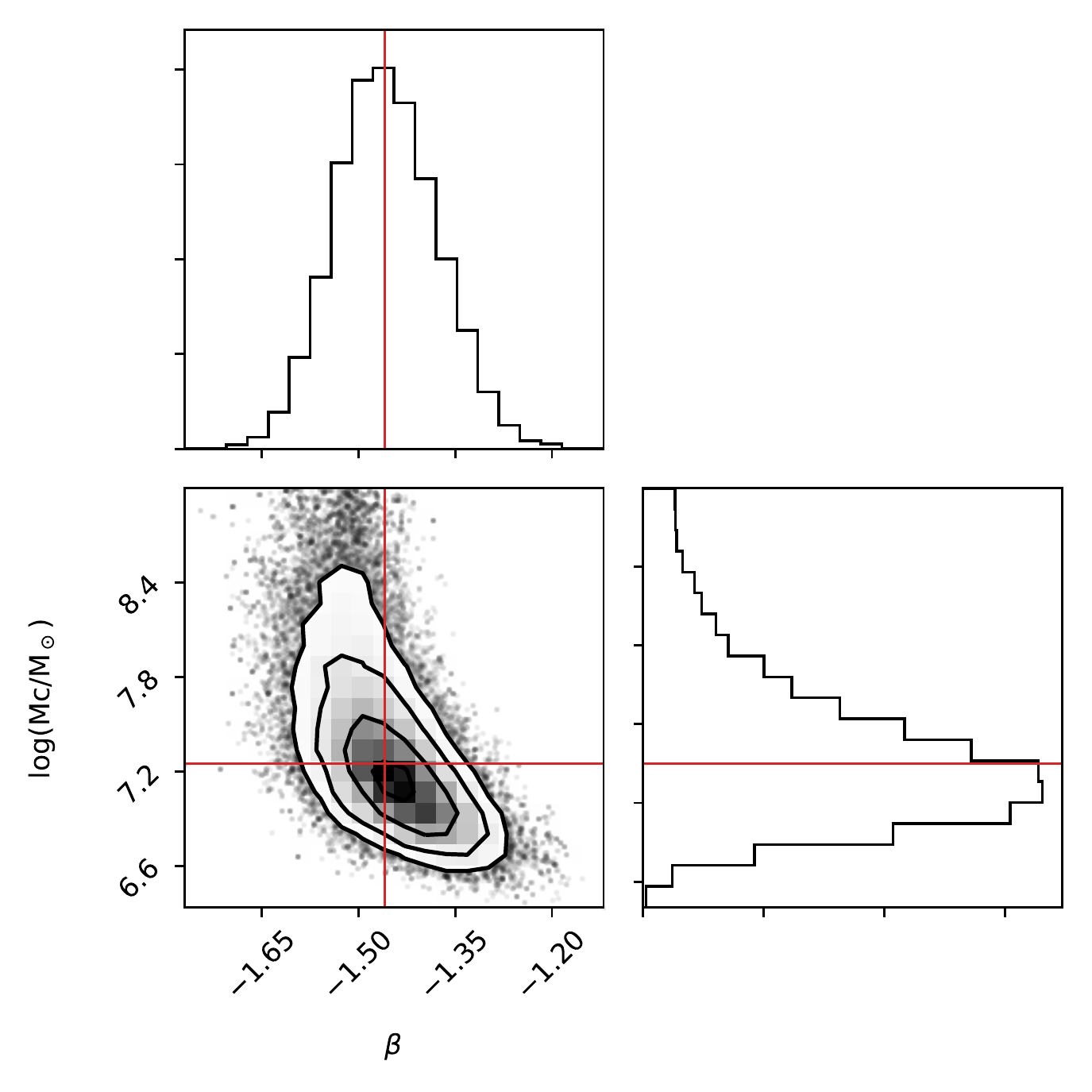}
    \includegraphics[width=0.32\textwidth]{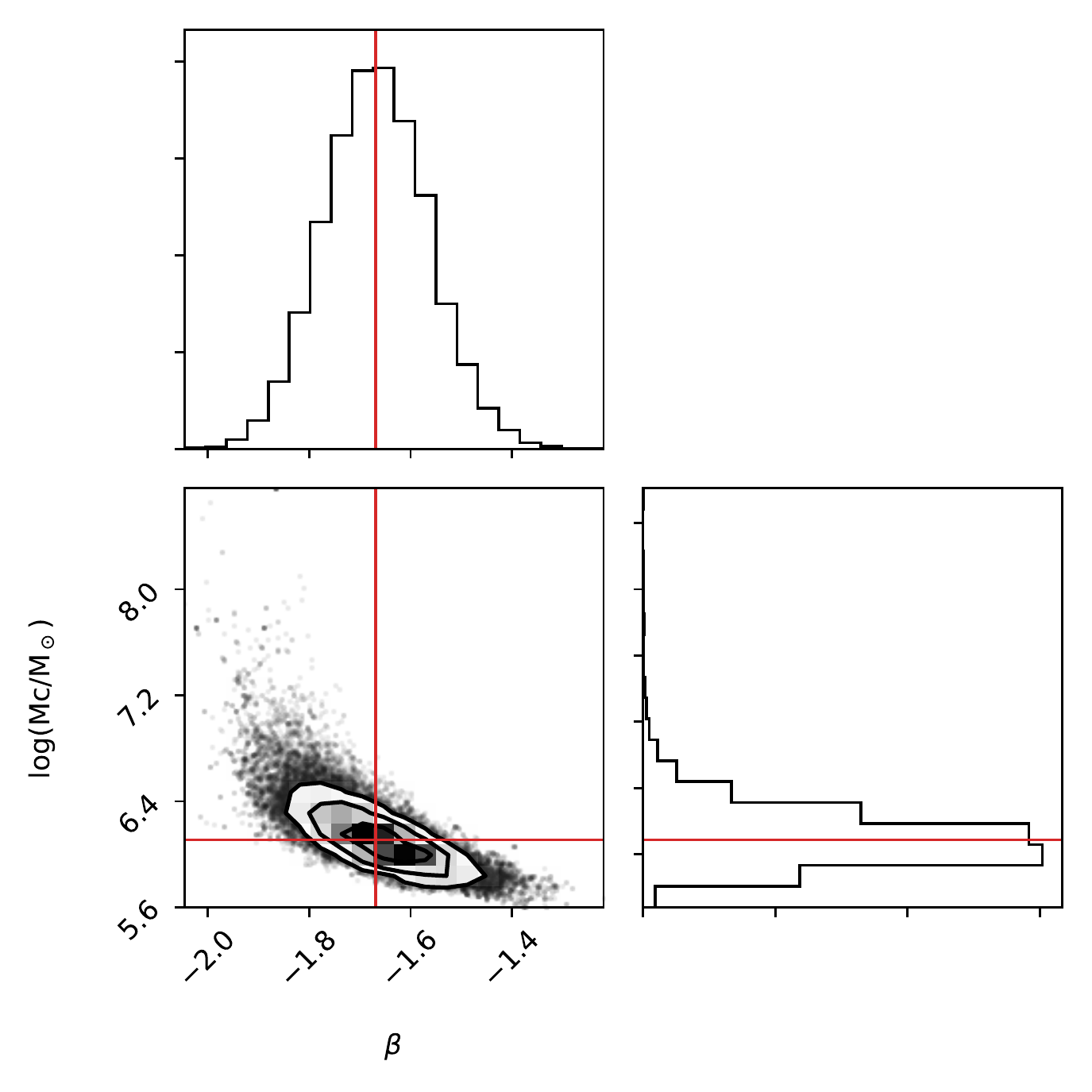}
 
    \caption{Bayesian Schechter function fit to the cluster mass function of NGC34 (top left), NGC1614 (top centre), NGC3256 (top right), NGC3690 (bottom left), NGC4194 (bottom centre), NGC6052 (bottom right) but in the age range 1--100 Myr. Values are reported in Table~\ref{tab:MF}. See main text for a discussion on the quality of the fit.}
    \label{fig:cornerplots}
\end{figure*}

\begin{table*}
    \centering
    \begin{tabular}{lccccccccc}
Galaxy & age & log(M$_{min}$) & N$_{\rm cl}$ & log(M$_c$)  & $\beta_{\rm Sch}$     &    $\beta_{\rm PL}$ & Fit & N$_{\rm cl, obs}$(M$>$M$_c$) & offset \\    
 & Myr &[\msun] & & [\msun] & & & & & factor  \\
 \hline
NGC34 & 1-10 & 4.9 &  13 &  6.83$^{+1.02}_{-0.5}$ & -1.36$^{+0.25}_{-0.38}$  &  -1.59$^{+0.17}_{-0.19}$ & PL & 0 & 0\\
 \hline
NGC34 & 1-100 & 4.9  & 34 &        8.11$^{+0.56}_{-0.53}$ & -1.48$^{+0.13}_{-0.15}$  & -1.52$^{+0.10}_{-0.11}$& PL & 0 & 0\\
 \hline
NGC1614 & 1-10 &   4.7  &69 &   8.46$^{+0.39}_{-0.52}$ &  -1.8$^{ +0.11}_{-0.36}$  &  -1.78$^{+0.09}_{-0.10}$& PL & 0 &  0 \\
 \hline
NGC1614 & 1-100 &  4.7  &185 &   7.86$^{+0.6}_{-0.41}$ &  -1.58$^{+0.06}_{-0.06}$  &  -1.62$^{+0.05}_{-0.05}$& PL & 1 &  2.0 \\
 \hline
NGC3256 & 1-10 &  4.7  &278&   6.59$^{+0.2}_{-0.16}$ &   -1.47$^{+0.08}_{-0.08}$  &  -1.72$^{+0.04}_{-0.04}$ &Sch & 3 &   4.0\\
 \hline
NGC3256 & 1-100 &  4.7  &363&   6.57$^{+0.17}_{-0.14}$ &  -1.49$^{+0.07}_{-0.07}$  &  -1.73$^{+0.04}_{-0.04}$& Sch & 4 &   3.9\\ 
 \hline
NGC3690A & 1-10 &  4.7  &123 &   7.0$^{+0.92}_{-0.43}$ &   -1.71$^{+0.13}_{-0.11}$  &  -1.82$^{+0.07}_{-0.08}$& PL  & &\\
 \hline
NGC3690A & 1-100 & 4.7  &199 &   6.66$^{+0.52}_{-0.28}$ &  -1.72$^{+0.12}_{-0.11}$  &  -1.88$^{+0.06}_{-0.06}$& Sch & &\\
 \hline
NGC3690B & 1-10 &   4.7 &75&   7.64$^{+0.76}_{-0.51}$ &  -1.6$^{ +0.11}_{-0.1}$  &   -1.66$^{+0.08}_{-0.08}$ & PL &  &\\
 \hline
NGC3690B & 1-100 &  4.7 &142&   7.36$^{+0.77}_{-0.41}$ &  -1.63$^{+0.09}_{-0.09}$  &  -1.72$^{+0.06}_{-0.06}$& PL &  &\\
 \hline
NGC3690 & 1-10 &   4.7  &198&   7.36$^{+0.71}_{-0.38}$ &  -1.67$^{+0.08}_{-0.08}$  &  -1.75$^{+0.05}_{-0.05}$ & PL & 1 &   2.0\\
 \hline
NGC3690 & 1-100 &  4.7  &341 &   7.04$^{+0.38}_{-0.25}$ &  -1.7$^{ +0.07}_{-0.07}$  &   -1.80$^{+0.04}_{-0.04}$& Sch & 1 &   4.5\\
 \hline
NGC4194 & 1-10 &   4.3  & 75&   7.21$^{+0.61}_{-0.35}$ &  -1.36$^{+0.1}_{-0.09}$  &   -1.47$^{+0.06}_{-0.06}$& Sch & 1 &    3.1\\
 \hline
NGC4194 & 1-100 &  4.3  &121&   7.25$^{+0.59}_{-0.35}$ &  -1.46$^{+0.08}_{-0.07}$  &  -1.55$^{+0.05}_{-0.05}$& Sch & 1  &   2.8\\
 \hline
NGC6052 & 1-10 &   4.5 &227&    5.99$^{+0.33}_{-0.21}$ &  -1.71$^{+0.14}_{-0.14}$  &  -1.99$^{+0.06}_{-0.07}$& Sch & 3 &   2.4\\
 \hline
NGC6052 & 1-100 &  4.5 &318&    6.11$^{+0.23}_{-0.17}$ &  -1.67$^{+0.1}_{-0.1}$  &   -1.93$^{+0.05}_{-0.05}$ & Sch  & 4 &  2.4\\
 \hline
Comb--all & 1-10 &     4.9 &487& 7.76$^{+0.54}_{-0.33}$ &  -1.79$^{+0.05}_{-0.05}$  &  -1.84$^{+0.04}_{-0.04}$& Sch  & &\\
 \hline
Comb--all & 1-100 &     4.9  &793&   7.78$^{+0.46}_{-0.28}$ &  -1.81$^{+0.04}_{-0.04}$  &  -1.86$^{+0.03}_{-0.03}$ & Sch  & &\\
 \hline
Comb--all  & 1-10 &   4.7  &769&  7.57$^{+0.39}_{-0.24}$ &  -1.73$^{+0.04}_{-0.04}$  &  -1.78$^{+0.03}_{-0.03}$& Sch & 1 &  4.5\\ 
\hline
Comb--all & 1-100 &  4.7  &1237& 7.44$^{+0.22}_{-0.16}$ &  -1.7$^{+0.03}_{-0.03}$  &  -1.77$^{+0.02}_{-0.02}$& Sch & 2  &  5.0\\
\hline
Comb--early& 1-100 &  4.7  &  934   & 6.82 $^{ +0.16}_{-0.13}$ &  -1.7$^{ +0.04}_{-0.04}$ & -1.83$^{+0.03}_{-0.03}$ &   Sch & 6 & 2.6 \\
\hline
Comb--advanced & 1-100 &  4.7  &  303   & 7.81$^{  +0.43}_{-0.28 }$ & -1.55$^{+0.05}_{-0.05}$ & -1.60$^{+0.03}_{-0.04}$ &  Sch & 1& 3.9 \\
 \hline

    \end{tabular}
    \caption{Mass function fitting analysis. For each galaxy (or pair component) and for the combined cluster sample we report the age range used to select clusters (second column); The minimum cluster mass applied (third column); number of clusters included in the fit (fourth column); the resulting M$_c$ and slope $\beta_{\rm Sch}$ including the 1 $\sigma$ errors (fifth and sixth columns), the resulting $\beta_{\rm PL}$ if a pure power-law function is fitted instead (seventh column); in the eight column we report the quality of the fit: "Sch" if the Schechter function fit converged, "PL" in the other case; the observed number of clusters with masses larger than the fitted M$_c$ (tenth column, if the number is $\geq 0$ the constraint is tighter); and in the last column we report the ratio between the number of clusters above M$_c$ predicted by a pure power-law fit and the the observed one. If the ratio is larger than 1 than the power-law fit overestimates that number. In the "Comb--all" fit we analyse the cluster mass function of the combined cluster catalogues of all the HiPEEC galaxies, while in the "Comb--early" and  "Comb--advanced" we combine cluster populations by separating the sample in galaxies in an early/intermediate stage of the merger (NGC3256, NGC3690, NGC6052) from those in a more advanced stage (NGC34, NGC1614, and NGC4194).}
    \label{tab:MF}
\end{table*}

\begin{figure*}
    \centering
    \includegraphics[width=0.48\textwidth]{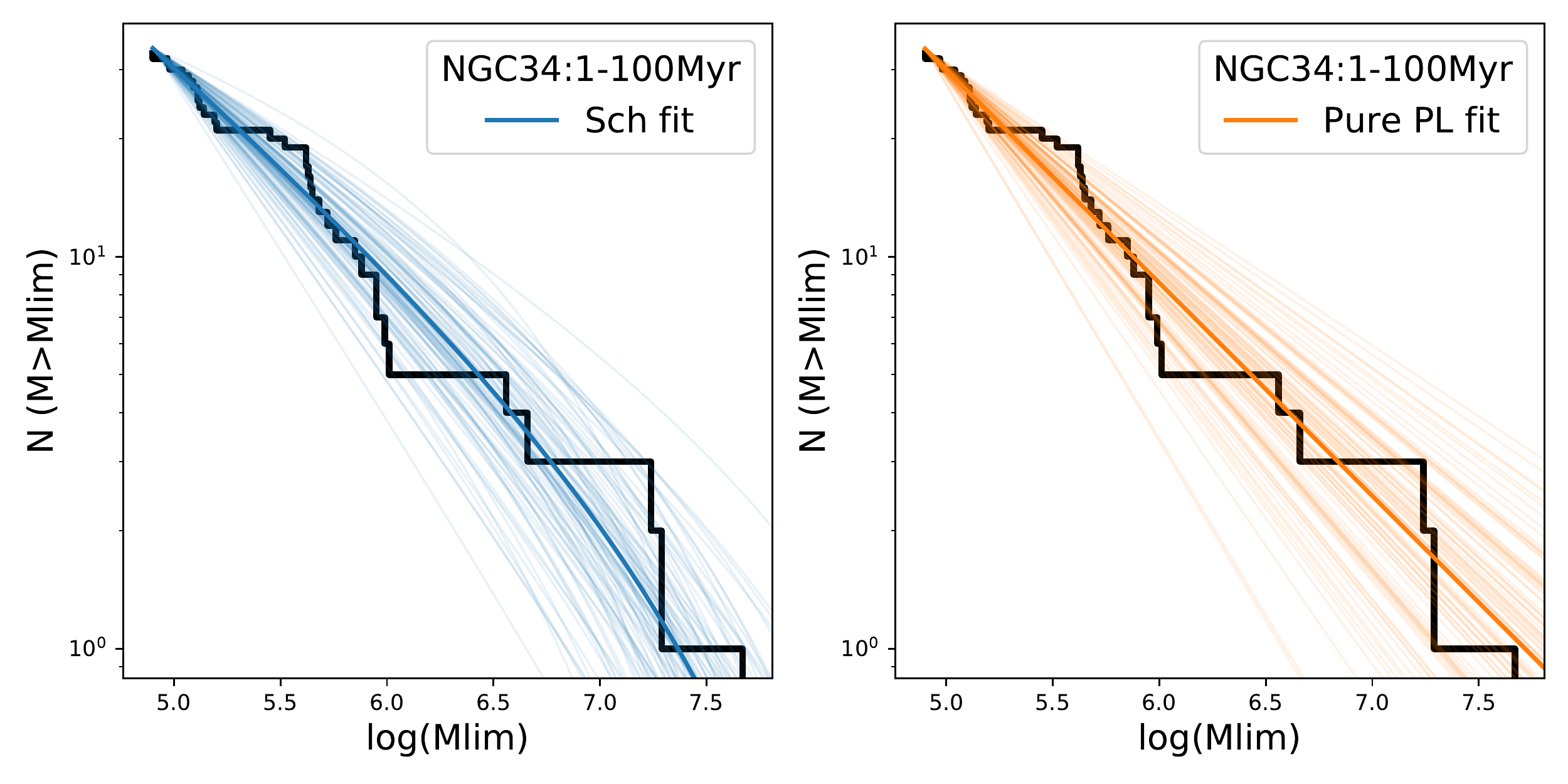}
    \includegraphics[width=0.48\textwidth]{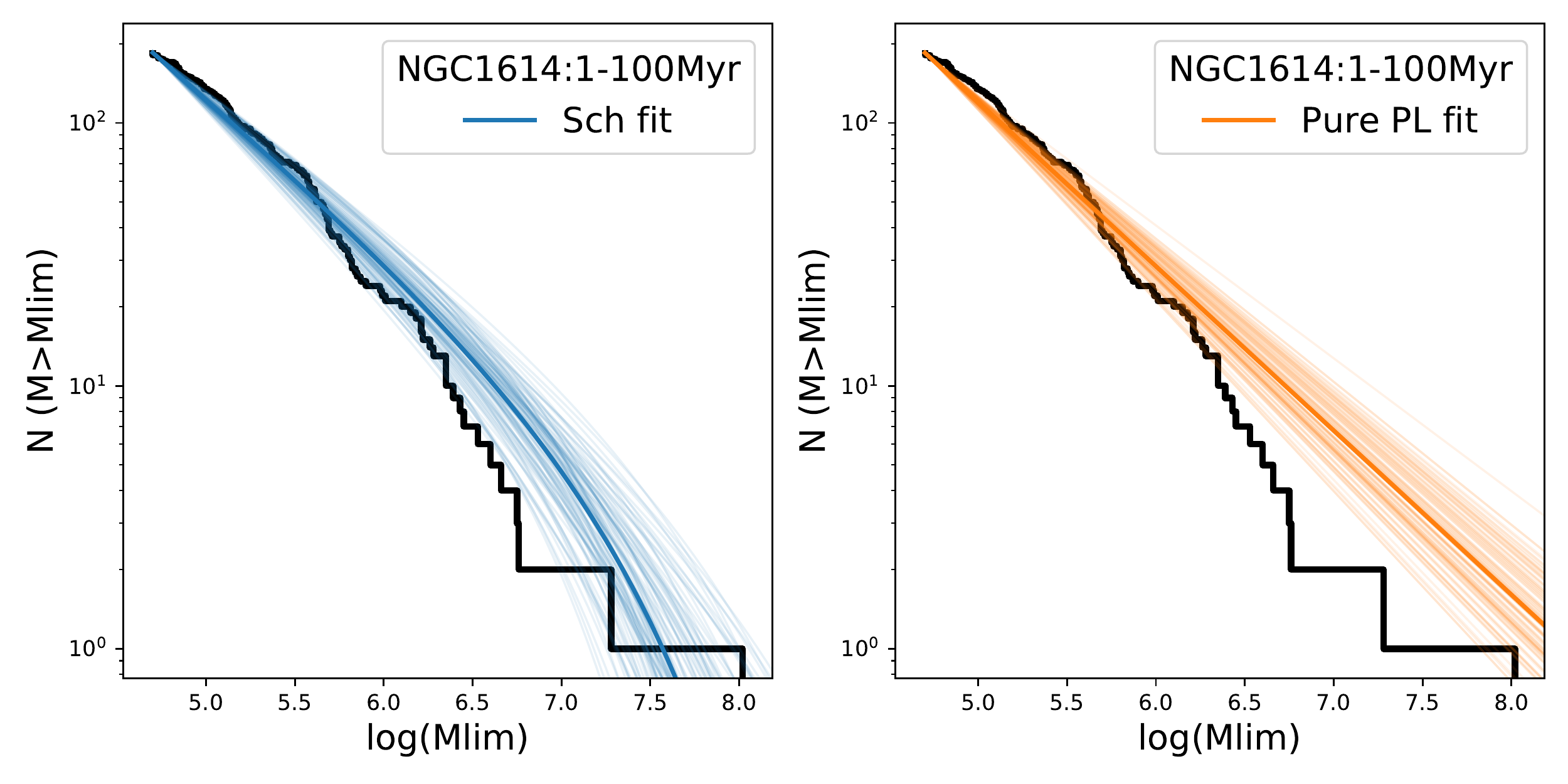}\\
    \includegraphics[width=0.48\textwidth]{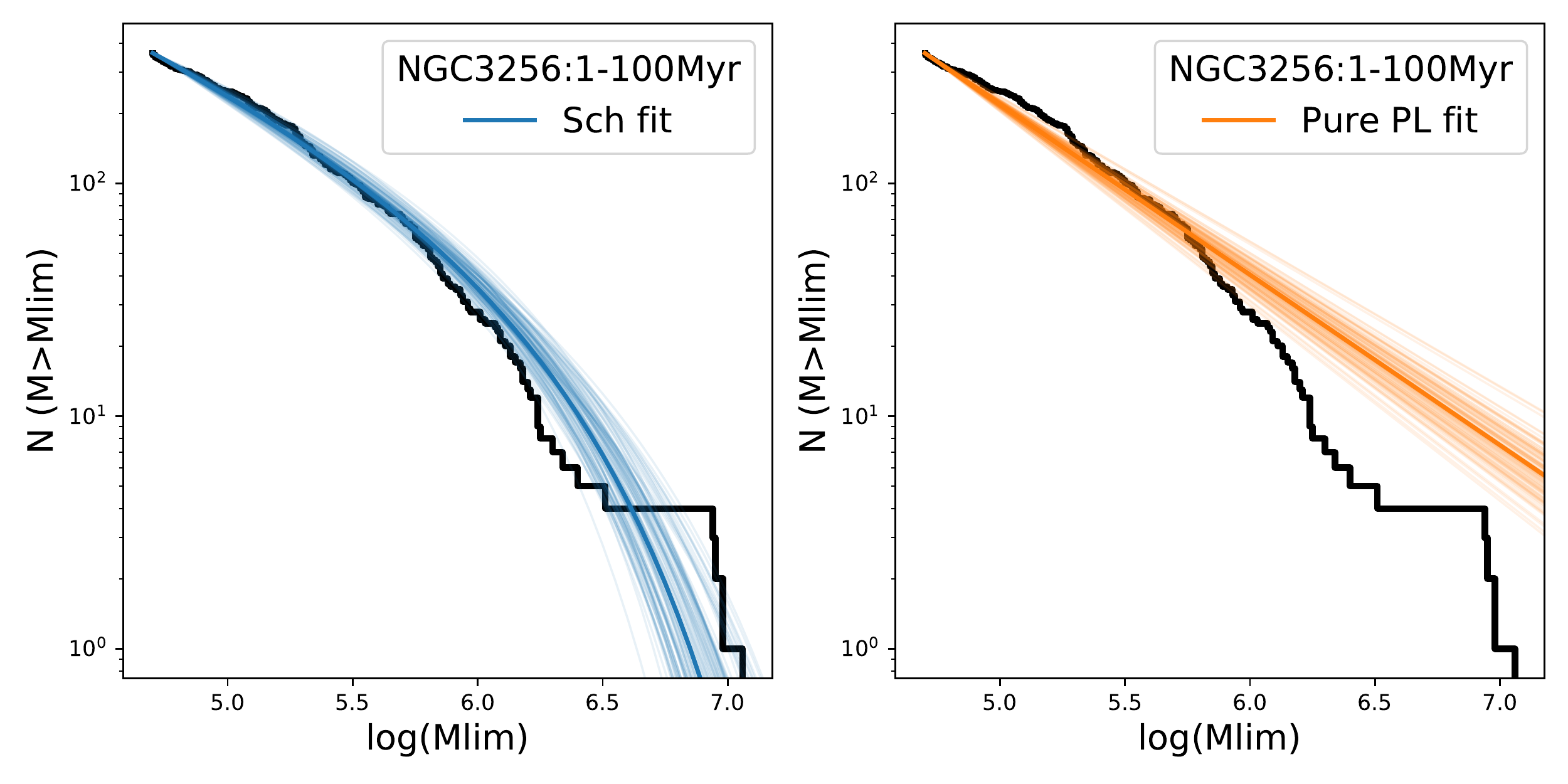}
    \includegraphics[width=0.48\textwidth]{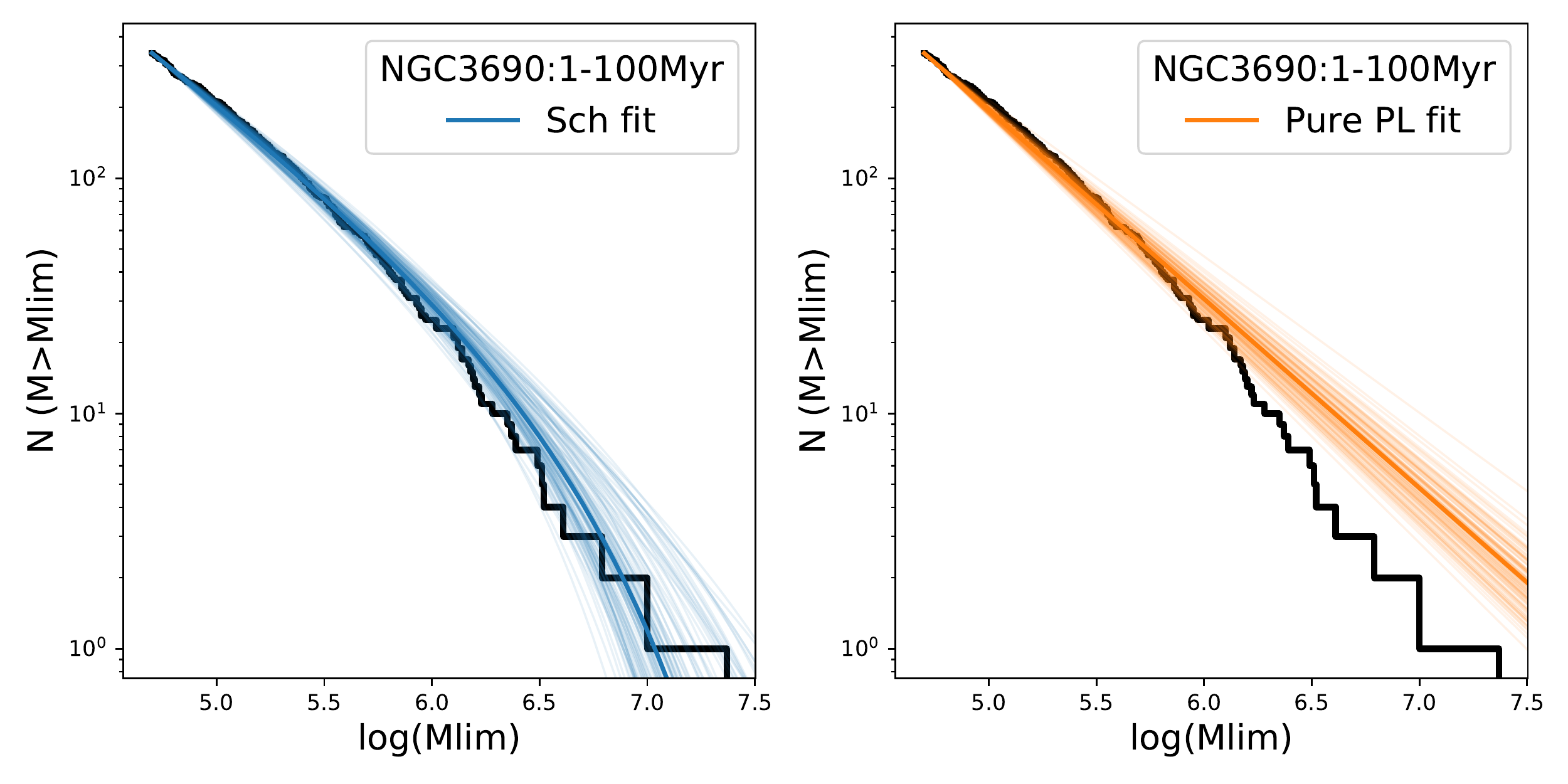}\\
    \includegraphics[width=0.48\textwidth]{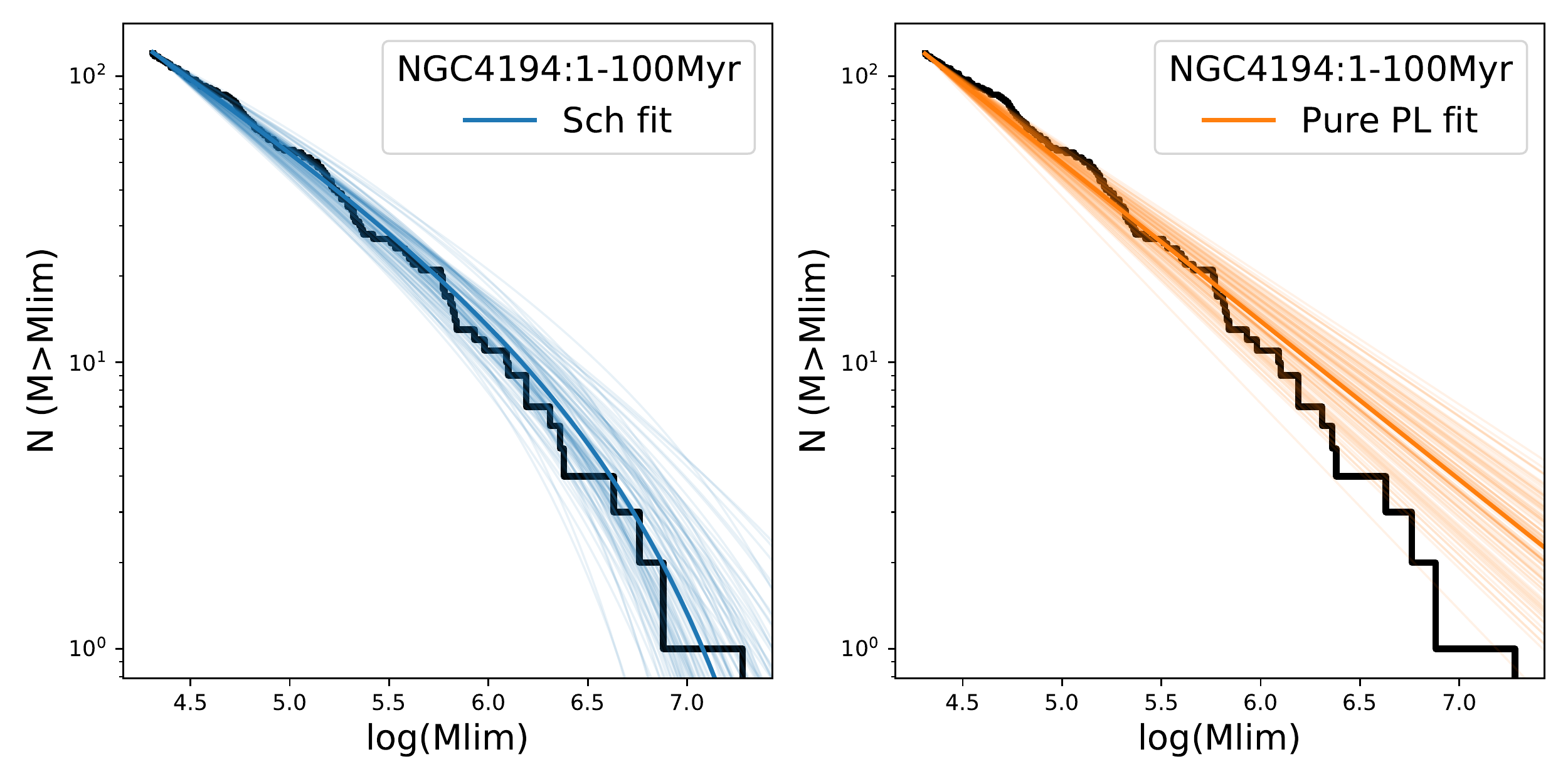}
    \includegraphics[width=0.48\textwidth]{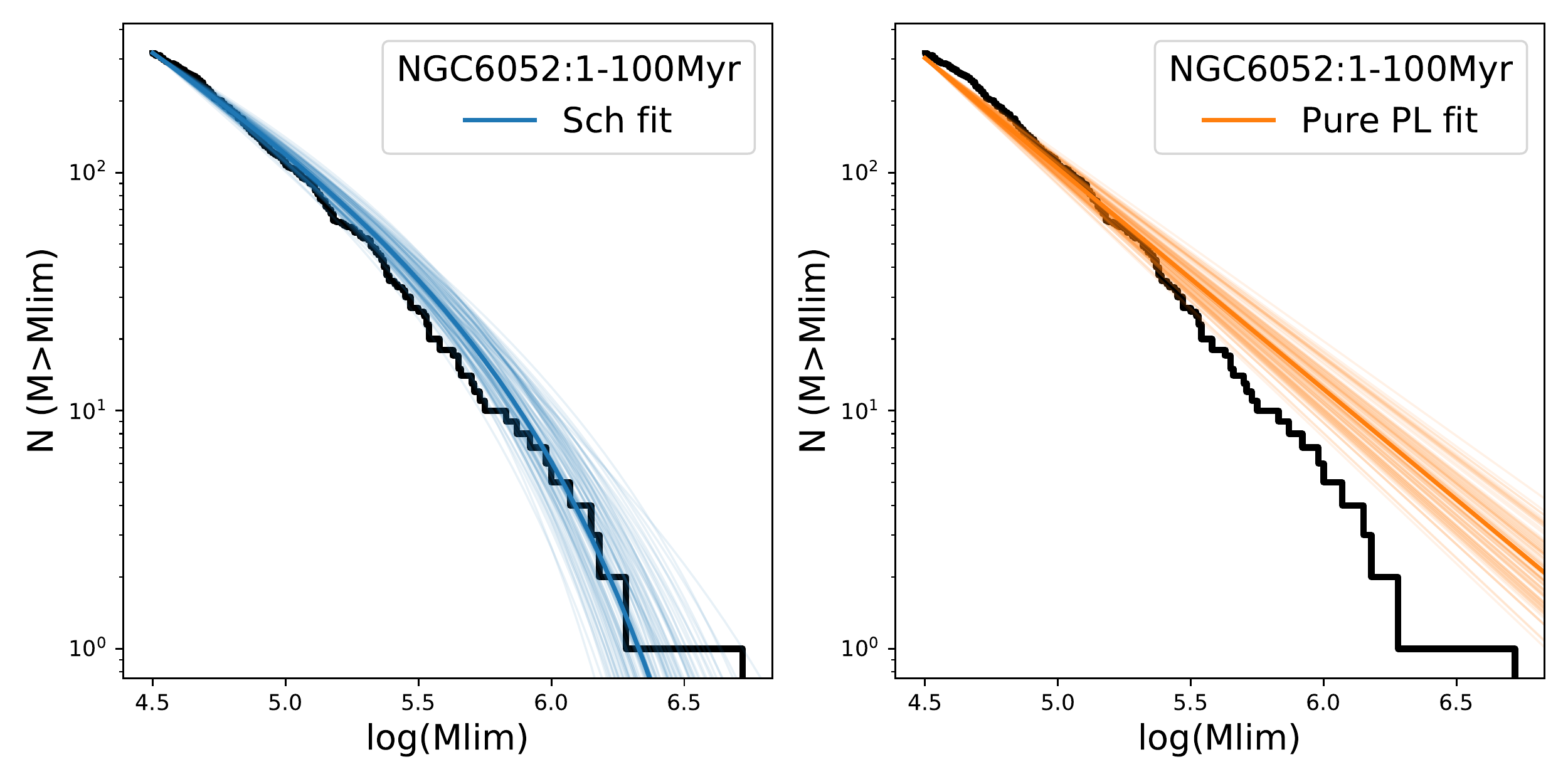}\\
    
    \caption{Observed cumulative cluster mass distributions (solid black line) of the HiPEEC sample in the age range 1--100 Myr. The same  plots but including the analysis of the mass functions in the age range 1--10 Myr is shown in Figure~\ref{fig:MF_app} of the Appendix.  The blue coloured solid and thinned associated lines show the best values and the family of solutions for M$_c$ and $\beta_{\rm Sch}$ (left panel, blue lines) contained within the 1$\sigma$ contours of the corner plots in Figure~\ref{fig:cornerplots}. Similarly we plot in the right panel the predicted cumulative mass distributions (orange lines) for the best value of $\beta_{\rm PL}$ and the associated 1$\sigma$. The predicted distributions contain the same number of clusters used to build the observed mass distributions. These plots help to visualise the goodness of the determined best fit values. Except in the cases where there is no convergence in the Schechter function fit (for NGC34 and NGC1614), an exponential cut off at the high-mass end is a better representation of the observed cluster mass function. See text for more details.}
    \label{fig:MF}
\end{figure*}

\begin{figure*}
    \centering
    \includegraphics[width=0.3\textwidth]{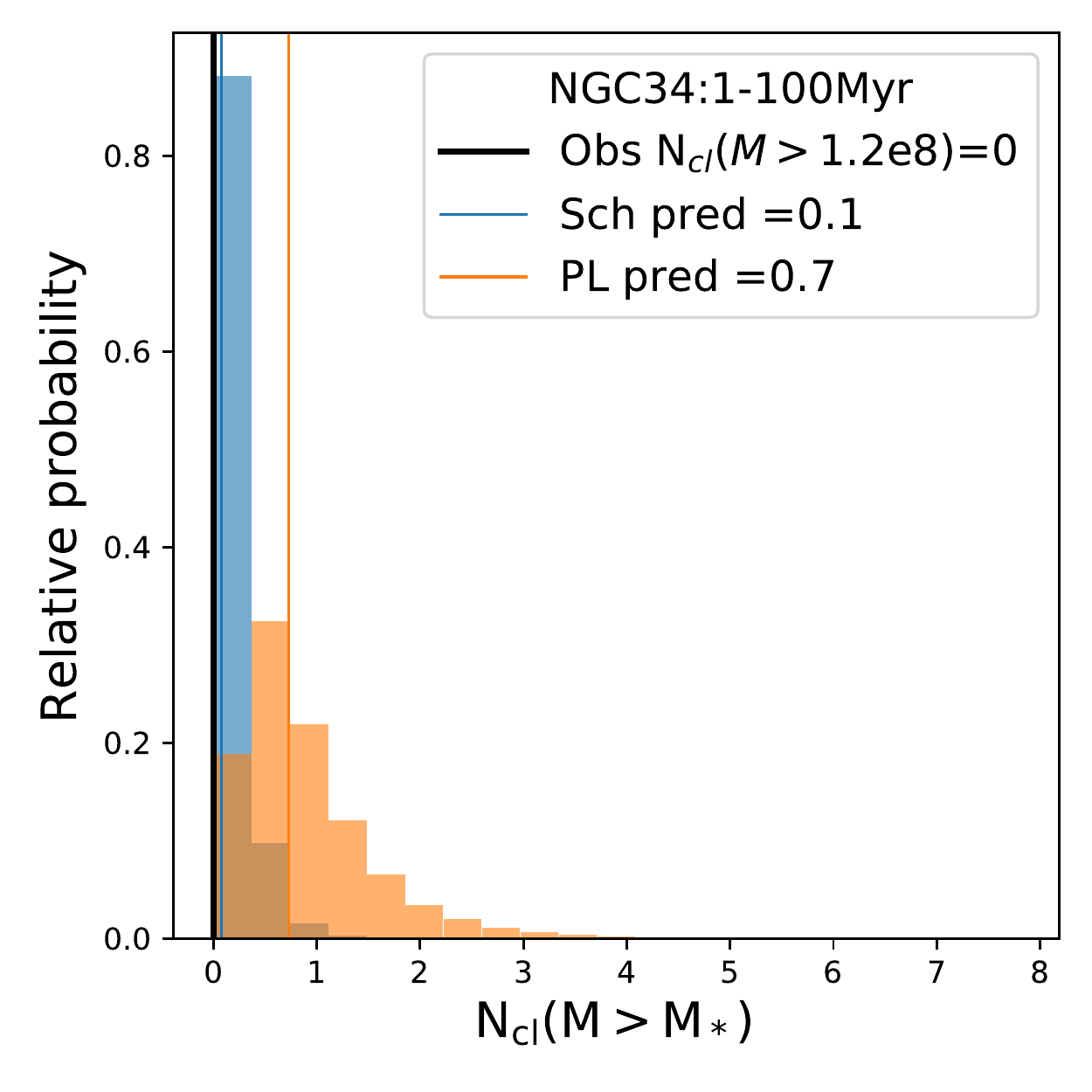}
    \includegraphics[width=0.3\textwidth]{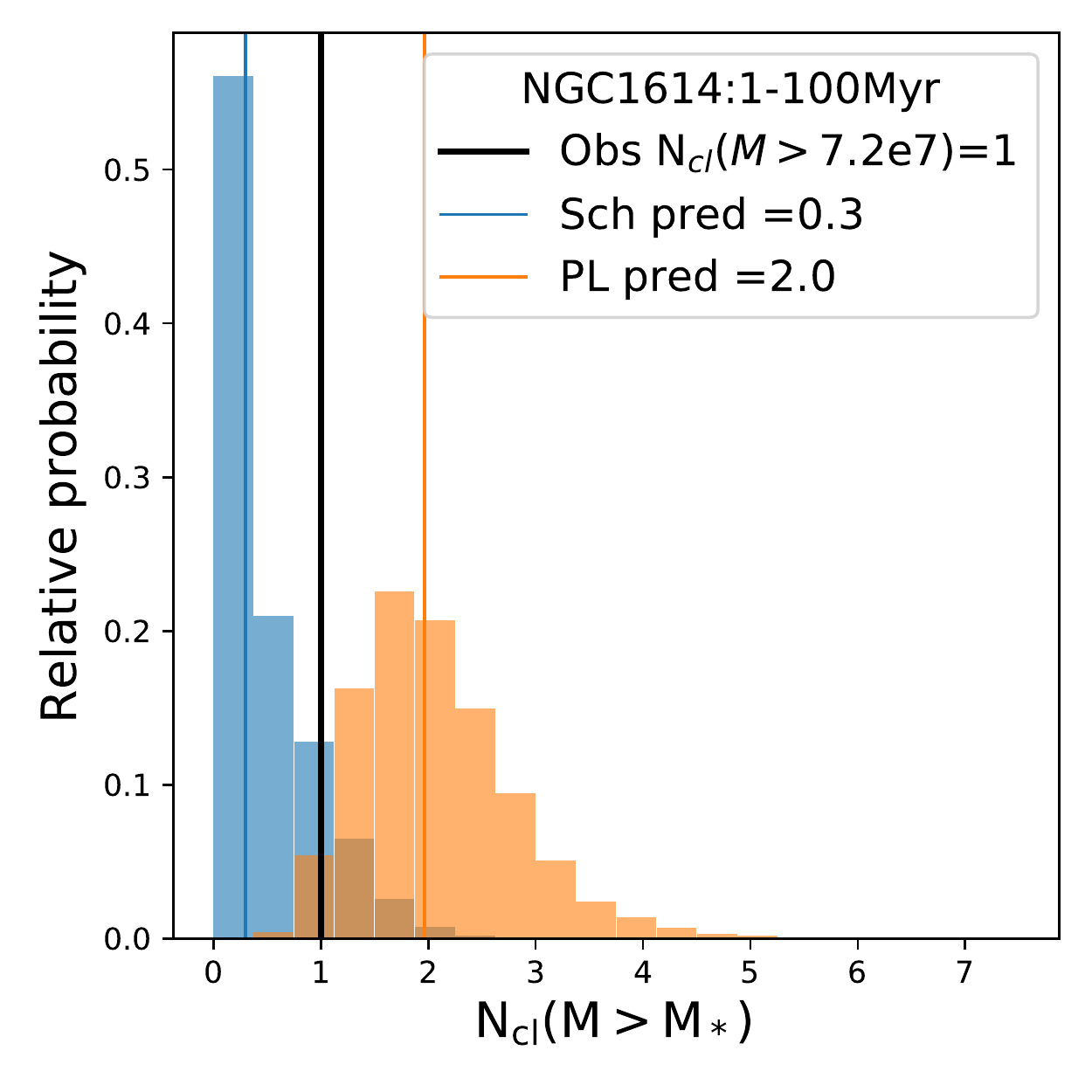}
    \includegraphics[width=0.3\textwidth]{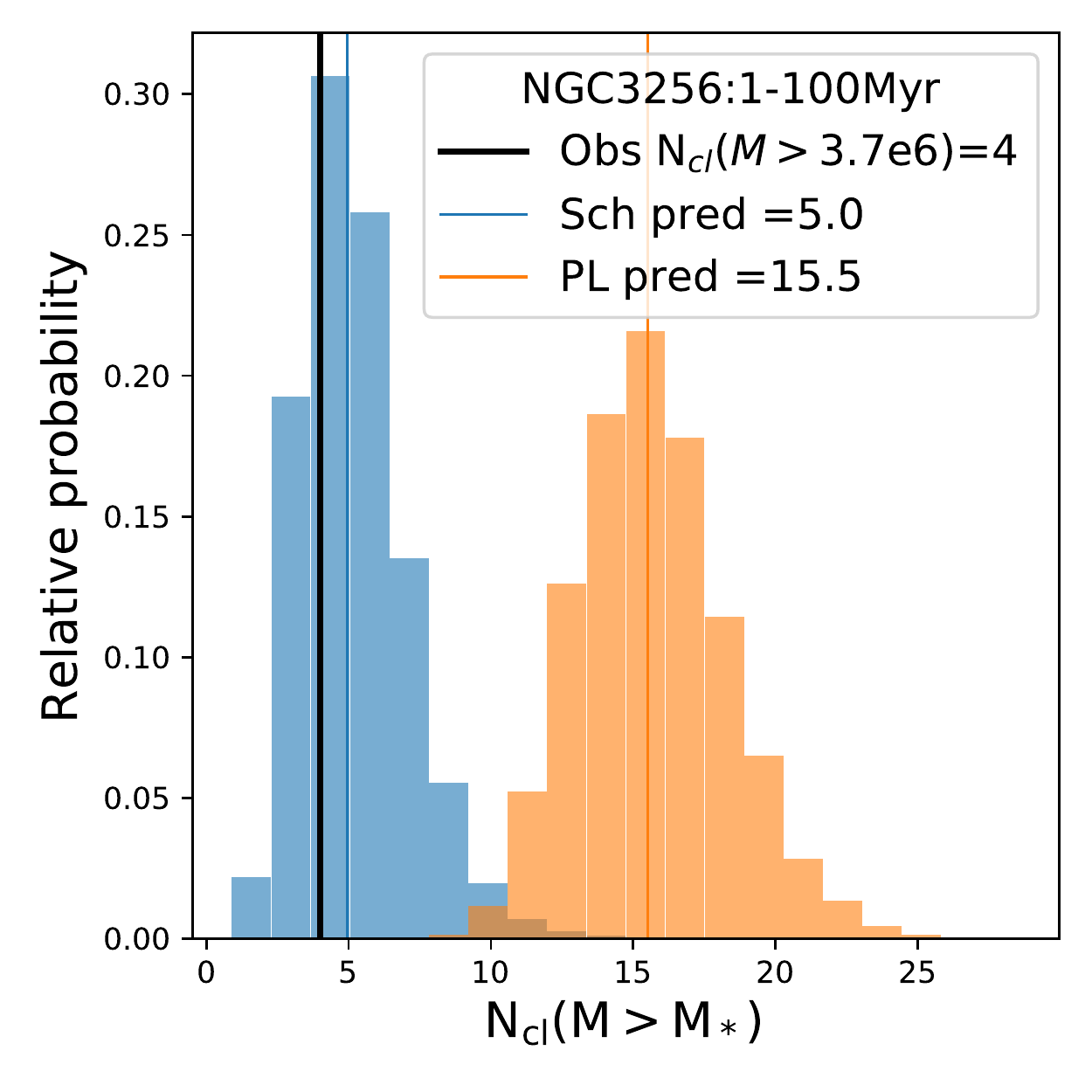}\\
    \includegraphics[width=0.3\textwidth]{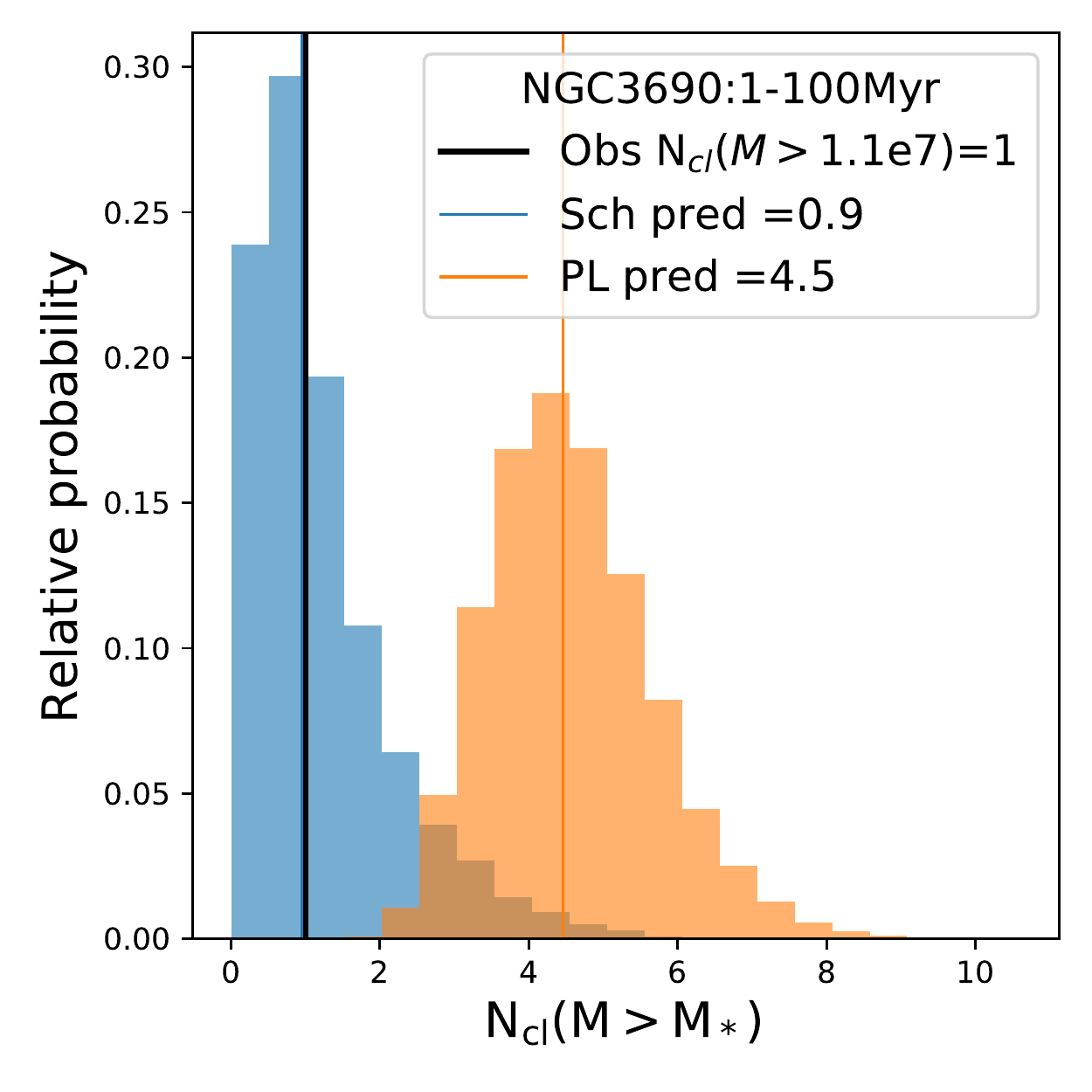}
    \includegraphics[width=0.3\textwidth]{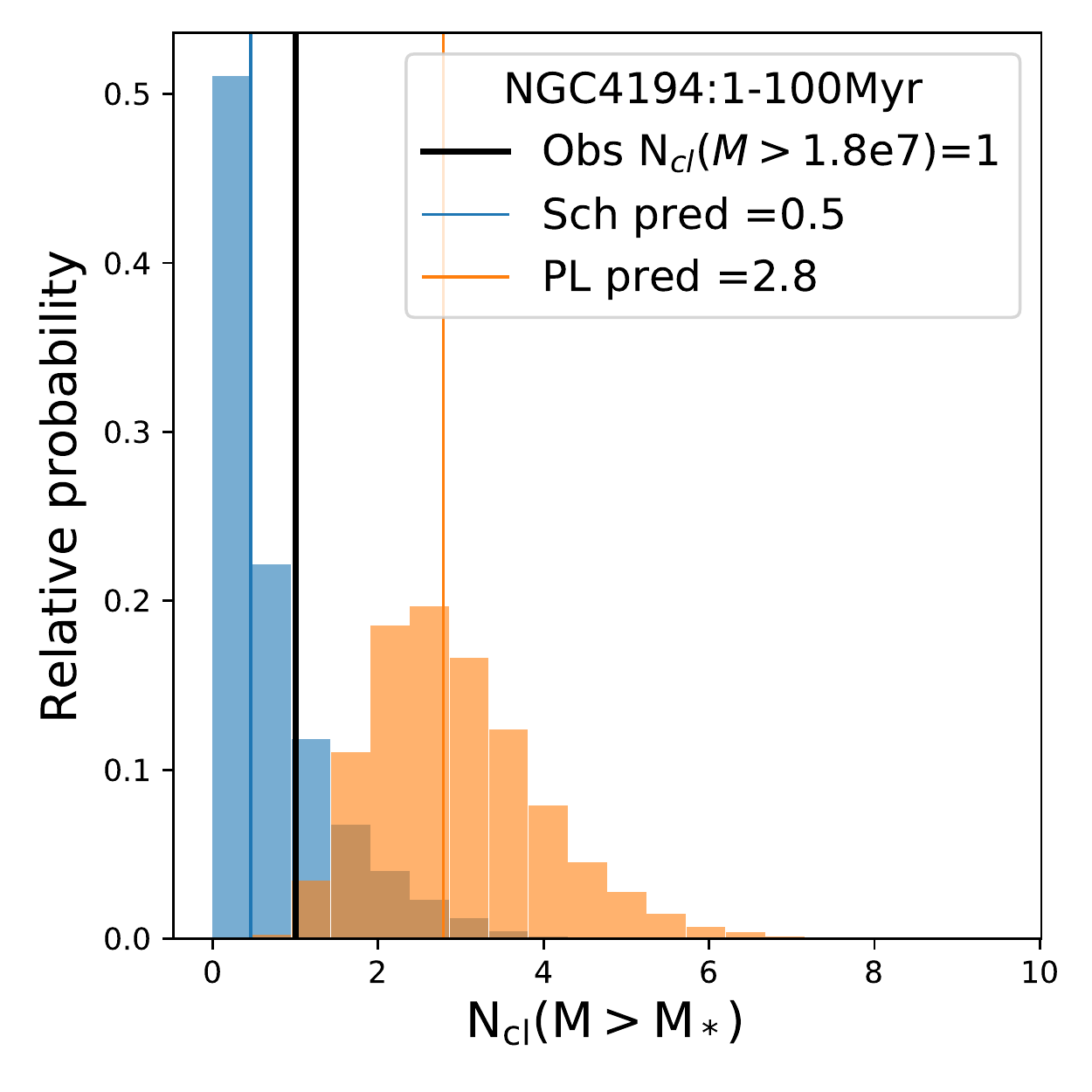}
    \includegraphics[width=0.3\textwidth]{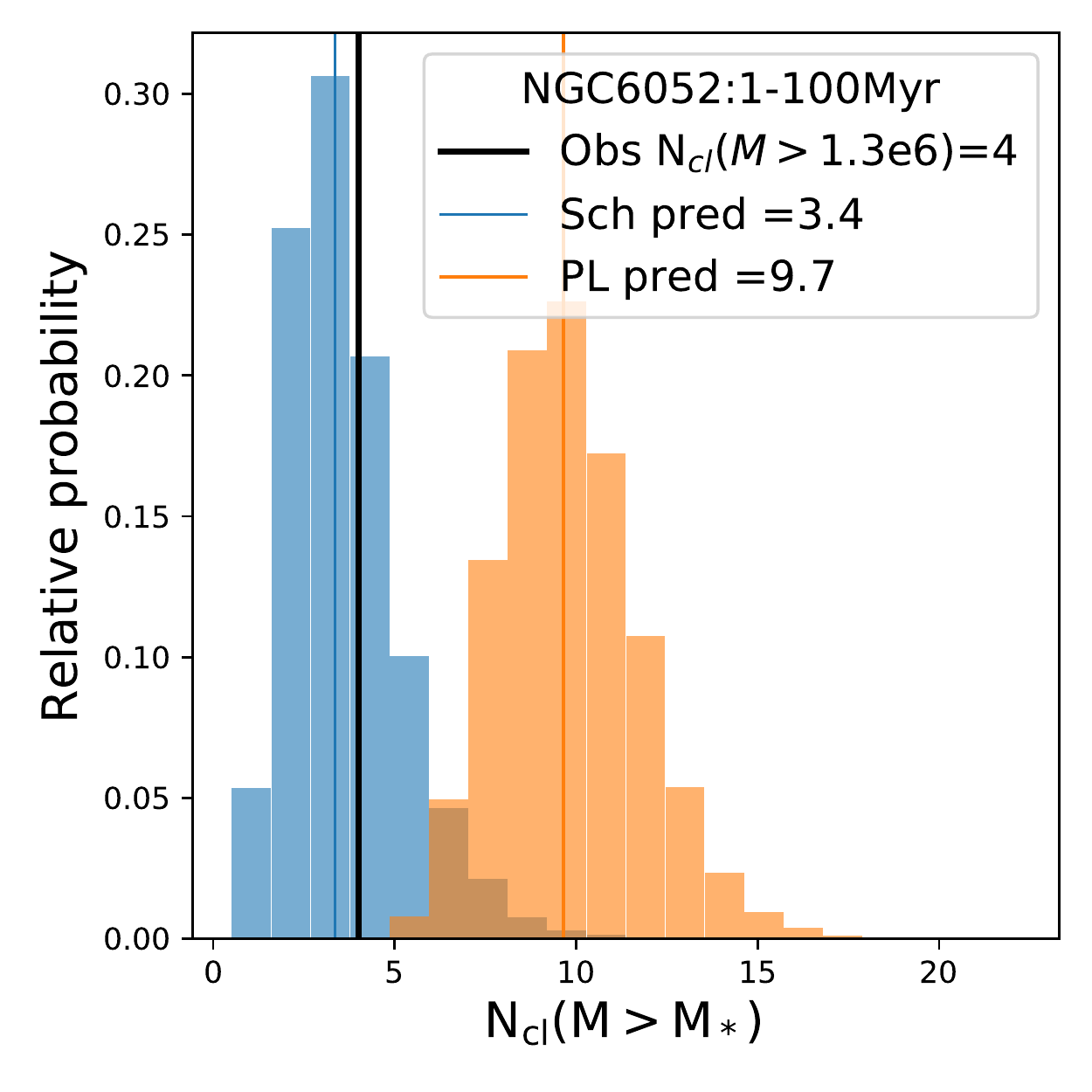}\\
    
    \caption{Normalised distributions of recovered number of clusters more massive than the determined truncation mass, M${_*}$, in each galaxy in the age range 1--100 Myr. The blue and orange histograms show the recovered number of clusters more massive than M${_*}$ in 1000 monte carlo runs of cluster populations with the same number of clusters used in the fit and mass distributions described by the best determined parameters of the fitted Schechter and power-law mass functions, respectively. The median of the two distributions are indicated by vertical lines and their numbers included in the insets. Observed number of clusters more massive than M${_*}$ are shown as a black vertical line and indicated in the inset for each galaxy. In general, except in the cases where the Schechter function fit doe not converge, the pure power-law function overpredict the number of expected clusters more massive than M$_c$. The same analysis but for clusters in the age range 1--10 Myr is shown in Figure~\ref{fig:predMF_app} of the Appendix.  }
    \label{fig:predMF}
\end{figure*}

\begin{figure*}
    \centering
    \includegraphics[width=0.45\textwidth]{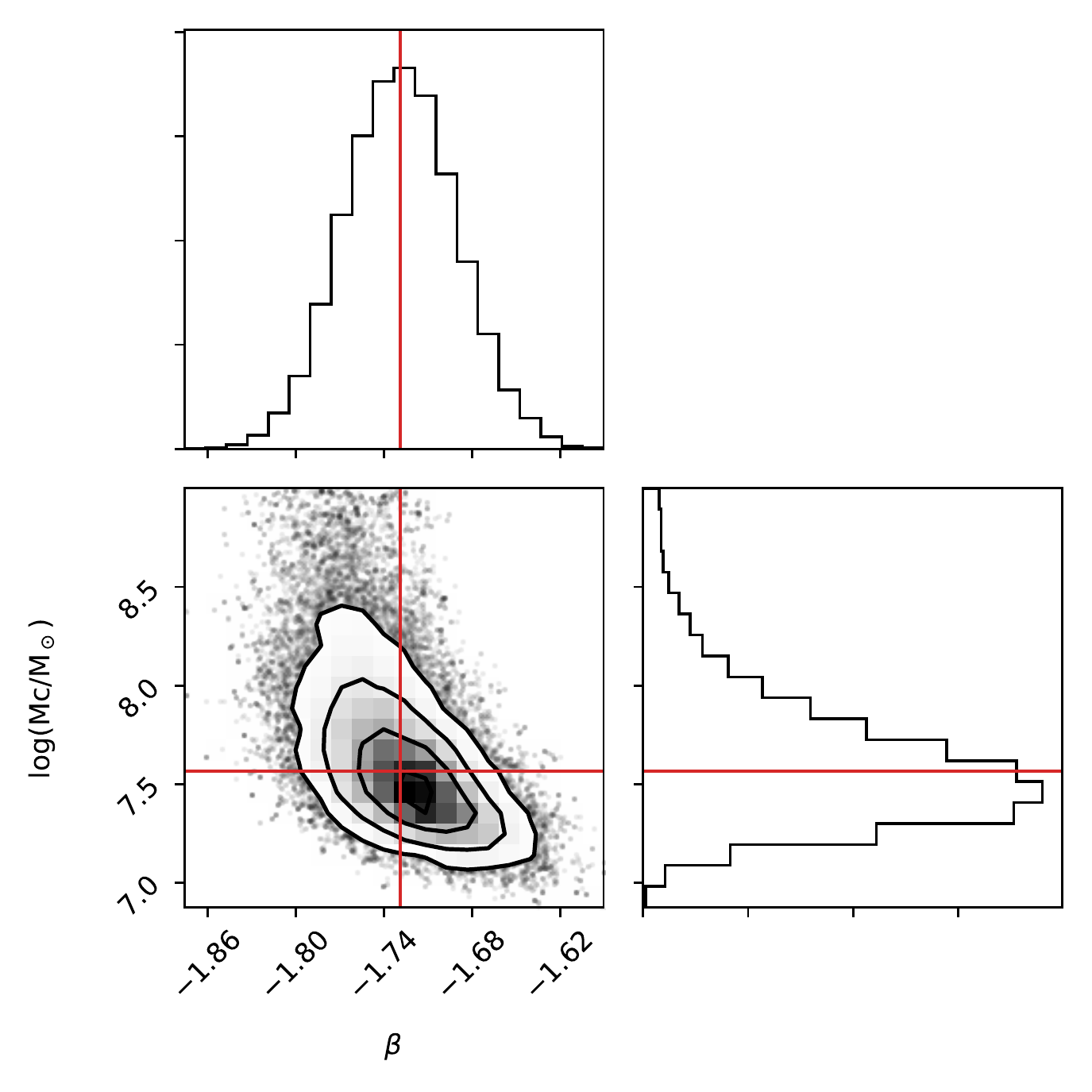}
    \includegraphics[width=0.45\textwidth]{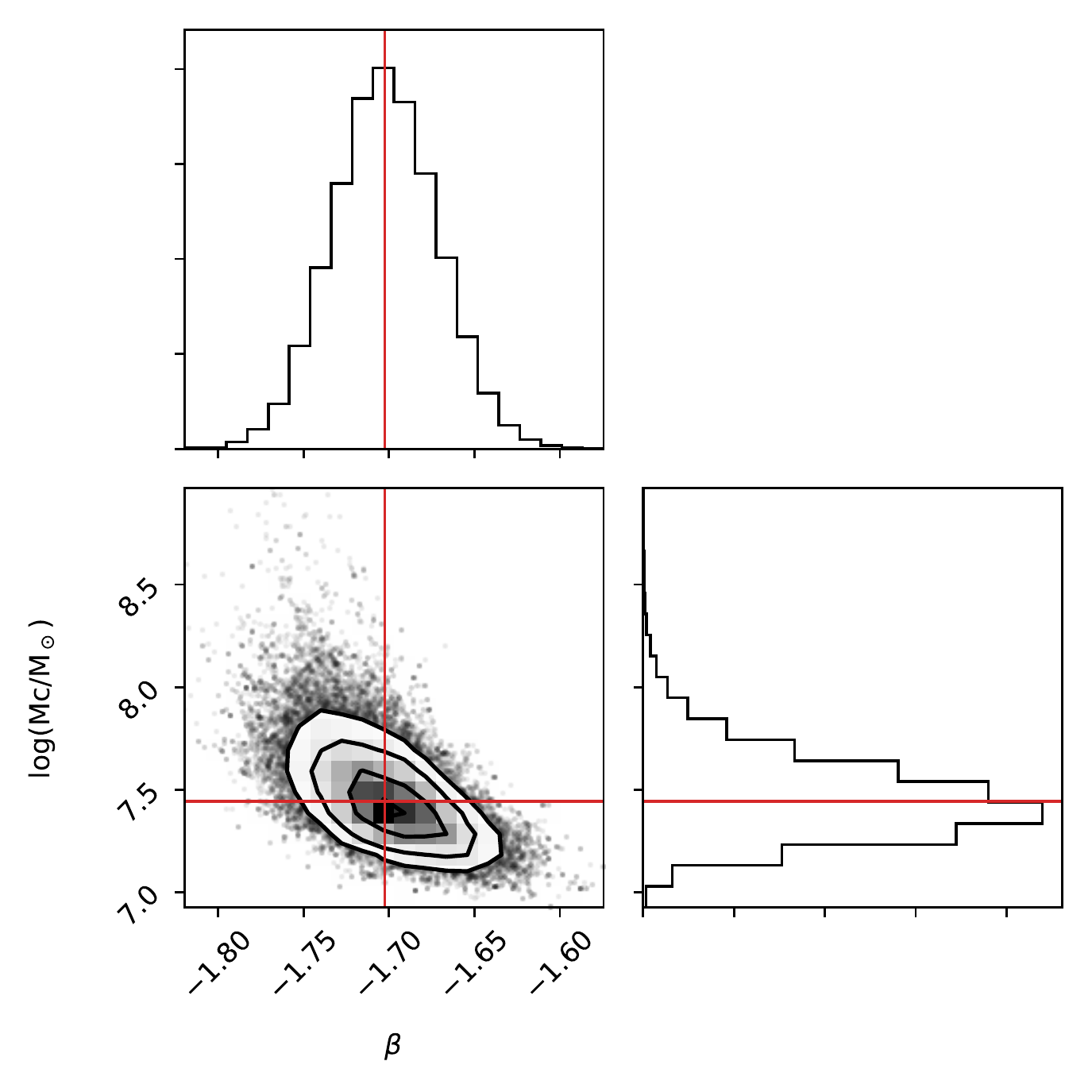}

    \caption{Bayesian Schechter function fit to the cluster mass function of the combined cluster population in the age range 1--10 Myr (left panels) and 1--100 Myr (right panels). We used clusters more massive than log(M)$\geq 4.7$ \msun. A stricter mass limits results in similar results (see Table~\ref{tab:MF}).  The corner plots show the parameter space visited by the walkers in our cluster mass function fitting analysis. The contours represent the 1, 2 and 3 sigma values of the density distribution around the two fitted parameters, M$_c$ and $\beta_{\rm Sch}$. The solid red lines show the median values of the mass and slope distributions reported in Table ~\ref{tab:MF}. Within the 1$\sigma$ uncertainties the best retrieved values of M$_c$ and $\beta_{\rm Sch}$ are very similar for the two age ranges used.}
    \label{fig:cornerplots_all}
\end{figure*}

\begin{figure*}
    \centering
    \includegraphics[width=0.66\textwidth]{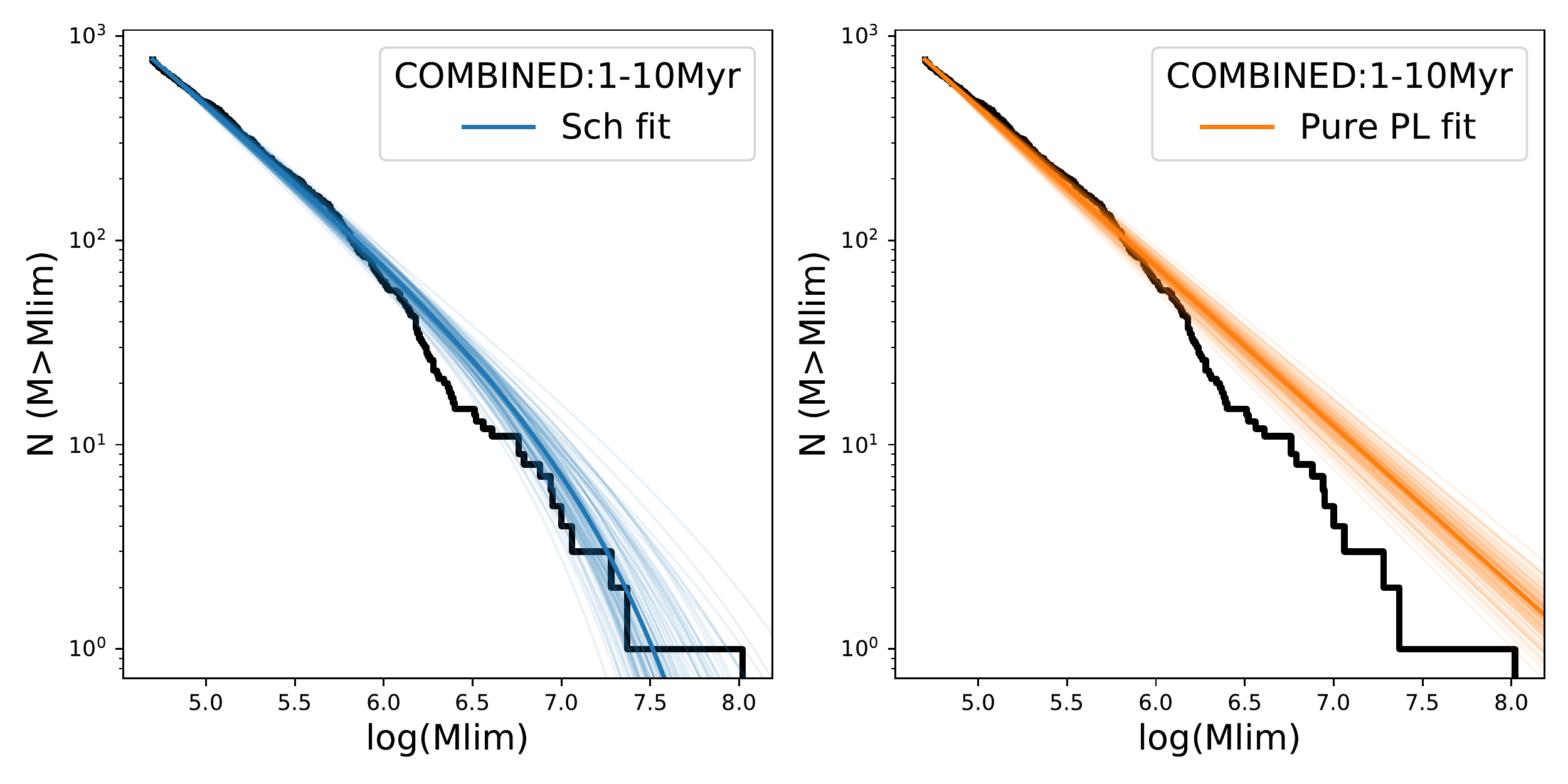}
    \includegraphics[width=0.33\textwidth]{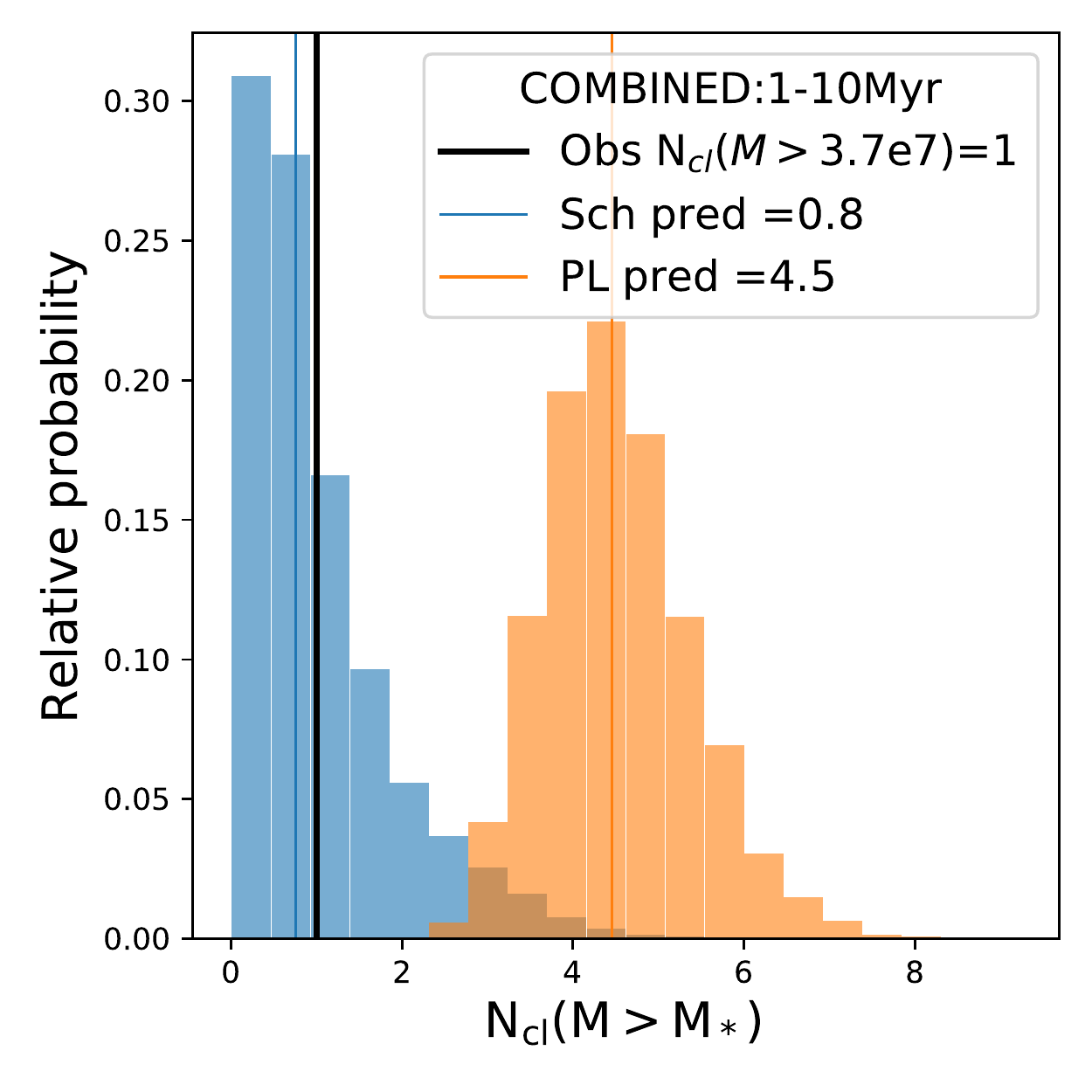}

    \includegraphics[width=0.66\textwidth]{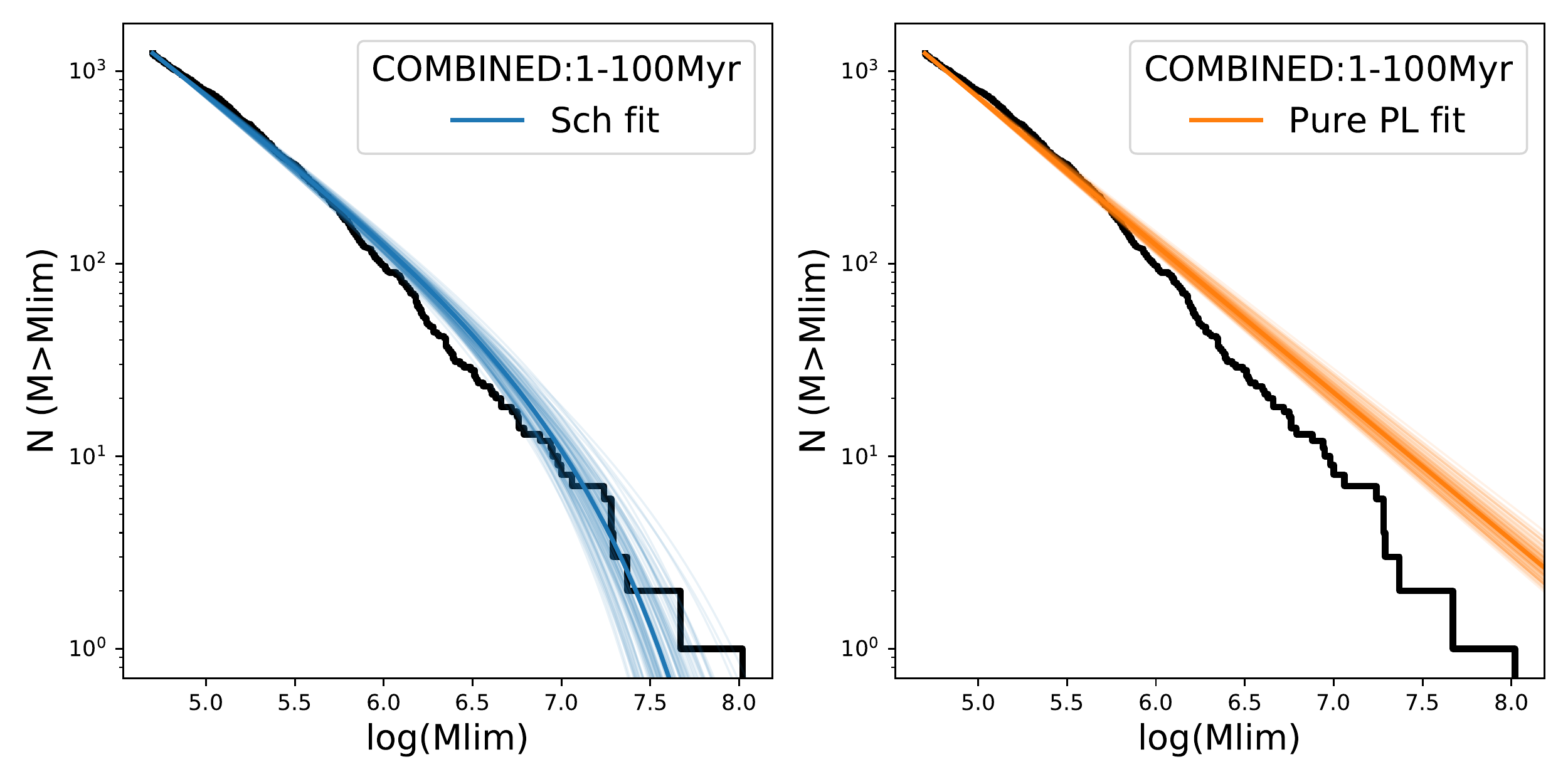}
    \includegraphics[width=0.33\textwidth]{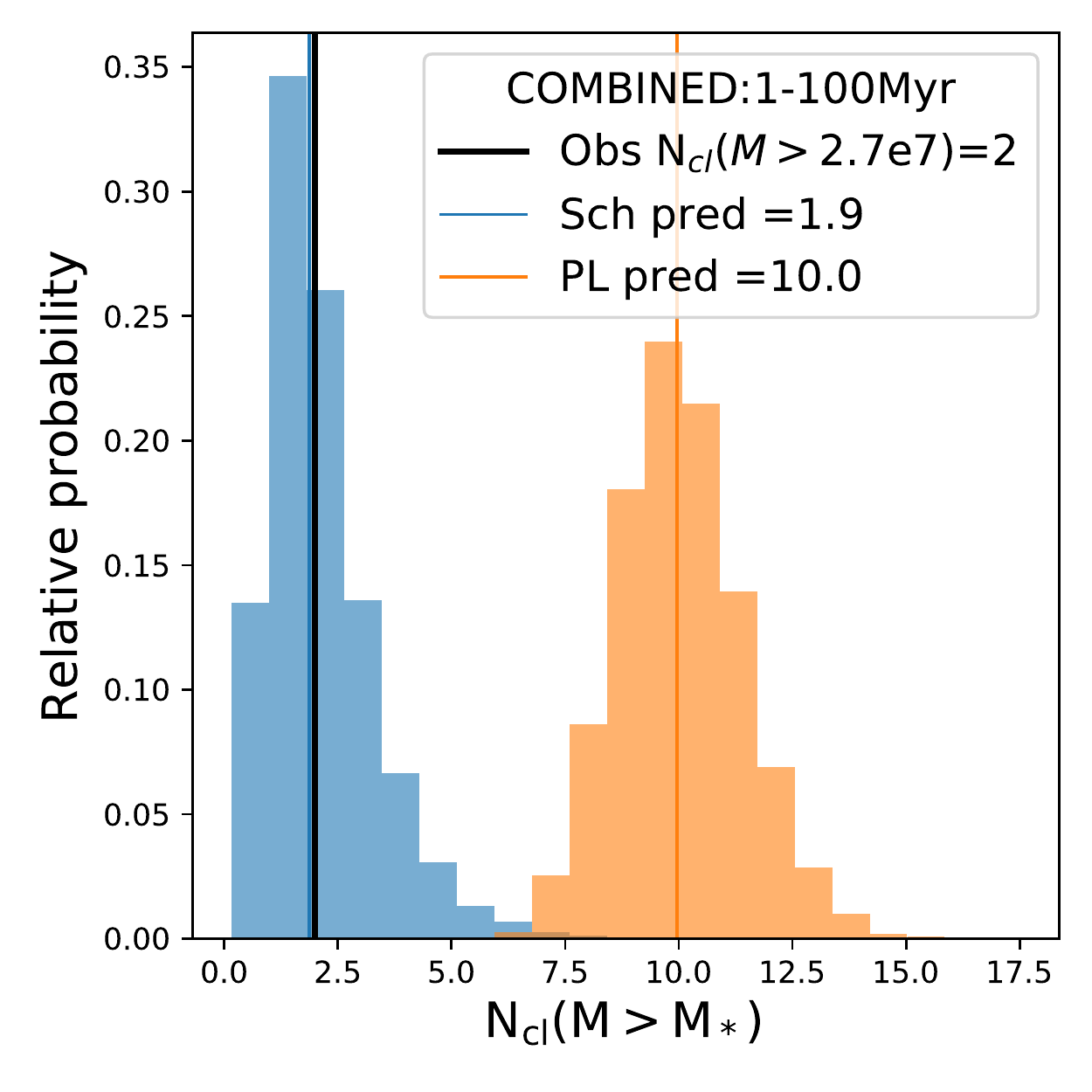}

    \caption{Same analyses as presented in Figure~\ref{fig:MF} and \ref{fig:predMF} but for the combined sample in the age range 1--10 Myr (top row), and 1--100 Myr (bottom row). The increased number of clusters used in the fit puts significantly better constraints to the determined M$_c$ and $\beta_{\rm Sch}$.}
    \label{fig:MF_all}
\end{figure*}

\begin{figure*}
    \centering
    \includegraphics[width=0.99\textwidth]{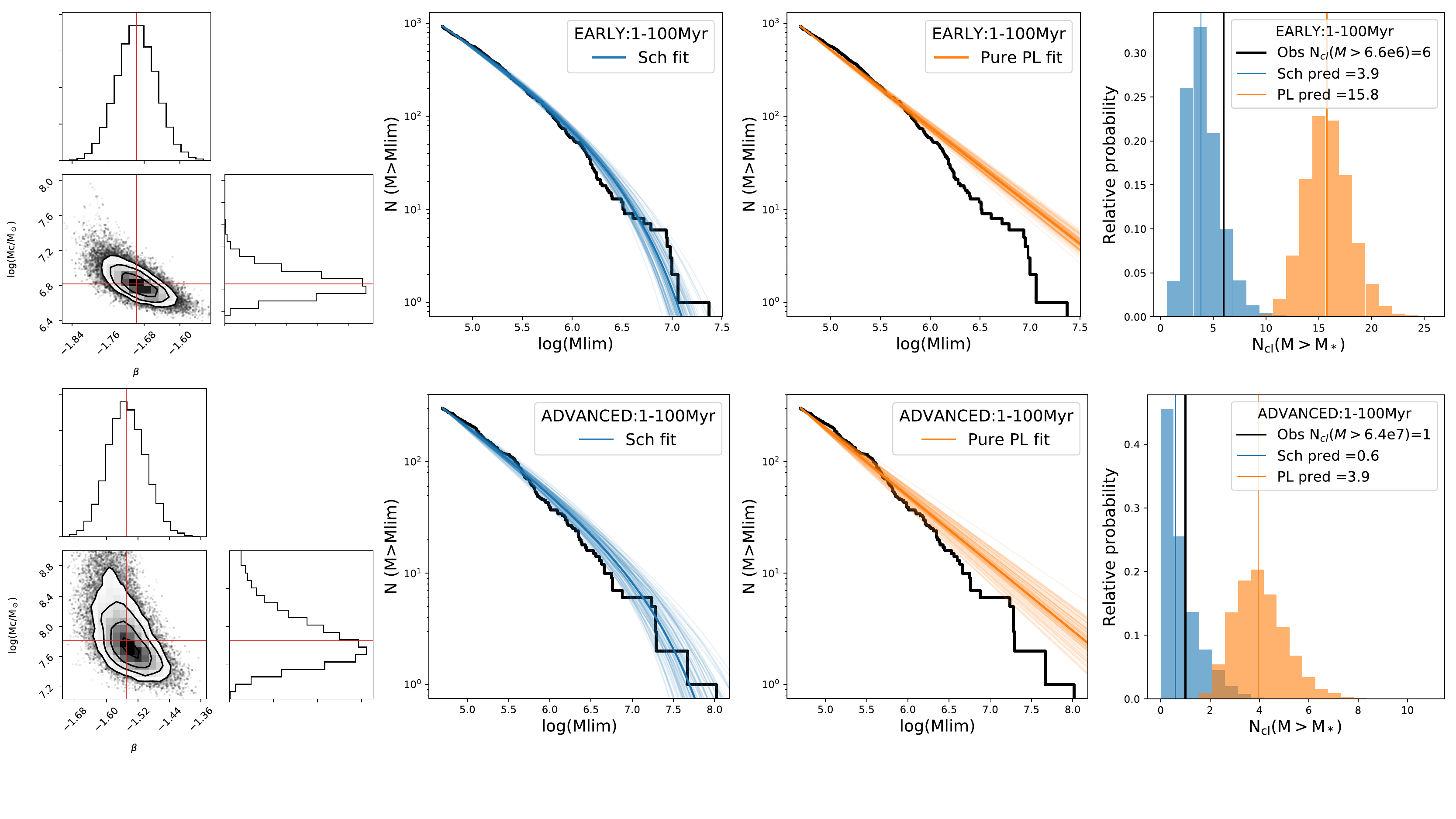}
    
    \caption{Bayesian analysis of the cluster mass function of the combined cluster populations of early merger stage (top row) systems, NGC3256, NGC3690, NGC6052, and late mergers (bottom row), NGC34, NGC1614, NGC4194. We used clusters masses above M$_{min} > 5\times10^4$ \msun\,  and age 1-100 Myr. The corner plots (left) show the parameter space visited by the walkers in our cluster mass function fitting analysis. The contours represent the 1, 2 and 3 sigma values of the density distribution around the two fitted parameters, M$_c$ and $\beta_{\rm Sch}$. The solid red lines show the median values of the mass and slope distributions reported in Table ~\ref{tab:MF}. On the centre panels we show the observed cumulative cluster mass distributions (solid black line) of the combined cluster sample. The blue coloured solid and thinned associated lines show the best values and the family of solutions for M$_c$ and $\beta_{\rm Sch}$ (left panel, blue lines) contained within the 1$\sigma$ contours of the corner plots. Similarly we plot the predicted cumulative mass distributions (orange lines) for the best value of $\beta_{\rm PL}$ and the associated 1$\sigma$. The predicted distributions contain the same number of clusters used to build the observed mass distributions. On the right plots we include the normalised distributions of recovered number of clusters more massive than the determined M${_*}$. The blue and orange histograms show the recovered number of clusters more massive than M${_c}$ in 1000 monte carlo runs of cluster populations with mass distributions described by the fitted Schechter and power-law mass functions, respectively. The median of the two distributions are indicated by vertical lines and their numbers included in the insets. Observed number of clusters more massive than M${_*}$ are shown as a black vertical line and indicated in the inset. }
    \label{fig:MF_mergerstage}
\end{figure*}

\begin{figure}
    \centering
    \includegraphics[width=8.5cm]{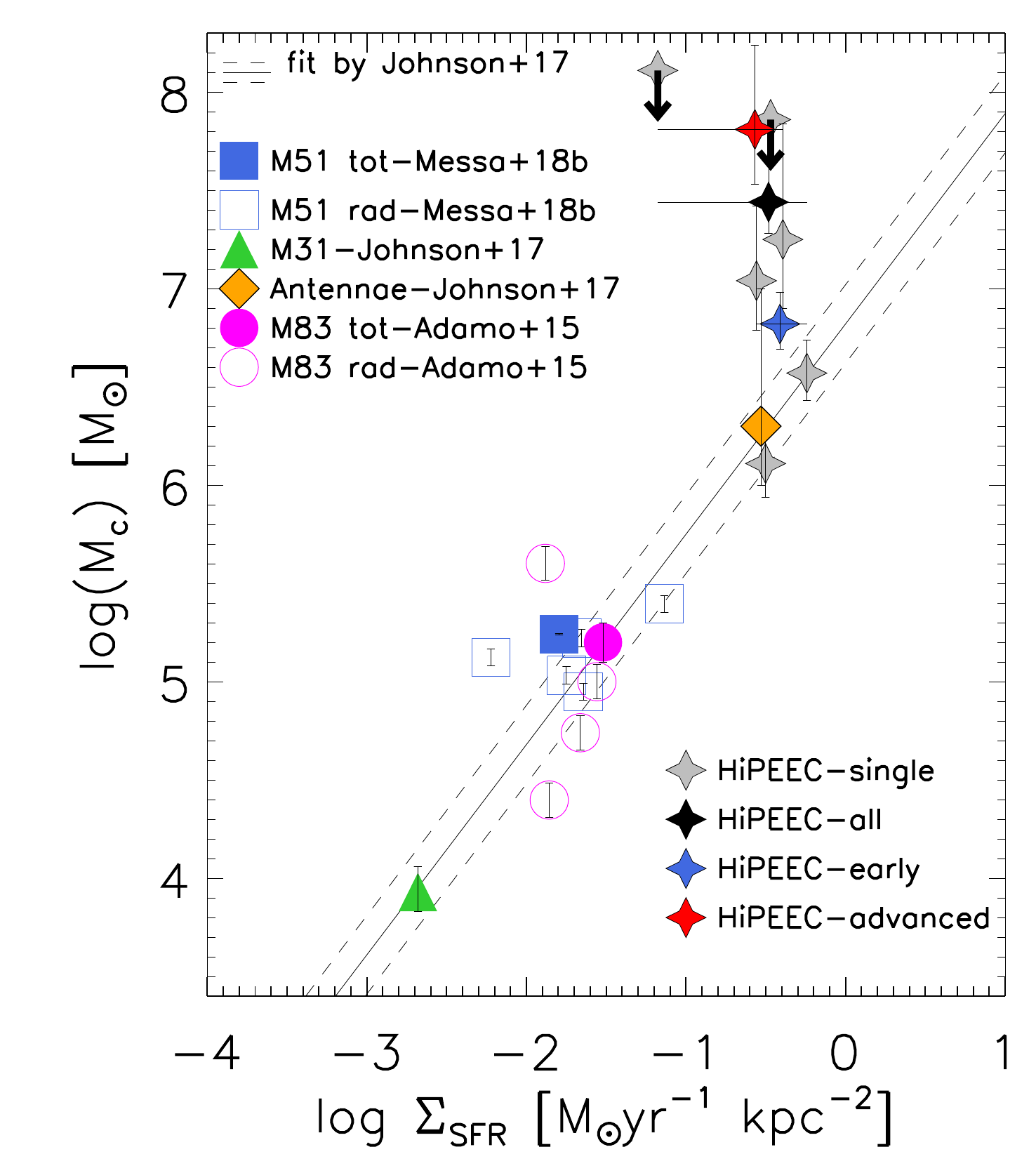}
    
    \caption{The recovered M$_c$ values for the HiPEEC sample (diamond symbols) plotted in the M$_c$ vs. $\Sigma_{\rm SFR}$ plane. Other M$_c$ values determined with high statistical significance in the literature are included. The M83, M51, Antennae and M31 M$_c$ was used by \citet{johnson17} to derive the relation plotted with solid line within dashed line uncertainties. The HiPEEC M$_c$ are among the highest observed M$_c$ reported in the literature. The black diamond show the determination of m$_c$ obtained with the HiPEEC combined cluster sample. The horizontal bar show the $\Sigma_{\rm SFR}$ over the HiPEEC sample. }
    \label{fig:truncmass}
\end{figure}

Table~\ref{tab:MF} contains a complete list of the recovered best fitted values of the mass function analysis, assuming both a Schechter and a power-law function. The analysis of the mass function has been performed for two age intervals 1--10 and 1--100 Myr for each galaxy and the combined sample. The mass limit has been extrapolated from the completeness limit analysis and listed in Table~\ref{tab:MF}.  In the table we also report the number of clusters that are included in the fit according to the selection criteria applied. In the pair system NGC3690, we have analysed the cluster mass function of each system separately, using the division shown in Figure~\ref{fig:ds9cont}, and of the combined cluster population of both systems. The results are all collected in Table~\ref{tab:MF}, while in general we show only the plots for the combined system.  

\subsubsection{The effect of distance on the cluster analyses}
At the distance range of the HiPEEC galaxies, cluster masses are derived from aperture photometry covering physical sizes of $\sim$40 pc (for the three closest targets, NGC3256, NGC4194, NGC3690), $\sim$60 pc (NGC6052, NGC1614), $\sim$70 pc (NGC34). Blending effects could therefore affect the resulting cluster masses and recovered mass function slopes producing a correlation with the distance of the targets. In Figure~\ref{fig:SR} we plot the recovered M$_c$, slopes, and $\Gamma$ (we report the measurements in Section~\ref{sec:gamma}) as a function of the distance. We do not see a systematic increase of M$_c$ and $\Gamma$ or a flattening of the recovered slopes with distance, as expected if blending would significantly affect our measurements. We test the observed distributions running a Spearman's rank analysis. The probabilities, p$_{SR}$, associated to the Spearman's rank correlation analysis  are reported in each panel. Values p$_{SR}>0.01$  are consistent with the null hypothesis of random draws between the two samples of measurements in the $x$ and $y$ axes.  Their values, significantly larger than 0, are in agreement with random draws, thus, confirming the lack of correlation. This is also in agreement with the blending analysis performed by \citet{randria13}, whom report that significant blending affect occurs at distances larger than 80 Mpc, hence, beyond the distance range covered by the HiPEEC galaxies.

\subsubsection{Schechter or pure power-law cluster mass function?}
In Figure~\ref{fig:cornerplots_app} and Figure~\ref{fig:cornerplots} we report the recovered corner plots of the parameters that define the Schechter function (power law slope and truncation mass) in the age range 1--10 and 1--100 Myr, respectively. We consider a fit to be converging (and therefore providing a strong constraint on the shape of the mass function) if the outermost contour (3$\sigma$) of both parameters are contained within the parameter spaces searched. This information is summarised in column 8 of Table~\ref{tab:MF}; where we indicate with the abbreviation {\it Sch} or {\it PL} the preferred fit to the shape of the mass function. In general, we observe that even for the same galaxy, the fit to the age range 1-10 Myr (Figure~\ref{fig:cornerplots_app}) provides poorer constraints to the mass function, with broader ranges of mass values searched, resulting in a large tail at the very high mass end. In these cases, the reported median values do not strictly coincide with the peak values (e.g. see red solid lines reported in the corner plots in Figure~\ref{fig:cornerplots_app}). Half of the targets do not have a converging constraint on the M$_c$ in the age range 1--10 Myr. In these cases, the power-law fit is preferred instead. The same fit performed on the age range 1--100 Myr, provides a tighter convergence on the resulting M$_c$ and $\beta_{\rm Sch}$, in four of the 6 targets. Except for NGC34 and NGC1614, where we do not get a statistically significant constraint of the M$_c$, we recover M$_c$ values that are  within 1$\sigma$ in the two age ranges for each target. The recovered $\beta_{\rm Sch}$ values are systematically shallower than $\beta_{\rm PL}$, but the difference is within the recovered 1$\sigma$ uncertainties between the two values. As already reported by \citet{messa18b} and discussed in \citet{adamo20},  the lack of strong constraints on the truncation mass is mostly limited to the low number of clusters relative to the sampling of the upper mass end of the mass function. The constraints on the fitted parameters improve when we extend the age range as it increases the number of clusters available for the fit. Therefore, the preference for a pure power-law function is driven by the lack of sufficient sampling and not by a realistic representation of the true cluster mass function in these systems. 

Taking this argument into account, we can directly look at the observed cumulative cluster mass distributions, plotted against realisations of the cluster mass functions obtained by the sampled solutions within the 16 and 84 \% of the marginalised posterior distributions in Figure~\ref{fig:MF}. In these panels, we show the cluster mass function in the age range 1--100 Myr, but trends do not change for the shorter age range (shown in Figure~\ref{fig:MF_app} of the Appendix). In the case of NGC34, the small number of clusters included in the fit does not allow to differentiate between a pure power-law or a Schechter distribution. In the case of NGC1614, the Schechter function is better at reproducing the distribution of observed cluster masses above 10$^6$ \msun. The number of clusters above 10$^6$ \msun\, is clearly overestimated by a simple power-law fit. This trend becomes more and more clear in the remaining targets. In all the cases, the number of clusters observed in the upper end of the mass function are overestimated in the pure-power law fit. These results reinforce the conclusions reached above, the lack of strong constraints on the mass function is driven by low number statistics and not by the real shape of the cluster mass function. Indeed, while it is easy to miss low mass (luminosity) clusters because of incompleteness issues, this becomes less likely at the very massive (luminosity) end of the distributions.  

Another way to quantify the differences observed in Figure~\ref{fig:MF} is to use the same family of solutions to estimate the number of clusters we should expect to observe with masses above the truncation mass. In Figure~\ref{fig:predMF} (see Figures~\ref{fig:MF_app} and \ref{fig:predMF_app} of Appendix~\ref{app:MF}  for the same prediction but in the age range 1--10 Myr), we report for each galaxy the number of clusters observed to have mass above M$_c$ (black solid line and value reported in each inset and in column 9 of Table~\ref{tab:MF}), and those expected if a Schechter or pure power-law fit solution is chosen (blue and orange distributions and median values reported in the inset). We outline that the numerical exercise uses the same number of clusters used to derive the fitted parameters (and listed in Table~\ref{tab:MF}) as starting point. In all cases, except in NGC34 where the differences remain elusive, the number of clusters observed and predicted by a Schechter function are always very close. The pure power law fit always over predicts the expected numbers and only marginally (less than 5\% of the realisations) overlaps with the observed values. We try to quantify the offset between the observed and predicted numbers in the last column of Table~\ref{tab:MF}, where we report the ratio of the numbers predicted by a power-law fit (median of the distributions) and the observed ones. In general, we observed that all the fits which have converging 3$\sigma$ contours in the corner plots have also higher significant discrepancies between the number of clusters more massive than M$_c$ observed vs. predicted by a pure power-law function. Discrepancies become smaller when the convergence becomes poorer (e.g. NGC1614, and NGC3690 in the age range 1-10 Myr). 

A different way to overcome the small number statistics is to combine the cluster populations into a single "super-merger galaxy". If the truncation at the high mass end is a result of the poor sampling due to stochastic effects, then the combination of several catalogues, should cancel out any constraint that is driven by low number statistics \citep{adamo20}. We, therefore, analyse the combined mass distribution of all the HiPEEC galaxies and report the results in Figures~\ref{fig:cornerplots_all} and \ref{fig:MF_all} and in Table~\ref{tab:MF}. When combining the cluster populations we need to take into account also the different mass limits determined for each galaxy. We perform the analysis imposing the highest mass limit from NGC34, log(M)$=4.9$ \msun\, (although, notice the very small number of clusters contributed by this galaxy to the combined sample) and the second highest, determined in 3 of the HiPEEC galaxies, log(M)$=4.7$ \msun. The recovered fitted values for M$_c$ and $\beta$ are, within 1$\sigma$ uncertainty, i.e., very similar. This conclusion is true irrespectively of the mass limit applied or age range used. In Figure~\ref{fig:cornerplots_all}, we see that there is a clear convergence for both M$_c$ and $\beta$, which becomes tighter, in the age range 1--100 Myr, were the number of clusters is much higher. In Figure~\ref{fig:MF_all}, we see that the combination of all the cluster populations does not mitigate but increases the discrepancies at the high-mass end between the observed cluster mass distributions and the values predicted by a pure power-law function. 

As a final step, we also analyse the cluster mass function combining the cluster catalogues of galaxies in an early/intermediate stages (NGC3256, NGC3690, NGC6052) versus clusters formed in galaxies in an advanced merger phase (NGC34, NGC1614, NGC4194). The division in the two sub-groups is done accordingly to the cluster age distributions of the galaxies (Figures~\ref{fig:agemass1} and \ref{fig:agemass2}), i.e. whether or not galaxies show enhanced cluster numbers at ages older than 100 Myr (see also Section~\ref{sec:discussion}). 
The analysis of the cluster mass function in these two sub-sample reveals some interesting findings. We report the fitted values in Table~\ref{tab:MF} and Figure~\ref{fig:MF_mergerstage}. In both sub-groups the cluster mass function is better described by an exponential truncation at the high-mass end. Interestingly, we observe a a factor of 10 difference between the recovered M$_c$ in galaxies at the early stage of their merging phase versus galaxies in a significantly advanced stage, which will be discussed below.

\subsubsection{The variation of M$_c$ as a function of star formation rate per unit area}

Already from early studies \citep{larsen09} it has been suggested that the M$_c$ could change as a function of host galaxy environment. These changes were better quantified by \citet{johnson17} who proposed that galaxies with higher $\Sigma_{\rm SFR}$ have larger truncation masses, i.e. can form more massive clusters. The HiPEEC sample is a perfect test-bench for the proposed relation, especially because it samples the locally more rare high $\Sigma_{\rm SFR}$ regimes. In Figure~\ref{fig:truncmass}, we show a revised Johnson's plot with the addition of literature data for local spiral galaxies at intermediate $\Sigma_{\rm SFR}$, and the HiPEEC sample with values of M$_c$ determined in the age range 1--100 Myr. We use the $\Sigma_{\rm SFR}$ reported in Table~\ref{tab:ha}, although variations should be expected over 100 Myr time scales. We include in the plot the HiPEEC galaxies with a secure determination of M$_c$ (gray diamonds), the upper limits derived for two of the galaxies (NGC34 and NGC1614), and the value derived for the combined population (black diamond). For the latter datapoint we include a horizontal bar indicating the $\Sigma_{\rm SFR}$ range spanned by the HiPEEC galaxies. The single galaxy M$_c$ values with secure determinations are all within 2$\sigma$ from the values predicted by the relation. The higher M$_c$ value obtained with the combined sample of all the galaxies, simply reflects the highest truncation masses in the sample, weakly constrained in the two nuclear starbursts NGC1614 and NGC34. We also include the M$_c$ values obtained for the two sub-samples, early/intermediate stage and advanced merger phase galaxies, as blue and red diamond respectively.  The M$_c$ determined from the early-stage mergers is within the error very similar to the position in the diagram of another merger system, the Antennae, that would fit within this category. On the other hand, we observe that the M$_c$ derived for the sub-sample of advanced mergers are significantly offset from the relation, in spite it covers similar $\Sigma_{\rm SFR}$.  The observed scatter may indicate that   $\Sigma_{\rm SFR}$ is not the only parameter that controls M$_c$. We will further discuss this point in Section~\ref{sec:discussion}.  

\subsubsection{Comparison with cluster mass function analyses available in the literature}

\citet{linden17}, using 3 broadband photometry (FUV  combined with B and I band) analysis, determined the cluster properties of a sample of 22 LIRG galaxies (some in common  with this work, e.g. NGC1614, NGC3256, NGC3690). They report the power law fit to the mass functions of NGC1614, NGC3690E (in this work NGC3690A) and NGC3690W (in this work NGC3690B) to be $-1.35\pm 0.23$, $-1.44\pm0.14$ and $-1.92 \pm 0.24$. Their fit was performed with a mass limit of M$>10^5$ \msun. Considering the incompleteness imposed by the detection in the FUV band, their derived values are well in agreement, within the uncertainties, to our power-law fits listed in Table~\ref{tab:MF} for the same galaxies. Their combined cluster populations of the 22 LIRGs result in a power law mass function of slope $\beta = -1.95\pm0.11$ for ages below 10 Myr (M$>10^5$\msun) and $\beta = -1.67\pm0.33$ for ages between 30 and 500 Myr (M$>10^6$ \msun). Although using different age intervals our combined cluster population power-law fits have similar slopes within the uncertainties. However, at odds with the \citet{linden17} study, where they conclude that their power-law slopes are still consistent with a canonical slope of $-2$, our results, with significantly reduced errors, suggest a flattening of the power-law slope for these efficient cluster formation environments. Interestingly, \citet{linden17} report that a Schechter function also produces a reasonable fit to their combined cluster population. Their best fitted parameters are M$_c\sim 10^7$ \msun\, and $\beta_{\rm Sch} \sim -1.8$, which is in very good agreement with our more detailed fit. Both \citet{goddard10} and \citet{mulia2016}, report for the NGC3256 cluster population a power-law mass function slope of $\beta = -1.85 \pm0.12$ and $-1.86 \pm 0.34$, respectively. Considering the more extended field of view we cover for NGC3256 in this work, our derived power-law slopes are in excellent agreement with these previous studies. Similar agreement is obtained with the reported slope obtained by fitting the inner youngest star-forming knots in NGC4194 by \citet{w04}.  Finally, \citet{randria19} report power-law slopes from $-2.61$ to $-1.68$ as a function of increasing age bins and a possible truncation mass at M$_c\sim2\times10^6$ \msun for the mass function of NGC3690. These values are quite different from our recovered parameters. However, we notice that in their luminosity function analysis the authors recover slopes that are much shallower and closer to the slopes we derive here.

\subsection{Cluster formation efficiency}
\label{sec:gamma}

The cluster formation efficiency, $\Gamma$, is effectively a measurement of the total stellar mass forming in bound star clusters with respect to the total stellar mass forming in the galaxy \citep{bastian08}. To estimate $\Gamma$, we use the same approach as developed in \citet{adamo15}. To derive the total stellar mass forming in clusters over an age interval (cluster formation rate, CFR) we integrate the total mass in clusters younger than 10 Myr above the established mass limit of each HiPEEC galaxy. We use that to derive the normalisation of the integral of the mass function and extrapolate the missing stellar mass in clusters less massive than the observational mass limit down to $10^2$ \msun. The error estimates take into account the uncertainties on the SFR and the poissonian error introduced by the uncertainties in the age and mass estimates of each cluster. These latter uncertainties propagate in the estimates of the observed CFR portion, because it relies on the number and total stellar mass in clusters more massive and younger than a certain age and mass limit. To estimate this part of the error we stochastically sample the mass function 1000 times and estimate the CFR at each realisation. The age range used to derive the CFR is limited by the time scales that our SFR tracer is sensitive to, i.e. 10 Myr. The resulting $\Gamma$ values are simply the ratio between the CFR and SFR. We report those values in Table~\ref{tab:gamma} estimated for the entire galaxy and for the clusters enclosed within the area defined by the R$_{80\%}$ (see Figure~\ref{fig:ds9cont}). Clusters younger than 10 Myr, located outside the simplified H$\alpha$ contours used to derive the total SFR, are excluded from the $\Gamma$ estimates.  

\begin{table}
\centering
\begin{tabular}{l c c }
\centering
Galaxy & $\Gamma_{\rm tot}$ & $\Gamma_{80}$ \\
 &[\%] & [\%]  \\
\hline
NGC 34	& 38.9$\pm$1.7 &	48.4$\pm$	3.0	\\
NGC 1614	& 83.1$\pm$15.2 &		99.8	$\pm$15.8	\\
NGC 3256	& 54.1$\pm$3.2 &		62.3$\pm$	4.2	\\
NGC 3690A & 	32.0$\pm$4.7 &		36.0$\pm$	5.3	\\
NGC 3690B & 	59.2$\pm$6.6 &		62.6$\pm$	5.7 \\
NGC 3690& 	41.4$\pm$3.8 &		49.0$\pm$	4.4 \\
NGC 4194	& 	69.8$\pm$4.9 &		84.3$\pm$	27.4 \\
NGC 6052 & 	40.8$\pm$2.0 &		50.7$\pm$	2.6 \\

\hline
\end{tabular}
\caption{Cluster formation efficiency estimated for the total galaxy (within the polygonal solid blue lines in Figure~\ref{fig:ds9cont}) and within the circular area defined by R$_{80\%}$ (shadowed circle in the same Figure). A reminder that NGC3690A is also referred to as IC694, while NGC 3690B is the south-west companion.}
\label{tab:gamma}
\end{table}

\begin{figure*}
    \includegraphics[width=0.7\textwidth]{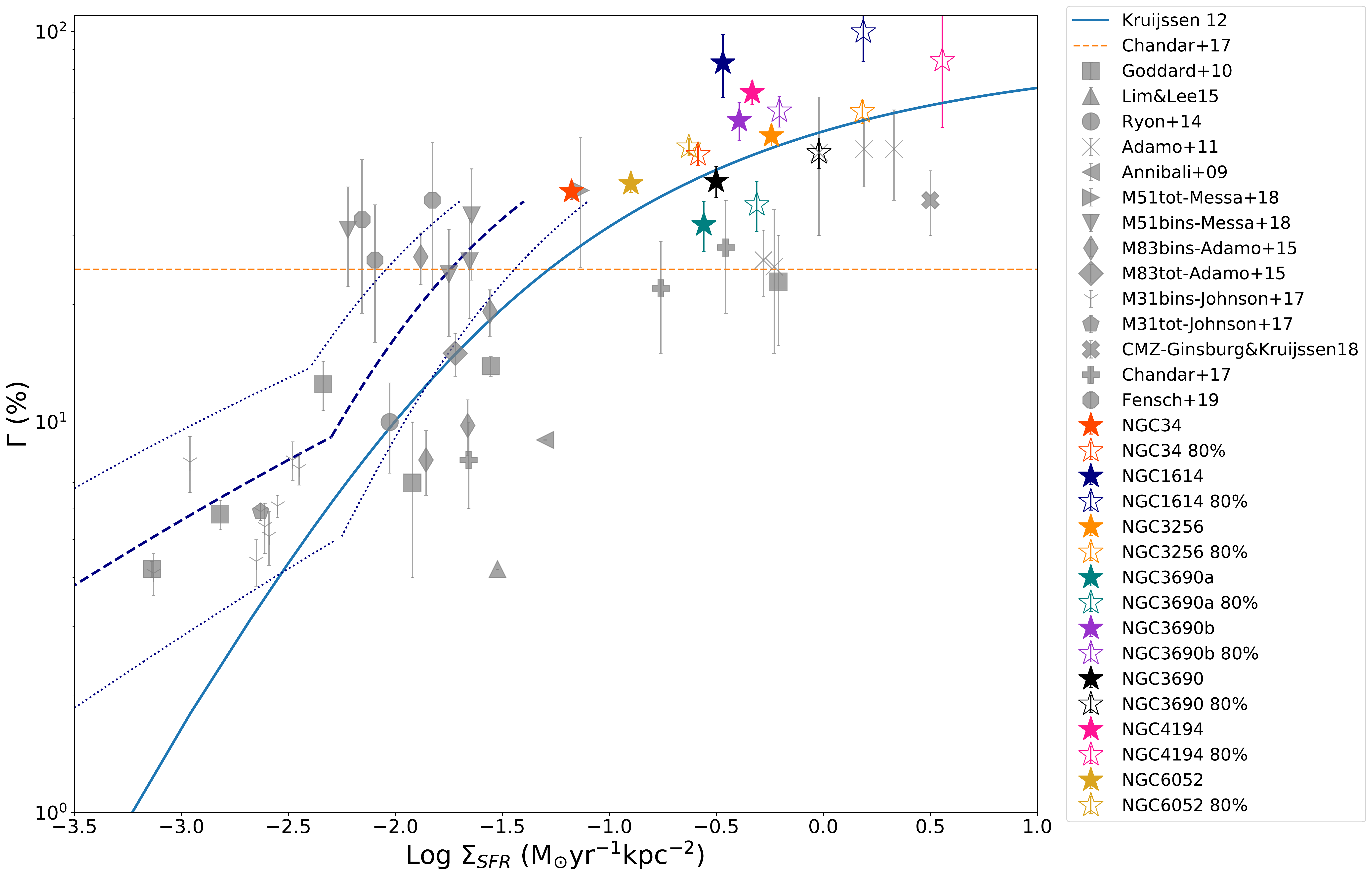}
    \caption{CFE vs. $\Sigma_{\rm SFR}$ diagram. Data in the literature are compiled by \citet{adamo20} and plotted with grey symbols (see legend). If the same galaxy has been studied by multiple authors we report it in the data only once. E.g., we report NGC4214, NGC4449, and the Antennae by \citet{chandar17}, LMC, SMC, NGC1569, M83-centre, NGC6946, MW by \citet{goddard10}. The solid and the dashed-dotted lines reproduce the \citet{kruijssen12b} fiducial model if the Schmidt-Kennicutt or Bigiel et al.'s $\Sigma_{\rm SFR}$ vs. $\Sigma_{gas}$ relation is used for the conversion between the two quantities. The orange horizontal line is the proposed constant CFE at 24 \% value by \citet{chandar17}. We use filled and empty star symbols to show the location in the diagram of the total and inner galactic values of $\Gamma$ for the HiPEEC galaxies. The HiPEEC $\Gamma$ and $\Sigma_{\rm SFR}$ are listed in Table~\ref{tab:gamma} and \ref{tab:sfr}.}
    \label{fig:gamma}
\end{figure*}

In Figure~\ref{fig:gamma} we plot the $\Gamma$ and $\Gamma_{80}$ recovered for the HiPEEC galaxies together with all the literature observed data collected in \citet{adamo20}. The solid line shows the analytical model solution proposed by \citet{kruijssen12b}. The model is based on galactic scale properties, such as the Toomre parameter $Q$, angular velocity, $\Omega$, and gas surface density $\Sigma_{\rm gas}$, and star formation efficiency of a few percent till SN feedback halts the star formation process. The model also takes into account the tidal fields exerted by the dense gas that will prevent stars from forming in bound clusters setting a limit on the resulting bound fraction. In particular, the model reproduced in Figure~\ref{fig:gamma} is considered a fiducial model and not tailored to the gas condition of each single galaxy or portion of galaxy plotted in the diagram. To convert  $\Sigma_{\rm gas}$ into $\Sigma_{\rm SFR}$ (plotted in the x-axes) we use the classic Schmidt-Kennicutt relation \citep[blue solid line,][]{KE12} or the formulation by \citet{bigiel08} (dashed line enclosed between the dotted line intervals) derived using sub-galactic scale regions, and therefore more sensitive to the dominant ISM phase (molecular vs. atomic). The orange dashed line shows the location of constant CFE at all $\Sigma_{\rm SFR}$ proposed by \citet{chandar17} and relevant for the age range used in this work to estimate $\Gamma$ values. Overall we observe that the HiPEEC $\Gamma$ values sit at the highest $\Sigma_{\rm SFR}$ regimes accessible in the local universe, populating a region of the diagram so far sparsely sampled. The scatter around the \citet{kruijssen12b} fiducial model is similar to what already found in the literature. Partially this scatter is caused by the intrinsic bias introduced by the data. As already discussed by \citet{adamo20}, the age range 1-10 Myr is more prone to contamination from unbound clusters that appear  morphologically compact. The second effect is due to the distance of the HiPEEC sample, our apertures correspond to diameters of $\sim$40 pc (for the three closest targets, NGC3256, NGC4194, NGC3690), $\sim$60 pc (NGC6052, NGC1614), $\sim$70 pc (NGC34) therefore,  blending might occurs at these young ages since clusters rarely form in isolation \citep[see discussion in][]{adamo20}. However, as shown in Figure~\ref{fig:SR}, we do not find a correlation between $\Gamma$ and distance suggesting that the trend observed in the $\Gamma$ vs.  $\Sigma_{\rm SFR}$ are not driven by systematics. In general, we consider the recovered $\Gamma$ values as upper limits to their true $\Gamma$. Our $\Gamma$ values are significantly above the constant $\Gamma \sim24$\% value proposed by \citet{chandar17} in the age range 1--10 Myr. We observe that the offset of the the HiPEEC sample from the constant reference value of $\Gamma=24$\% increases with $\Sigma_{\rm SFR}$ and it does not depend on the distance of the galaxies, suggesting that the offset is not driven by increasing blending effects, but it is physical.

If we limit our considerations to the HiPEEC sample we observe some interesting trends which reinforce the idea that the fraction of star formation in clusters is truly dependent on the properties and conditions of the dense gas where star formation will take place. In general, CFE within the inner regions of the galaxies are significantly higher, as expected, because of the increase observed in the $\Sigma_{\rm SFR}$(80\%). The two minor merger systems, NGC4194 and NGC1614, have record values of $\Gamma$ with respect to the rest of the HiPEEC targets, for both the total galaxy and within their central regions. Their elevated $\Gamma$ could possibly be the result of the rapidly changing conditions in the gas, not truly traced by our SFR tracer. We observe for the pair system NGC3690, that CFE is higher in the south west component (NGC3690B). This difference is also reflected in the properties of the clusters, with the most massive youngest clusters ($<$10 Myr) detected in NGC3690B,  as well as in the SFR and SFR densities, higher in NGC3690B. These findings are at odds with the FIR and radio view of the system where the brightest region coincide with the nuclear starburst in NGC3690A (see Section~\ref{n3690}).  However it is important to keep in mind that our study is biased to the temporal window accessible to optical wavelengths, while FIR and radio are sensitive to the next generation of stars and clusters that are on the making. The offset between the two tracers clearly suggests that starburst conditions within NGC3690 are rapidly changing on scales of ~10 Myr.   

\section{Discussion \& Conclusions}
\label{sec:discussion}
As discussed in the Introduction, rare local (U)LIRGs represent a unique chance to understand star cluster formation mechanisms and conditions at the peak of the cosmic formation history. The merger event is recreating in these local galaxies similar gas conditions experienced by normal "main-sequence" disk galaxies at redshift $\sim 2-3$. It is in this redshift range that a large fraction of GC populations surviving today and observed around local L$^*$ galaxies have formed \citep{reina19}.  

In Section~\ref{sec:cluster_pos}, the analyses of the position of cluster ages and masses within the galaxies reveal two main different stages in the cluster formation process of these merging systems. The most advanced merger stages, i.e. NGC34, NGC1614, NGC4194, are mostly forming massive clusters in their most inner regions, where gas pressures are high. Cluster formation is also happening in tidal features but under totally different gas conditions, resulting in significantly smaller cluster masses (at least of a factor of 10) as observed in the analyses shown in Figure~\ref{fig:agemass1} and \ref{fig:agemass2}. The masses we retrieve in these streams are indeed comparable with the masses and ages reported by \citet{fensch19} in a small sample of tidal dwarfs. The remaining 3 merger galaxies in our sample are forming massive clusters across the whole galaxy body and do not show confinement of very young clusters ($<$10 Myr) in their inner regions or in tidal streams. In the latter systems, the differences of cluster masses within and outside R(80\%) at very young ages are not as pronounced. Detailed high-spatial resolution hydrodynamic simulations of the Antennae system merger by \citet{renaud15} can also help to understand the progression of cluster formation. They find that, during the first passages, cluster formation is extended to the entire system and that clusters can form with masses up to 30 times more massive than in local spirals like the Milky Way, similarly to what we see in half of our sample (e.g. NGC3256, NGC3690, NGC6052). During the final coalescing phase, they observe enhanced star formation only in the inner region of the galaxy, where gas is driven in via dynamical mechanisms. However, they do not see significant cluster formation in these final phases. More recent numerical simulations by \citet{lahen20}, confirm the general trends observed by \citet{renaud15} but find that the CFR follows very closely the enhanced SFR peaks, including the final merger stages. Observationally, the confinement of the starburst within the inner regions of NGC34, NGC1614, NGC4194, would suggest that these systems are approaching their final merging phase, and that during these final phases very massive clusters can still form. Studies of GMC populations in local nuclear starbursts and merger systems (e.g. NGC253, the Antennae), as well as in high-redshift galaxies \citep{DZ19}, show that GMCs can easily reach $10^7-10^8$ \msun\, and live in an almost stable equilibrium under ambient gas pressure reaching $10^7-10^8$ [K/cm$^3$]. We know very little of the GMC properties in our HiPEEC sample, however for those systems for which GMCs have been detected, e.g. NGC4194 and NGC1614 \citep{koenig14, koenig13}, it confirms the general scenario we have described. 

We do not observe significant differences between the maximum masses of clusters located within or outside R(80\%) at ages older than 10 Myr. This evidence could be connected to orbital migration of clusters after their formation as observed in numerical works \citep[e.g. see simulations by][]{kruijssen12a, renaud17}. During the merger event, clusters are affected by strong tidal fields, which can easily change the orbits of the clusters and mix clusters formed in different episodes of star formation. 

The analysis of the cluster mass function of the HiPEEC galaxies, presented in Section~\ref{sec:massfunc}, can help to shed light on the formation mechanism of massive star clusters. We observe that in 4 of the 6 galaxies a Schechter function (i.e., a power law distribution exponentially truncated at the high-mass end above a certain characteristic mass, M$_c$) is statistically a better representation of the observed cluster mass function. The statistical significance of the presence of a truncation mass increases when we combine all the cluster catalogues of our sample, reinforcing the idea that the lack of convergence in the fit is driven by small number statistics, allowing only weak constraints on M$_c$. We observe M$_c$ to vary from $10^6$ up to a few times $10^7$ \msun\, and slopes between $\sim -1.5$ and $-2.0$. The fit to the combined sample results in a $\log($M$_c)= 7.44^{+0.22}_{-0.16}$ \msun\, and a slope $\beta = -1.70 \pm0.03$. These are among the most massive M$_c$ yet determined in local galaxies. The analysis of the combined cluster populations of advanced versus early/intermediate stage galaxies show significant differences in the recovered truncation masses. Galaxies in a more advanced merger stage have $\log($M$_c) =  7.81^{  +0.43}_{-0.28 }$ and slope of $-1.55^{+0.05}_{-0.05}$, while in early/intermediate merger stage galaxies we obtain $\log($M$_c)=6.82^{ +0.16}_{-0.13}$ and slope of $-1.7^{ +0.04}_{-0.04}$. These differences suggest that condition for cluster formation  are changing during the merger stage and become quite extreme in the final coalescing phase, when the starburst has a very compact morphology and is confined in the centre of the galaxy.

We fit the mass function for cluster ages of 1--10 and 1--100 Myr and we do not observe significant differences between the retrieved M$_c$ and $\beta$ values. The latter evidence suggests that the observed mass functions are quite close to the initial cluster mass function and have not yet been severely affected by cluster disruption, expected to further flatten the mass distributions \citep{kruijssen12a}. 

The origin of M$_c$ is possibly linked to the galactic environment and studying YSC populations in the local universe may help to shed light on the formation process of GC populations. \citet{johnson17} propose a positive empirical relation between M$_c$  and $\Sigma_{SFR}$. We notice that the single HiPEEC galaxies with well constrained M$_c$ values scatter around the expected values suggested by the relation, which was derived by using a single determined M$_c$ value at the high $\Sigma_{SFR}$ obtained from the fit to the mass function of the Antennae. Indeed, our derived M$_c$ value for the combined cluster population of the early/intermediate merger galaxies, is very close to the value obtained for the Antennae system, which would belong to this sub-sample. \citet{elmegreen18}, via theoretical arguments, suggests that the observed relation between  M$_c$ and  $\Sigma_{SFR}$ is tracing a more fundamental relation between M$_c$ and gas pressure (density). Only galaxies that can reach high gas pressure, like starburst dwarf systems and merger/interacting galaxies in the local universe can form massive clusters (M$\sim10^6$\msun), while typical gas pressure in local spirals would have a much lower M$_c$. 

However, the retrieved truncation mass for the advance stage sample, as well as for the combined HiPEEC sample, results in larger M$_c$ values that scatter away from the relation, towards higher masses but still same $\Sigma_{SFR}$ range. This offset would suggest that other physical parameters, not captured by the observable $\Sigma_{SFR}$, govern the formation of M$_c$.   \citet{RCK17} propose a theoretical model where the resulting M$_c$ for a galaxy or a sub-region of a galaxy, is proportional to the Toomre mass multiplied by a star formation efficiency and cluster formation efficiency, weighted by a times scales parameter that controls whether feedback or shear halts star formation within a GMC. Interestingly, the Toomre mass is proportional to both the gas density and the angular velocity, therefore, this model predicts that M$_c$ depends also on the dynamical conditions of the gas in the disk, described by the angular velocity, which traces the stability of the gas in the disk. Cosmological simulations incorporating sub-grid analytical models \citep[which included the analytical model by][among others]{RCK17} for GC formation and evolution \citep[E-MOSAICS,][]{pfeffer18, kruijssen19} can reproduce the M$_c$ vs. $\Sigma_{SFR}$ relation observed by \citet{johnson17}. In particular, \citet{pfeffer19} finds increasing normalisations in the retrieved M$_c$ vs. $\Sigma_{SFR}$ relation as a function of redshift. They explain this offset towards higher M$_c$ for similar $\Sigma_{SFR}$ as an effect of increasing turbulence and gas instability in the disk of high-redshift galaxies. Similarly, the offset we observe between M$_c$ in early/intermediate stages and in advanced merger stages over the same  $\Sigma_{SFR}$ in the HiPEEC sample, suggests a dramatic change in the dynamical conditions of the gas during the final evolutionary phases of the merger. The similarity of the trend observed in the \citet{pfeffer19} simulations and the scatter in the HiPEEC sample towards higher M$_c$ for a given $\Sigma_{SFR}$ would reinforce the suggestions that indeed cluster formation during these merging phases is quite close to cluster formation happening in normal main-sequence disk galaxies at redshift $z \sim 2$.

It is important to point out that, observationally, determining the CFE is very challenging and several uncertainties affect the measurements \citep[e.g.][]{KMK18, adamo20}. At the average distance of the HiPEEC sample ($\sim60$ Mpc) we are not able to ensure the ideal conditions for single cluster detection, thus our derived cluster formation efficiency should be considered upper-limits to real values, when compared to $\Gamma$ determined in much closer galaxies ($< 20$ Mpc). However, galaxies with elevated SFRs are very rare in the local universe and no galaxies with SFR$>10$ \msunyr\, are found within 20 Mpc (i.e. the Antennae system is the closest system). The same limitations also affect the majority of the data collected in the literature at $\Sigma_{SFR}\geq -1$ [\dsfr], except the central molecular zone (CMZ) in our own Milky Way \citep{GK18} and the nuclear starburst in M83 \citep{goddard10}. Overall, the scatter we observe around the \citet{kruijssen12b}  fiducial model is similar to the one found at  $\Sigma_{SFR}\leq -1$ [\dsfr] for more nearby galaxies, suggesting that while the fiducial model reproduces the trends observed in the data, it is not tailored to the physical properties of each galaxy, which drive $\Gamma$ in the model. We observe an increasing offset of our measured $\Gamma$ from the constant $\Gamma\sim24\%$ proposed by \citet{chandar17}. The offset positively correlate with increasing $\Sigma_{SFR}$ values. In a recent numerical work by \citet{lahen20}, the authors report that the CFE during the merger phase of two low mass galaxies, changes between 20 and 80 \% at the peak of the merging phase. We observe the same trend in the most advanced mergers in our sample, with the two nuclear starburst systems, NGC1614 and NGC4194, having the highest $\Gamma$ reported in the literature. Detailed study of the dense molecular gas in these two nuclear regions \citep{koenig14, koenig16}, show that large flows of dense gas are conveyed within regions of a few 100 pc in radius, most likely forming extreme conditions for cluster formation. It is interesting to notice that our optical study is biased against the extreme nuclear starburst arbored in major merger systems like NGC3690 and NGC3256, suggesting that star and cluster formation in these systems will rapidly evolve.

While historically we have been anchored by the HST sensitivity to cluster population analyses at UV and optical wavelengths, this electromagnetic window is far from ideal when studying cluster populations in merging/interacting systems, usually affected by large extinction variations. The combination of medium and broad band red optical HST data and NIR and MIR imaging with the upcoming James Webb Space Telescope and Extremely Large Telescope will provide a much needed leap forward into studying clustered star formation in these systems. When comparing the properties of star clusters studied at optical wavelengths with studies tracing the dense gas conditions in these galaxies, we clearly see a temporal delay and spatial displacement between the two tracers. A dust-free view into these galaxies, with comparable spatial resolution as HST and ALMA, will provide a unique chance to bridge the gap in our view of these rapid evolving starburst phases and derive tighter constraints of the properties of cluster formation and evolution in these rare local analogues of main sequence galaxies at  cosmic noon.

\section*{Acknowledgements}
We thank Nate Bastian and Florent Renaud for providing useful comments to the manuscripts. A.A., M.H. and G.\"O acknowledge the support of the Swedish Research Council, Vetenskapsr{\aa}det, and the Swedish National Space Agency (SNSA). M.H is a Fellow of the Knut and Alice Wallenberg Foundation. J.M.D.K. gratefully acknowledges funding from the Deutsche Forschungsgemeinschaft (DFG, German Research Foundation) through an Emmy Noether Research Group (grant number KR4801/1-1) and the DFG Sachbeihilfe (grant number KR4801/2-1), as well as from the European Research Council (ERC) under the European Union's Horizon 2020 research and innovation programme via the ERC Starting Grant MUSTANG (grant agreement number 714907). Based on observations obtained with the NASA/ESA Hubble Space Telescope, at the Space Telescope Science Institute, which is operated by the Association of Universities for Research in Astronomy, Inc., under NASA contract NAS 5-26555. This research has made use of the NASA/IPAC Extragalactic Database (NED), which is operated by the Jet Propulsion Laboratory, California Institute of Technology, under contract with the National Aeronautics and Space Administration. This research has made use of NASA’s Astrophysics Data System Bibliographic Services (ADS). This research made use of Astropy, a community-developed core Python package for Astronomy (Astropy Collaboration et al. 2013, 2018).


\section*{Data Availability}

Reduced HST data and photometric cluster catalogues are associated to the DOI 10.17909/t9-cn0b-ht83 and will be made available at the URL http://dx.doi.org/10.17909/t9-cn0b-ht83. 






\appendix

\section{Completeness limits}
\label{app:completeness}
We report here the plots illustrating the detection rates of mock clusters per luminosity bin in the 4 reference broadbands used for the cluster selection in all the HiPEEC galaxies.
We refer to Section~\ref{sec:completeness} and Table~\ref{tab:completness} for the description of the analysis and the adopted completeness limits in our sample.

\begin{figure*}
    \centering
    \includegraphics[width=0.45\textwidth]{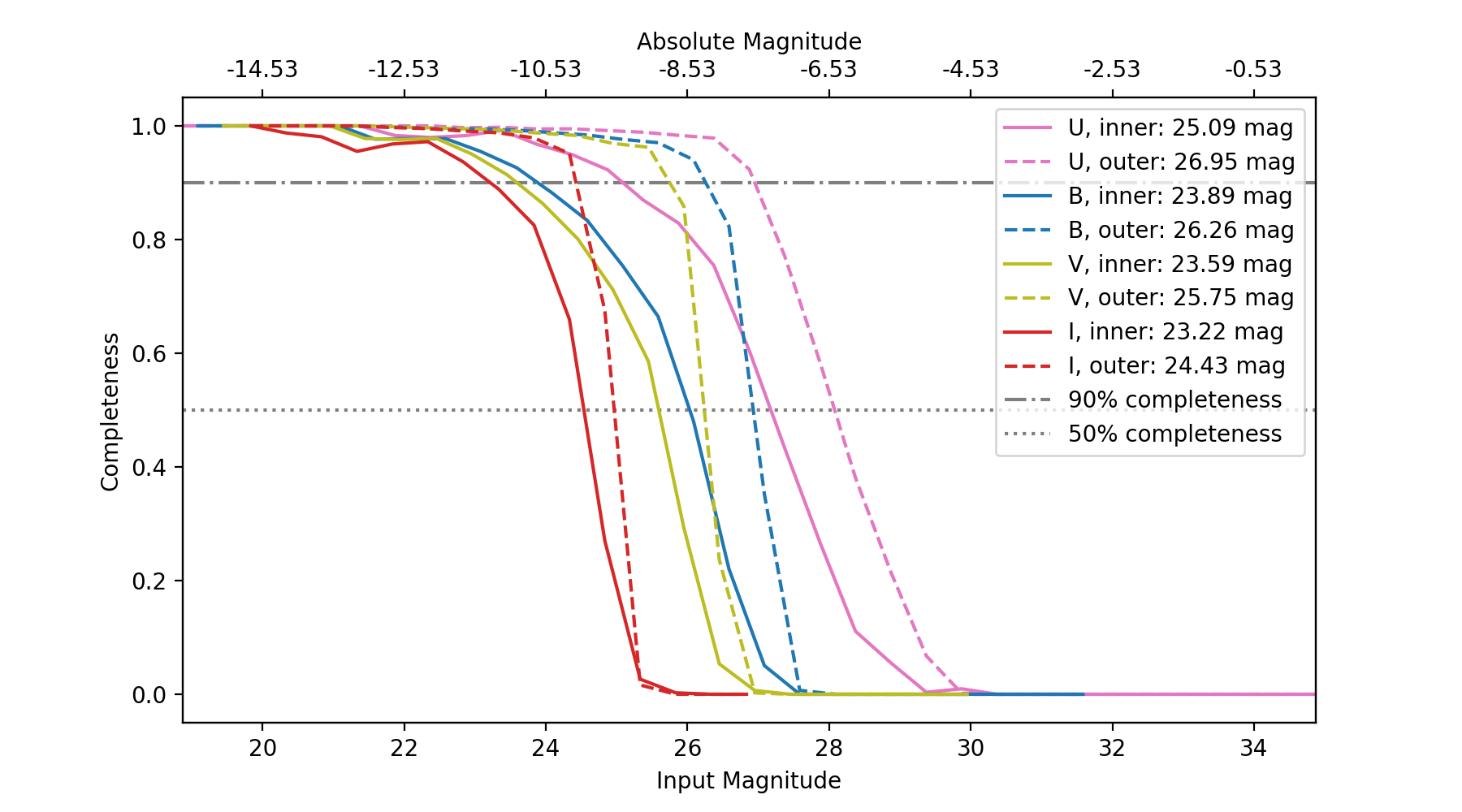}
    \includegraphics[width=0.45\textwidth]{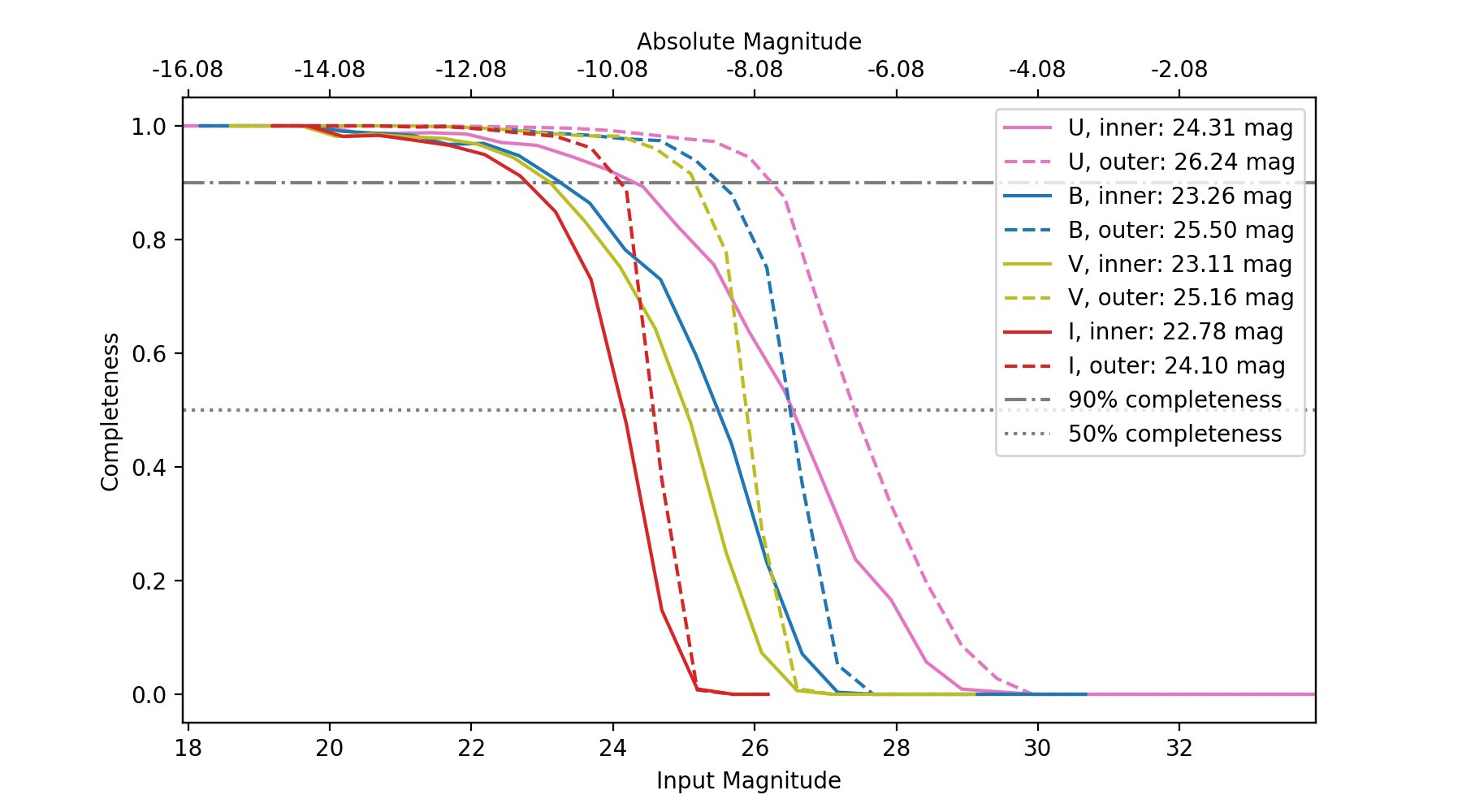}
    \includegraphics[width=0.45\textwidth]{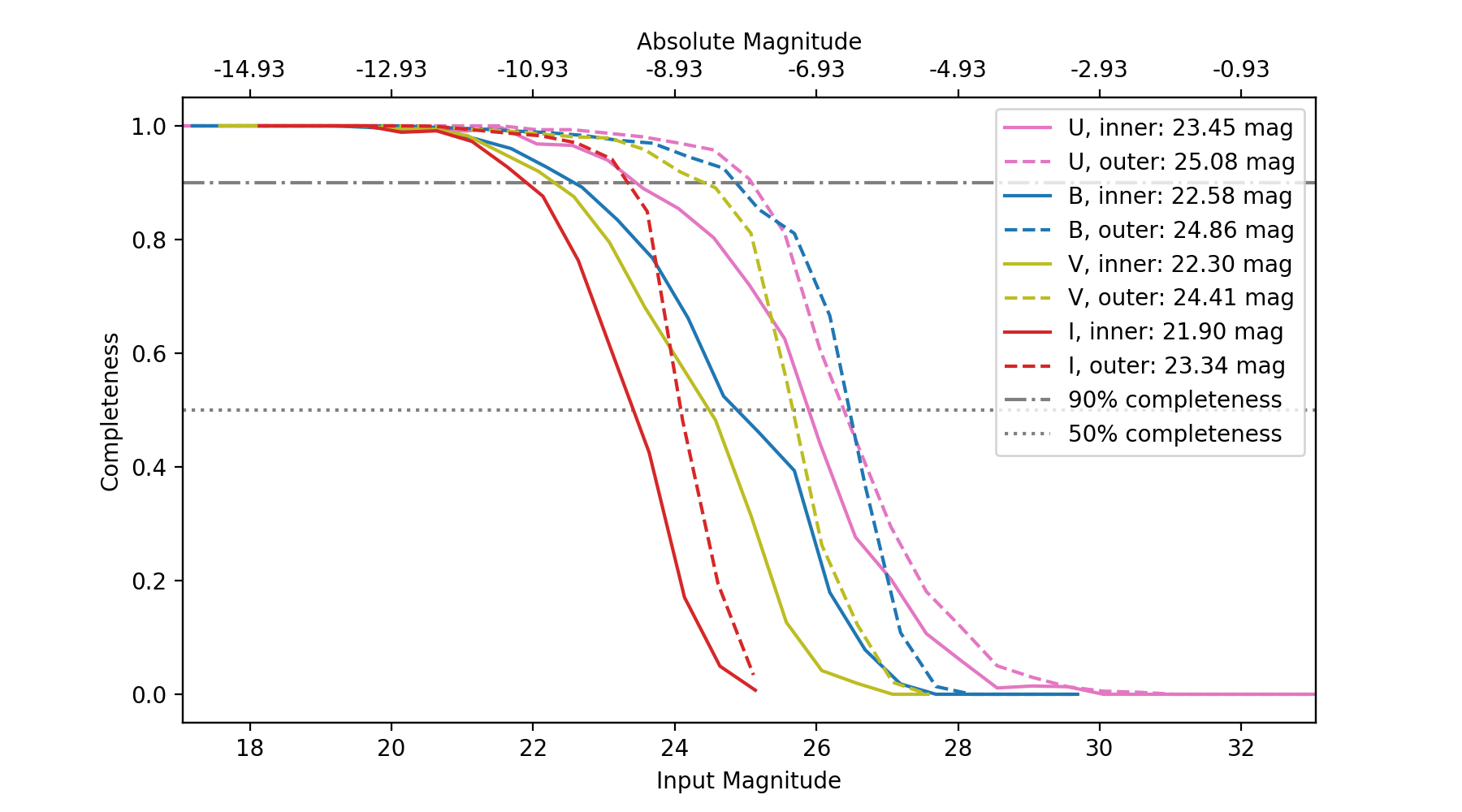}
    \includegraphics[width=0.45\textwidth]{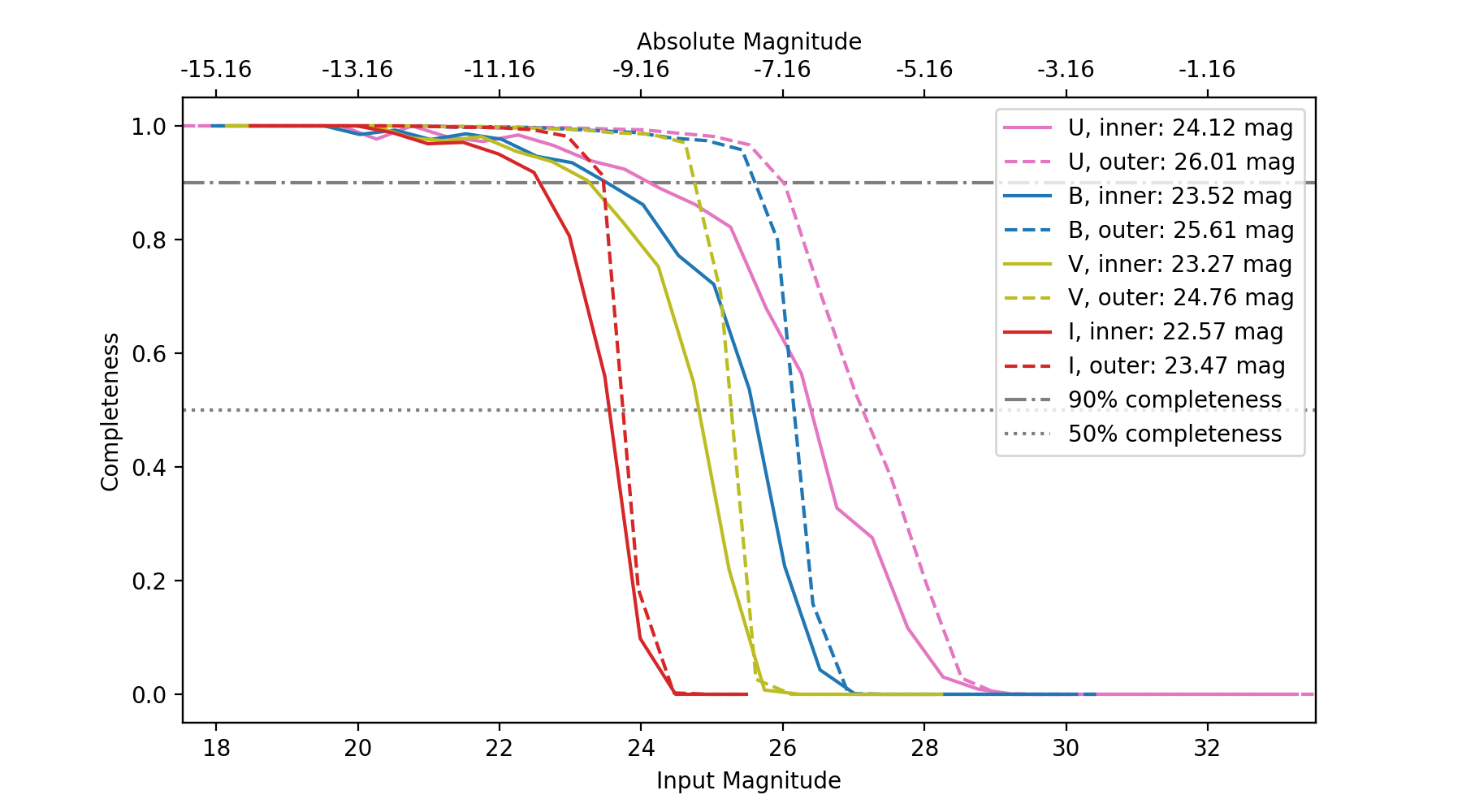}
    \includegraphics[width=0.45\textwidth]{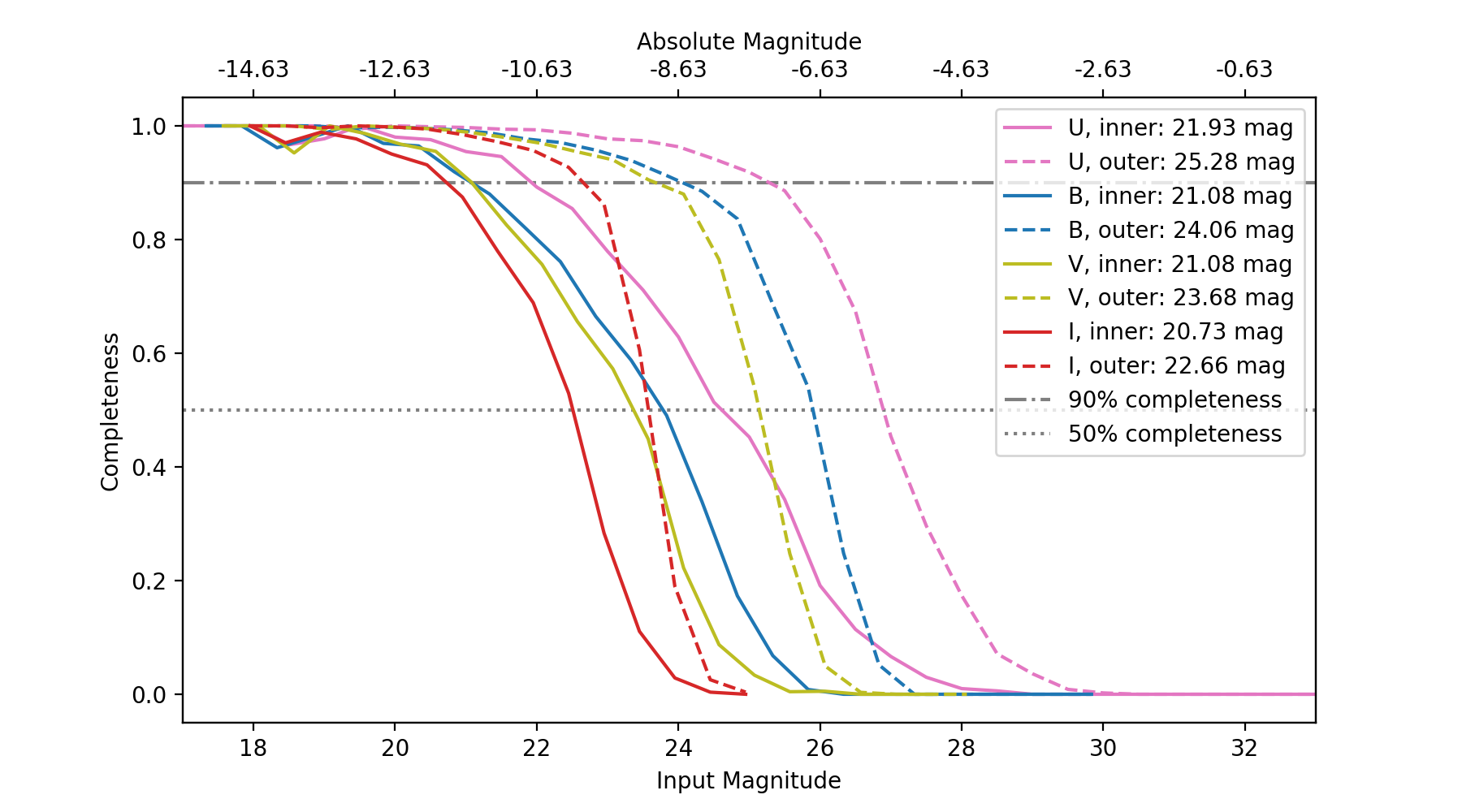}
    \includegraphics[width=0.45\textwidth]{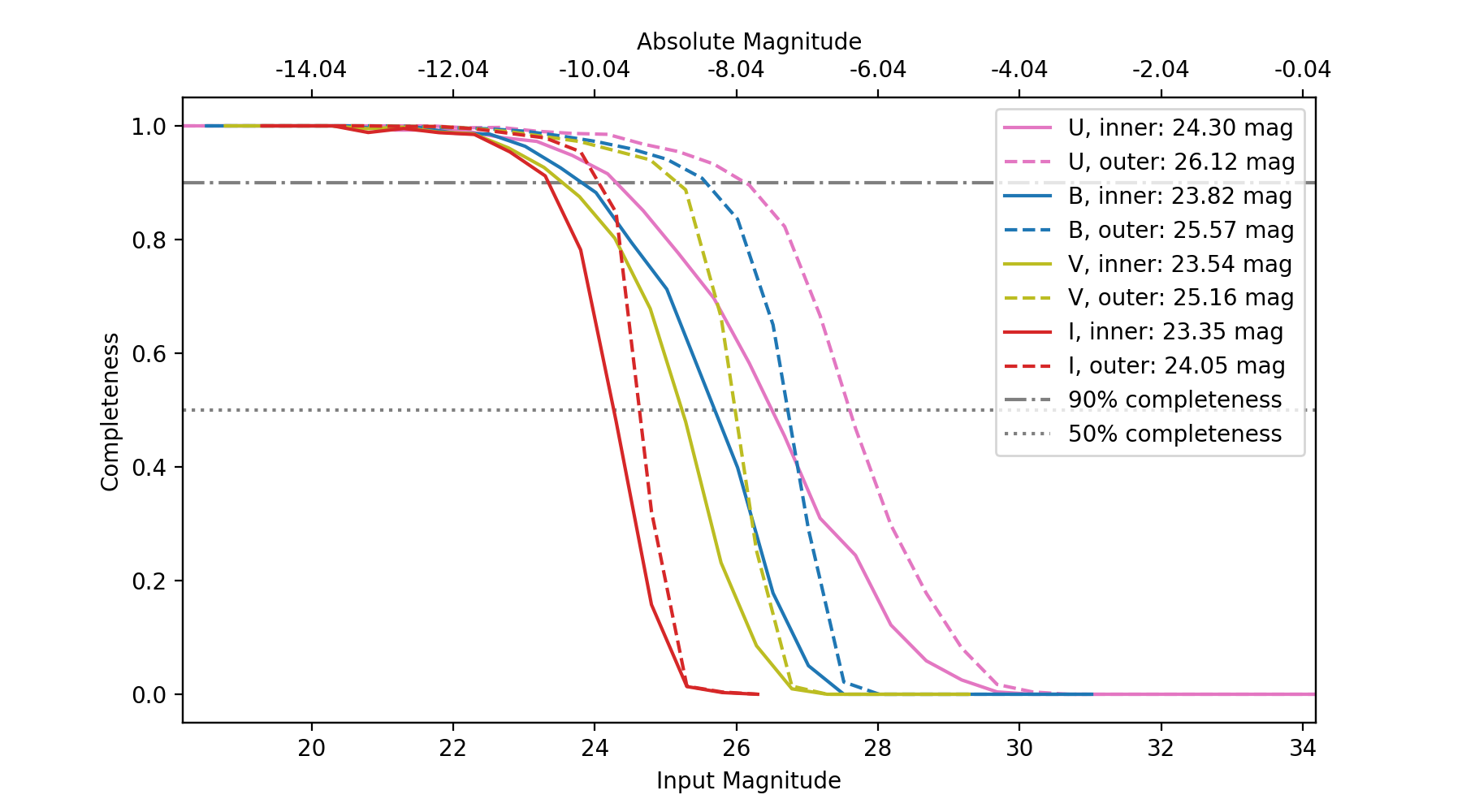}
      \caption{Fraction of mock clusters recovered by the HiPEEC pipeline, or completeness, as a function of input magnitude. Completeness was calculated in 0.5 magnitude bins. The line colour corresponds to filter, and the line style corresponds to the inner or outer region of the galaxy. The legend tabulates the 90\% completeness limits for each filter and galaxy region combination.}
    \label{fig:completeness}
\end{figure*}

\section{Mass function analysis: supplementary material}
\label{app:MF}
We report here, the cluster mass function analysis performed on the cluster age range 1--10 Myr and on the combined cluster sample of 4 galaxies with converging Schechter function fits presented and discussed in Section~\ref{sec:massfunc}. We refer the reader to that Section of the draft for more details.

\begin{figure*}
    \centering
    \includegraphics[width=0.49\textwidth]{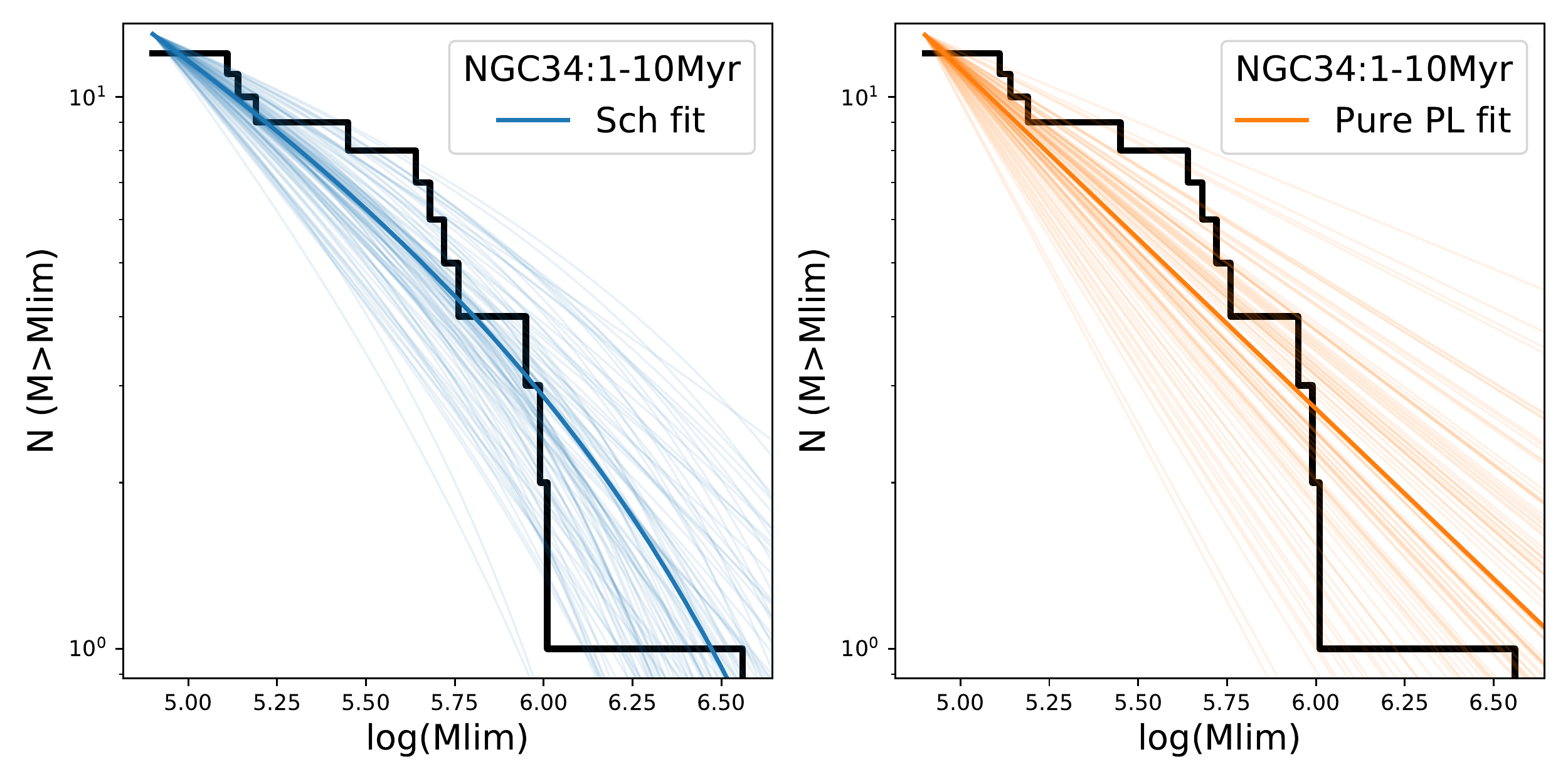}
    \includegraphics[width=0.49\textwidth]{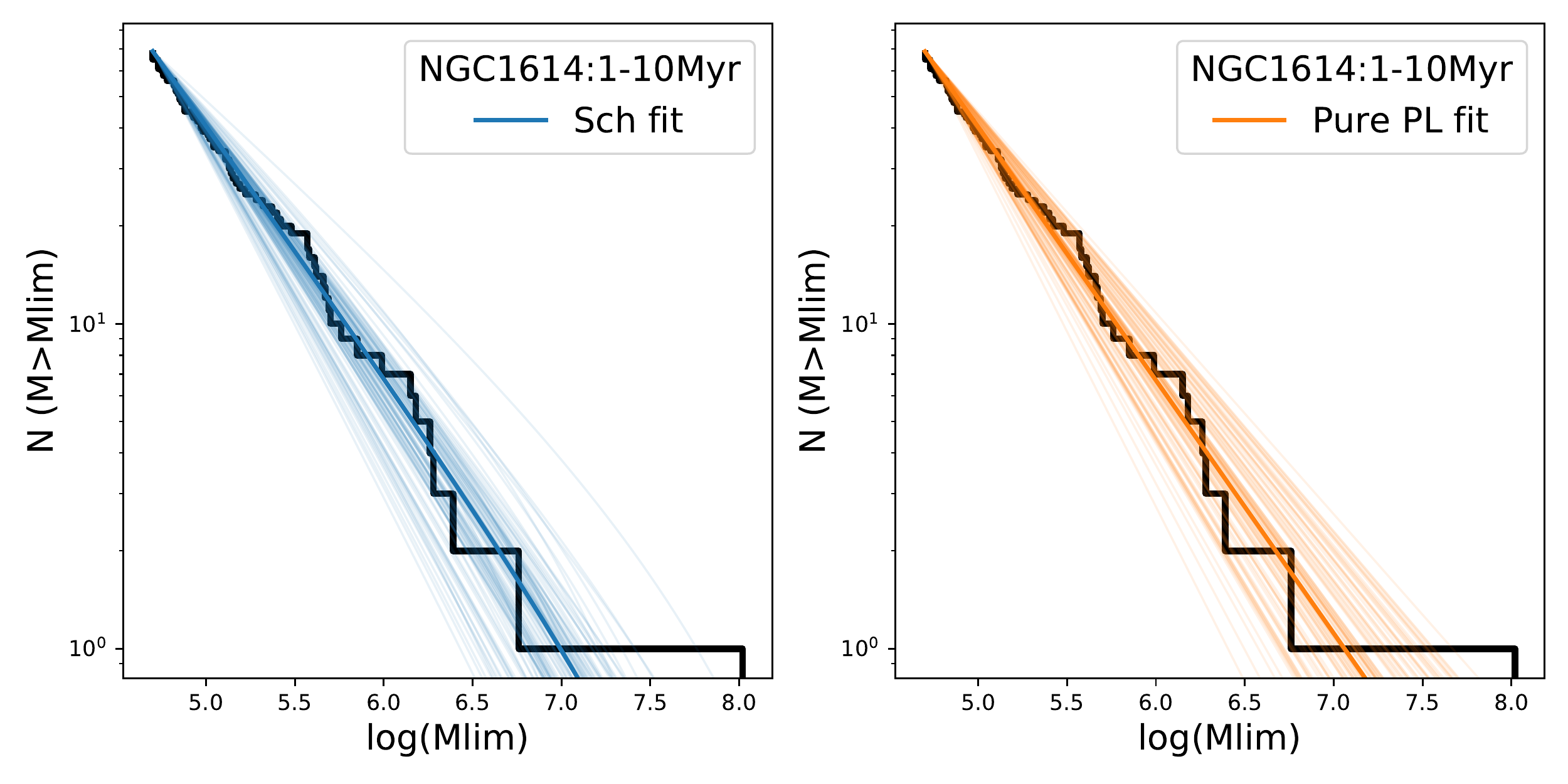}\\
    \includegraphics[width=0.49\textwidth]{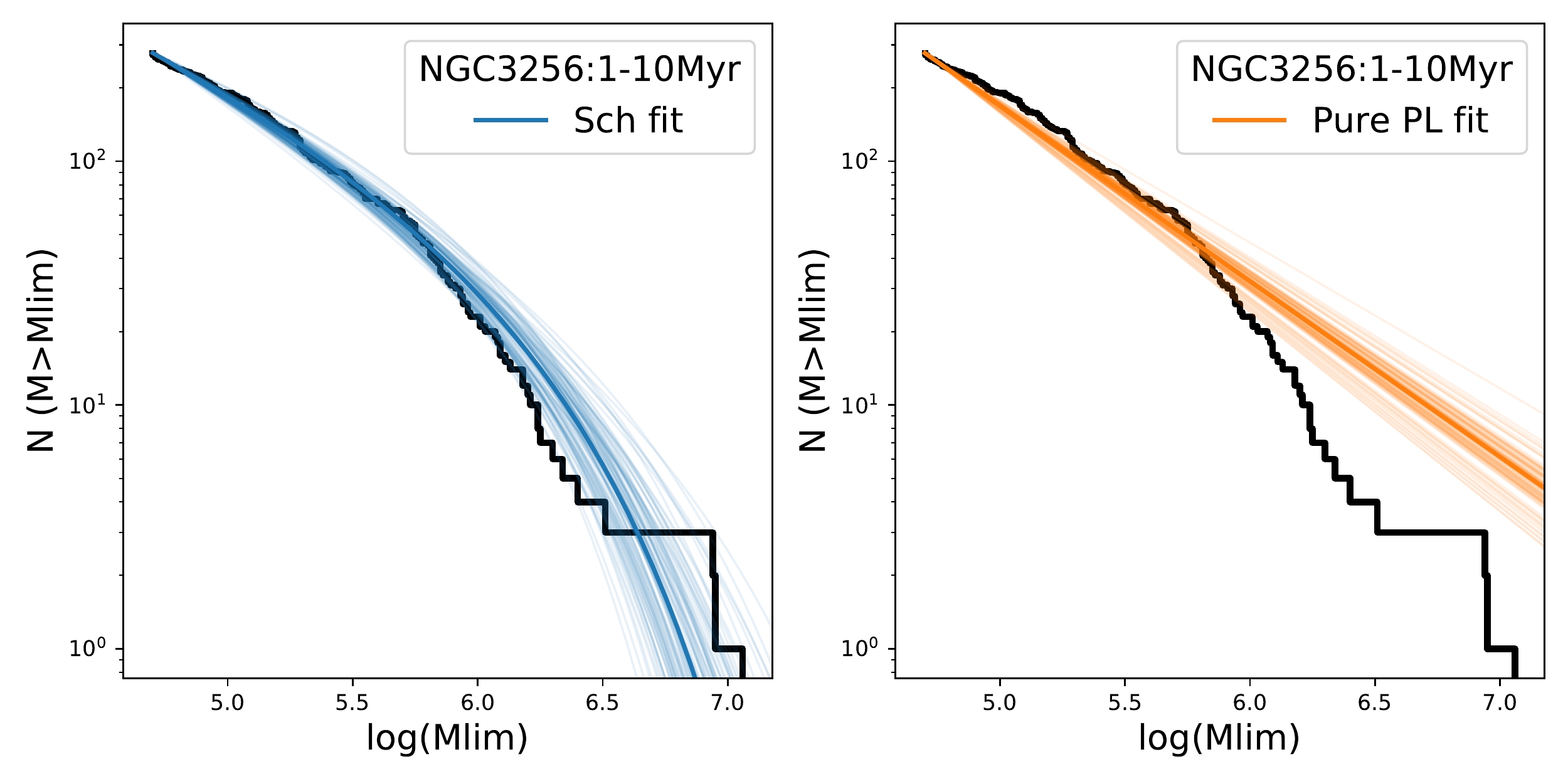}
    \includegraphics[width=0.49\textwidth]{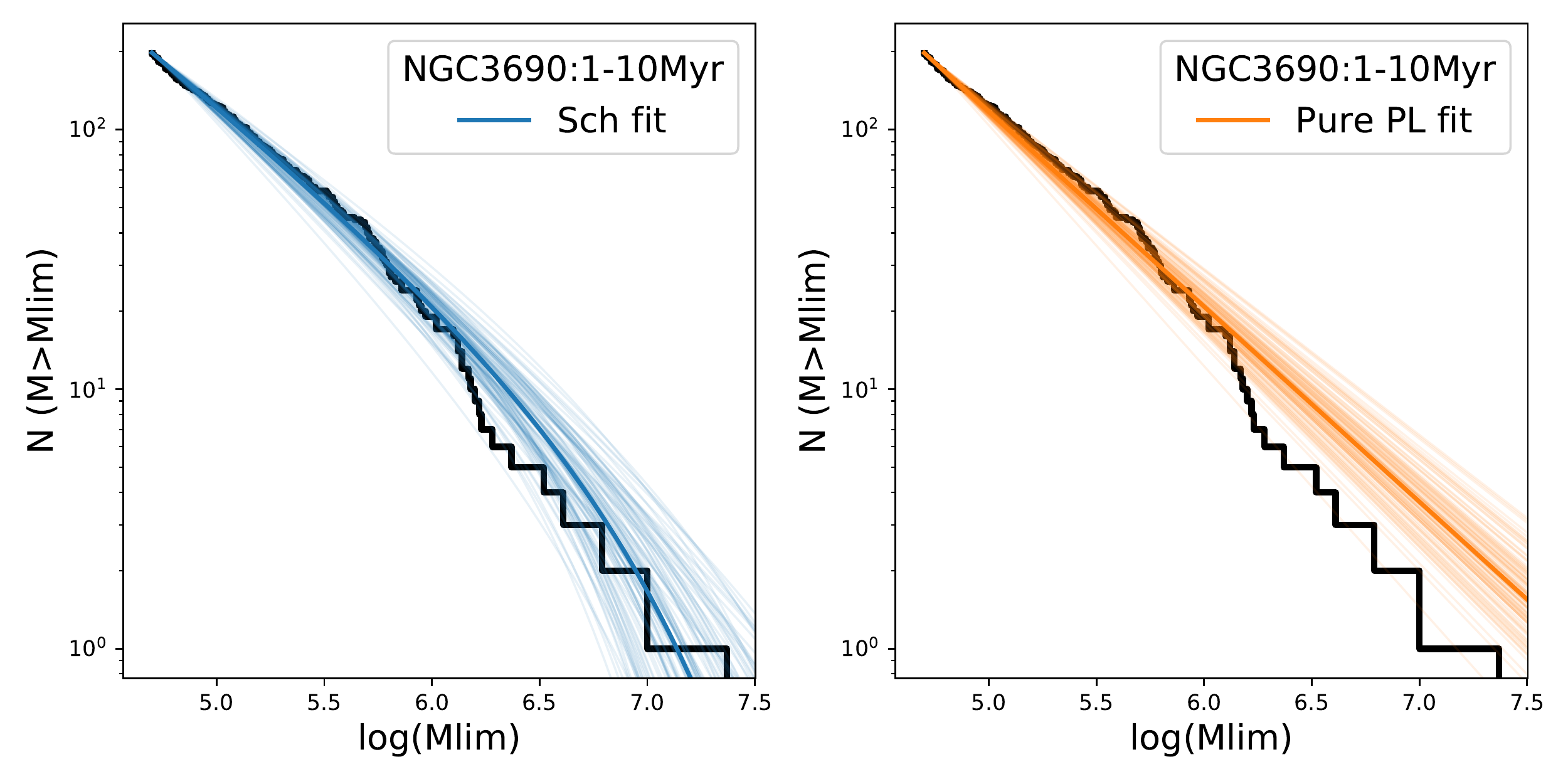}\\
    \includegraphics[width=0.49\textwidth]{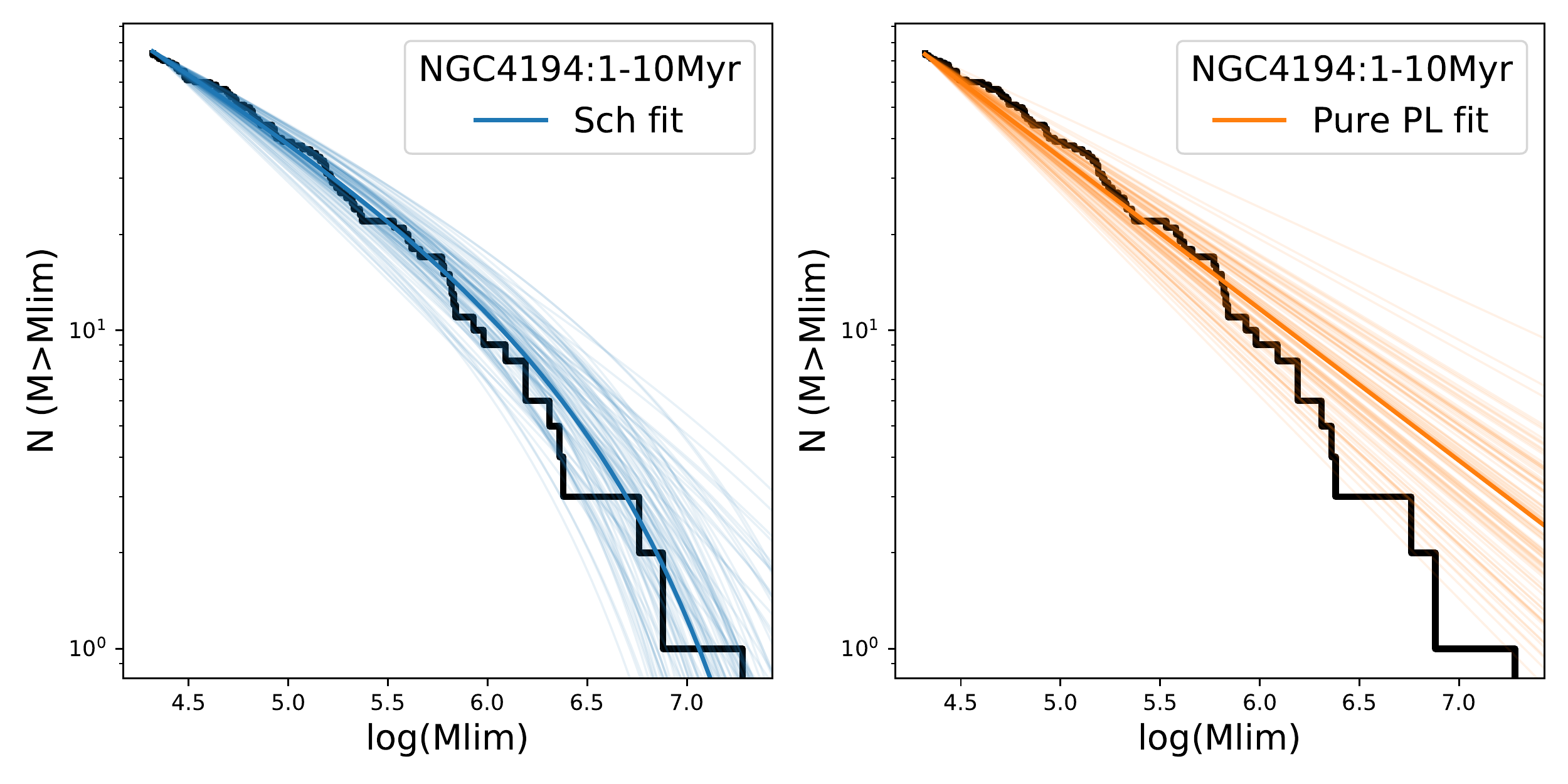}
    \includegraphics[width=0.49\textwidth]{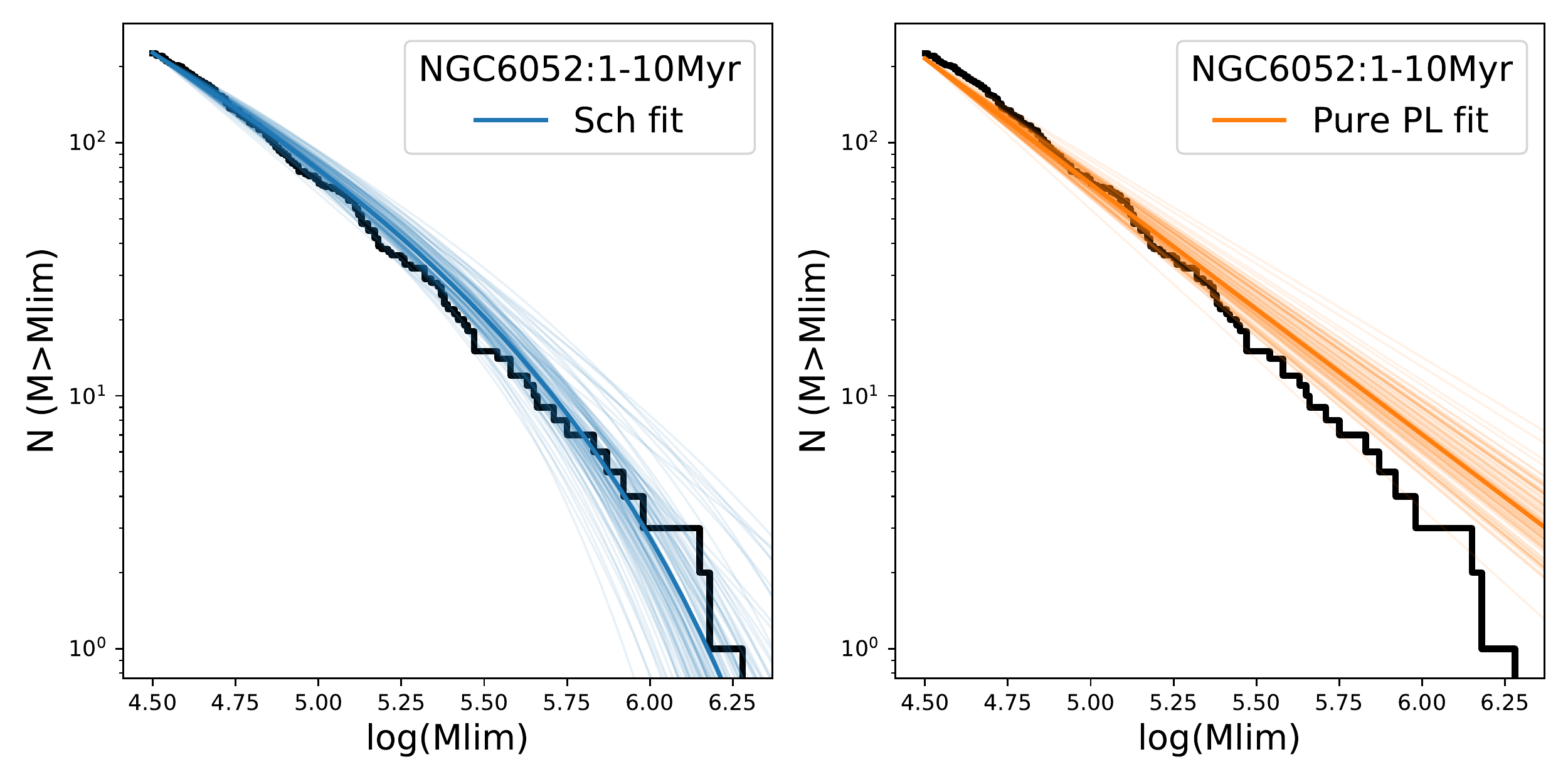}\\
    
    \caption{Observed cumulative cluster mass distributions (solid black line) of the HiPEEC sample in the age range 1--10 Myr. The blue coloured solid and thinned associated lines show the best values and the family of solutions for M$_c$ and $\beta_{\rm Sch}$ (left panel, blue lines) contained within the 1$\sigma$ contours of the corner plots in Figure~\ref{fig:cornerplots_app}. Similarly we plot in the right panel the predicted cumulative mass distributions (orange lines) for the best value of $\beta_{\rm PL}$ and the associated 1$\sigma$. The predicted distributions contain the same number of clusters used to build the observed mass distributions. See Figure~\ref{fig:MF} for the same analysis but performed on clusters in the age range 1--100 Myr. We refer to Section~\ref{sec:massfunc} for a detailed discussion of these results.}
    \label{fig:MF_app}
\end{figure*}

\begin{figure*}
    \centering
    \includegraphics[width=0.3\textwidth]{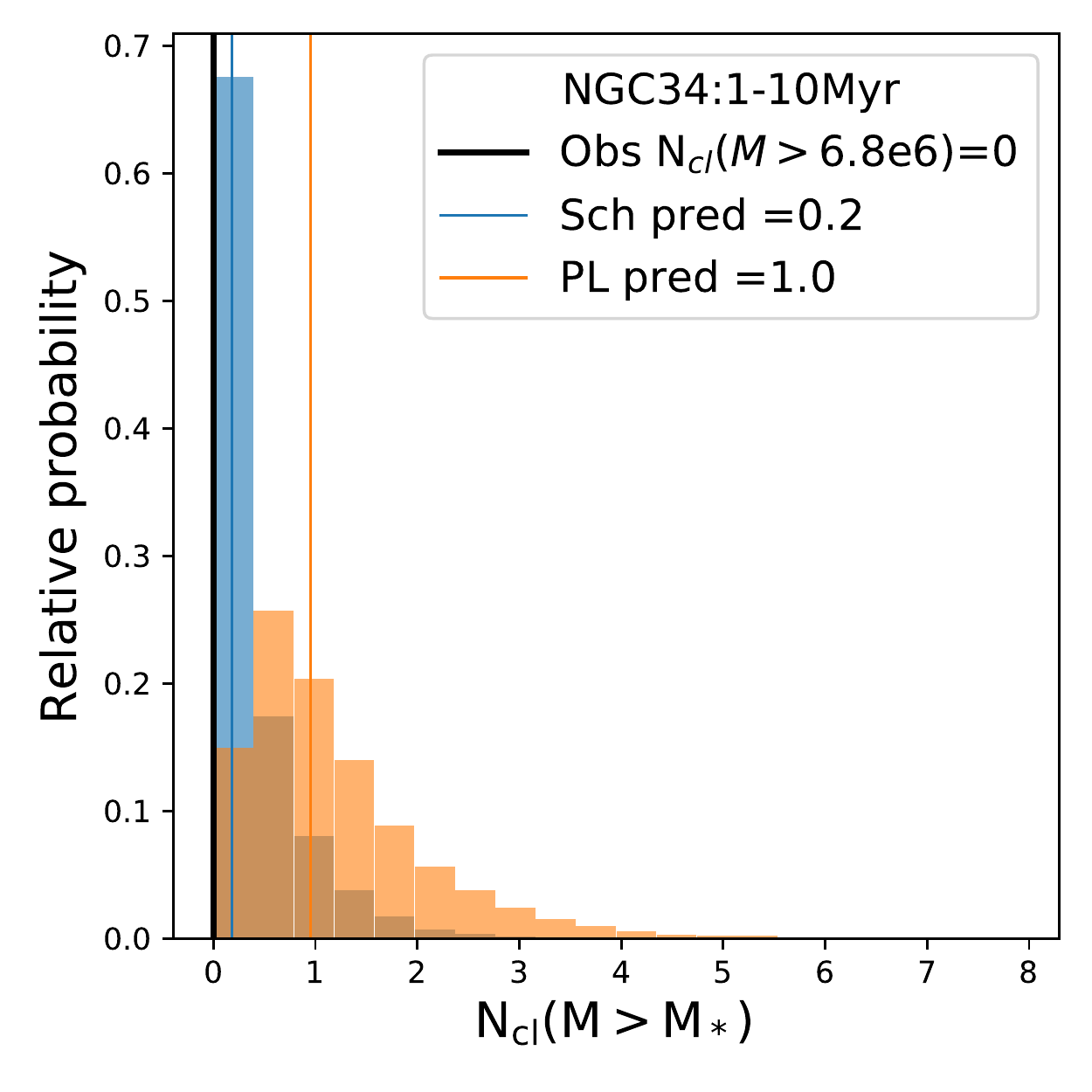}
    \includegraphics[width=0.3\textwidth]{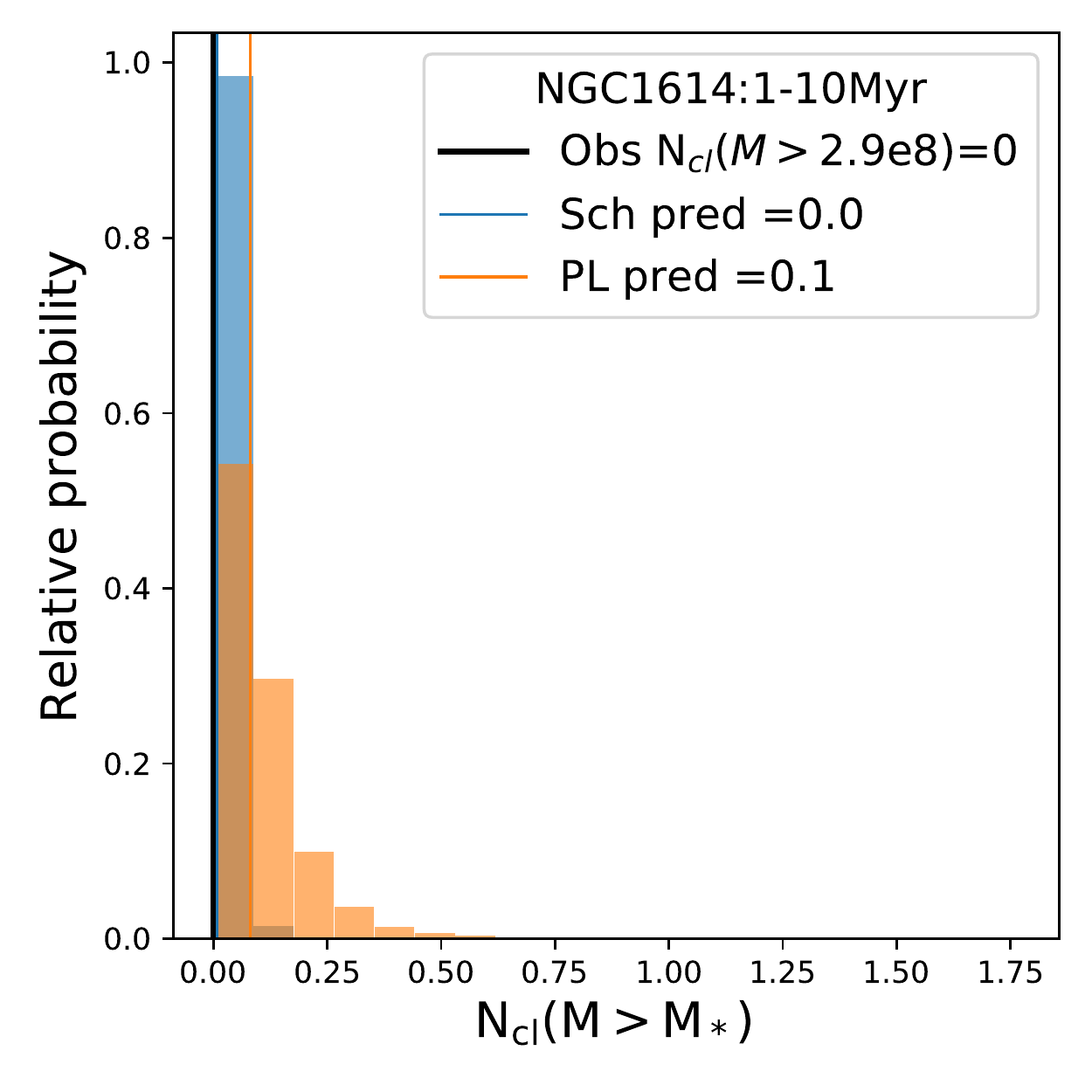}
    \includegraphics[width=0.3\textwidth]{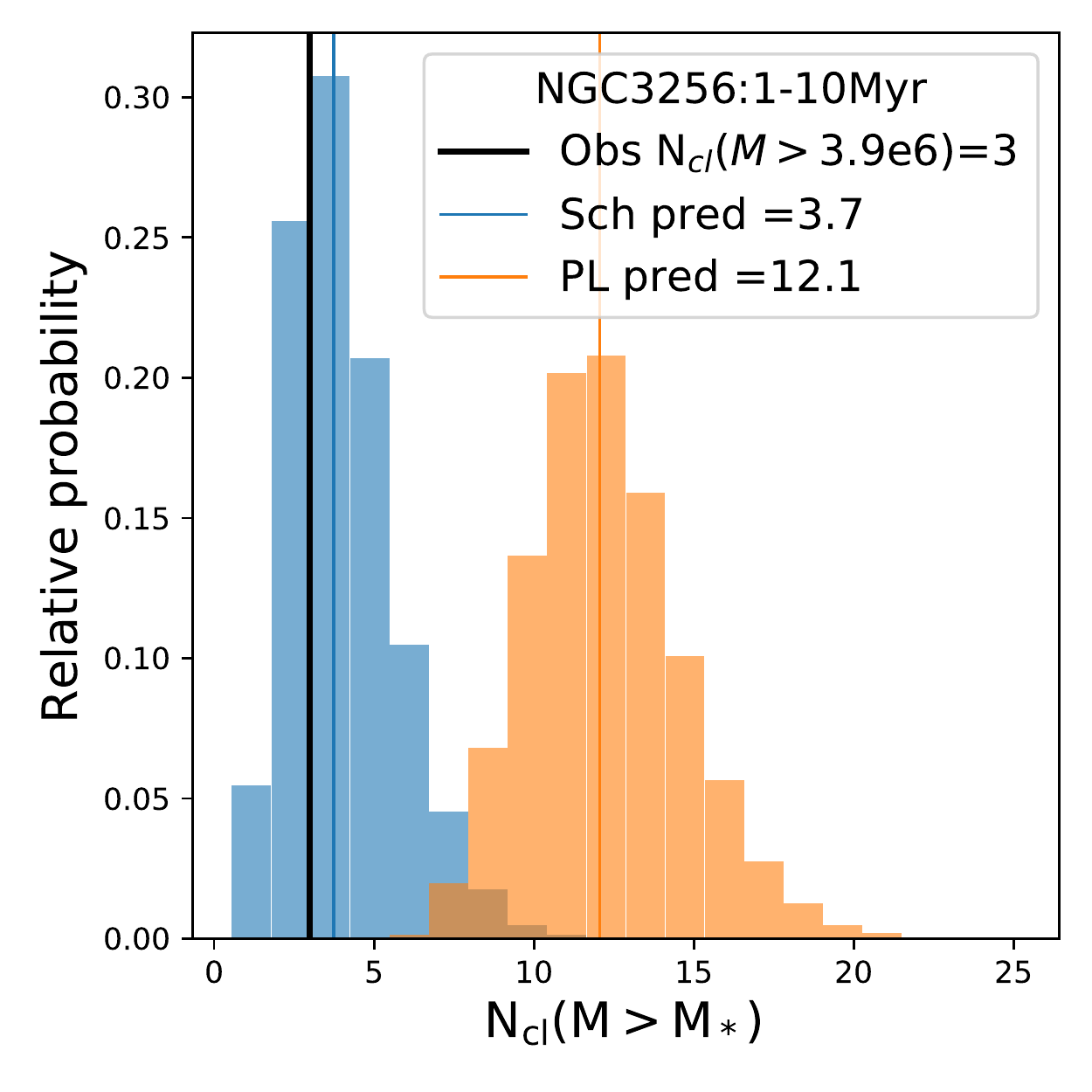}\\
    \includegraphics[width=0.3\textwidth]{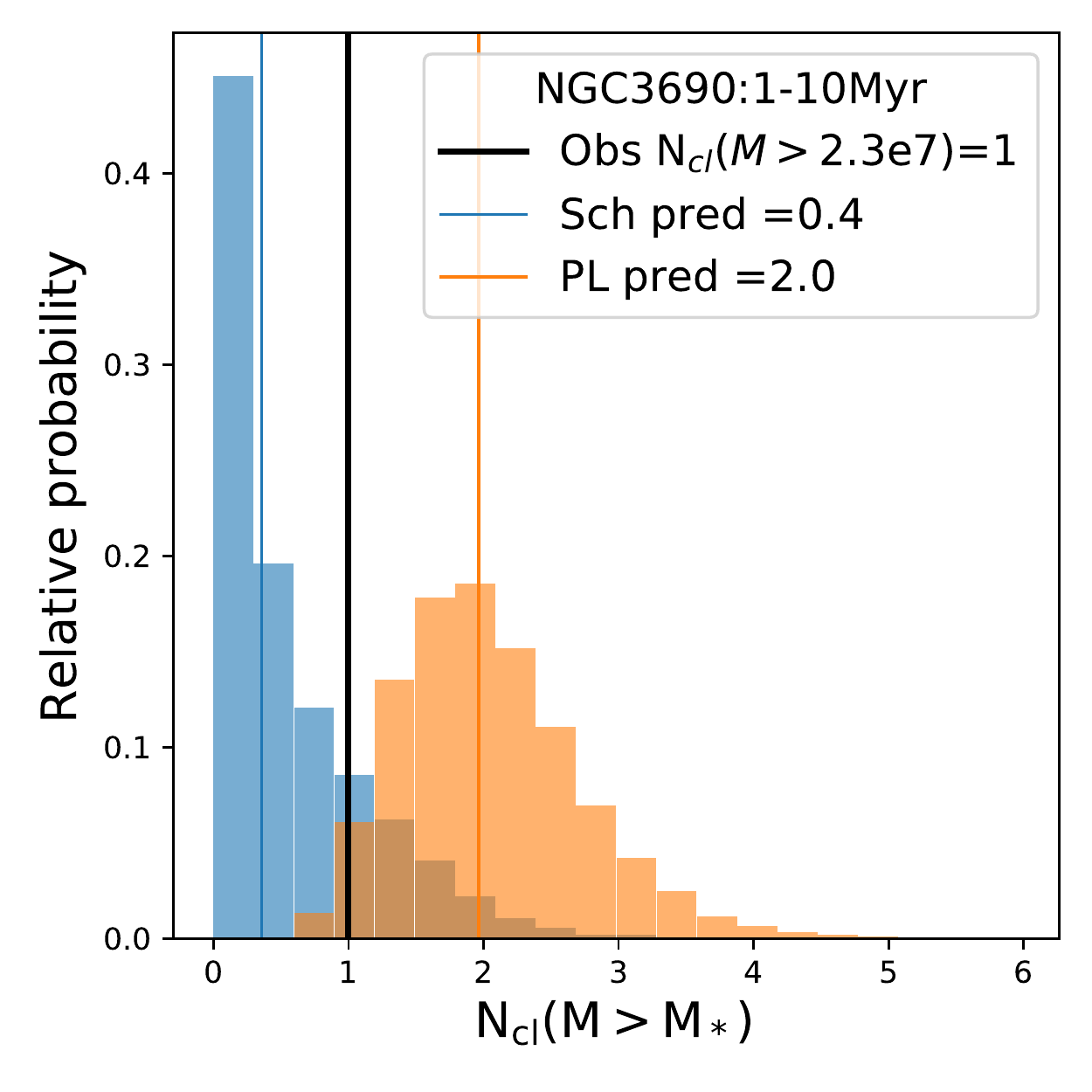}
    \includegraphics[width=0.3\textwidth]{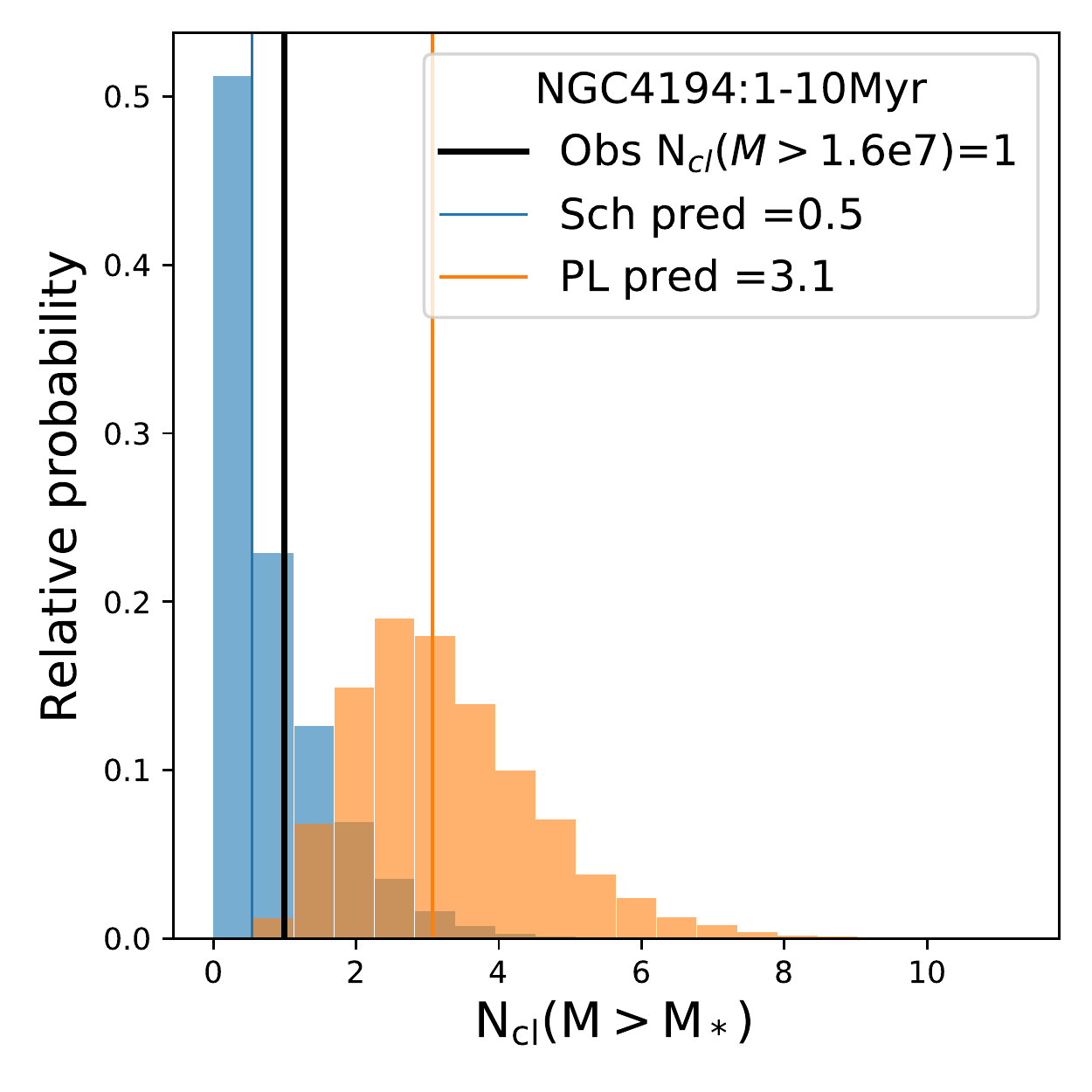}
    \includegraphics[width=0.3\textwidth]{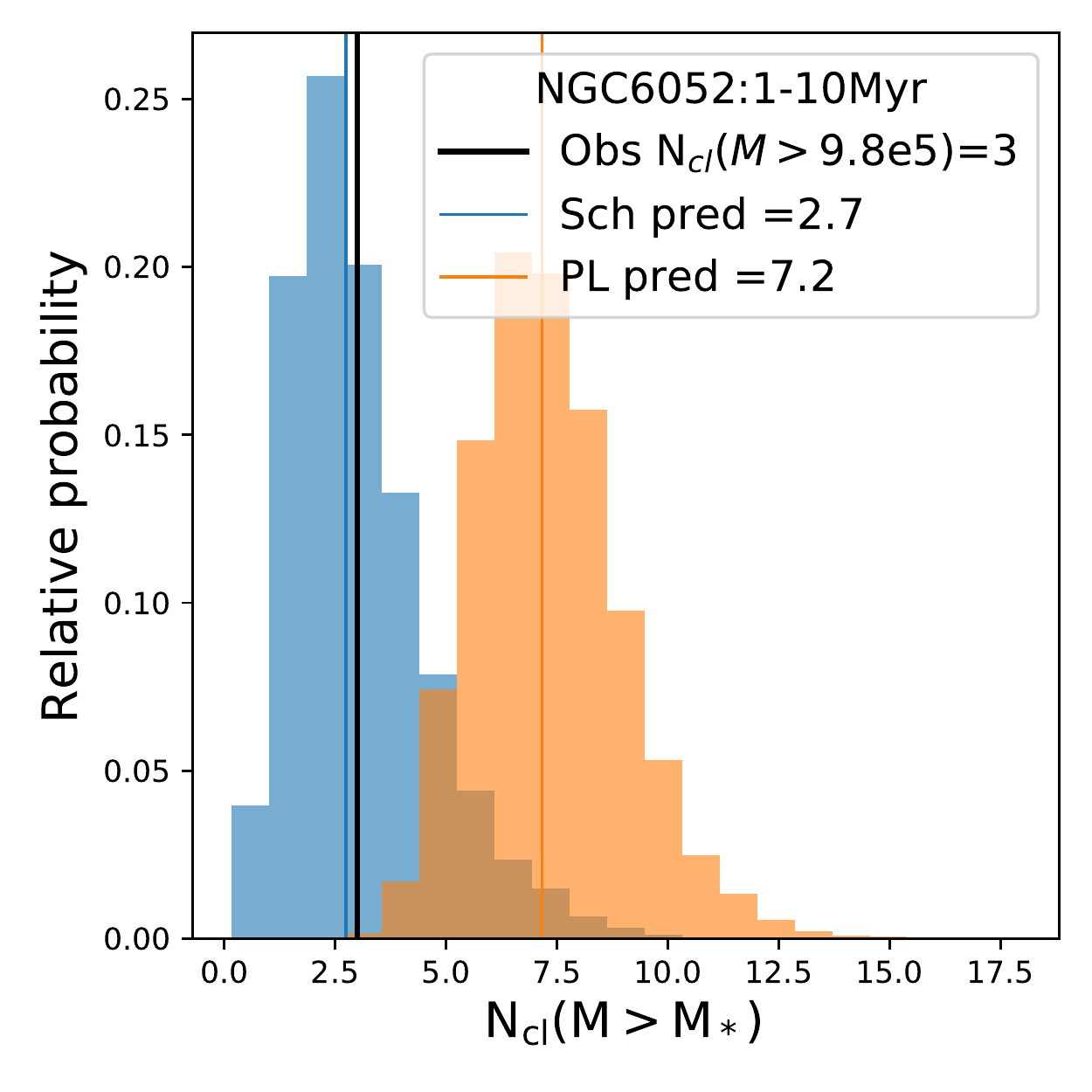}\\
    
    \caption{Normalised distributions of recovered number of clusters more massive than the determined M${_*}$ in each galaxy. The blue and orange histograms show the recovered number of clusters more massive than M${_*}$ in 1000 monte carlo runs of cluster populations with mass distributions described by the fitted Schechter and power-law mass functions, respectively. The median of the two distributions are indicated by vertical lines and their numbers included in the insets. Observed number of clusters more massive than M${_*}$ are shown as a black vertical line and indicated in the inset for each galaxy. See Figure~\ref{fig:predMF} for the same analysis but performed on clusters in the age range 1--100 Myr. We refer to Section~\ref{sec:massfunc} for a detailed discussion of these results.}
    \label{fig:predMF_app}
\end{figure*}


\bsp	
\label{lastpage}
\end{document}